\pdfoutput=1
 \documentclass[twocolumn,showpacs,preprintnumbers,amsmath,amssymb]{revtex4}
 
 \usepackage{graphics}
 \usepackage[dvips]{epsfig,rotating}
 \usepackage{here}
 \usepackage{mathrsfs}
 \usepackage{graphicx}

 \newcommand{\tpx}{\tilde{\varphi}_x}
 \newcommand{\tpy}{\tilde{\varphi}_y}
 \newcommand{\tax}{\tilde{a}_x}
 \newcommand{\tay}{\tilde{a}_y}
 \newcommand{\hax}{\hat{a}_x}
 \newcommand{\hay}{\hat{a}_y}
 \newcommand{\ya}{\mathscr{Y}}

 \newcommand{\Dx}{\mathscr{D}_x}
 \newcommand{\Dy}{\mathscr{D}_y}

 \newcommand{\ex}{\mathpzc{x}}
 \newcommand{\ey}{\mathpzc{y}}
 \newcommand{\epx}{\mathpzc{x}_\mathpzc{p}}
 \newcommand{\epy}{\mathpzc{y}_\mathpzc{p}}

\DeclareMathAlphabet{\mathpzc}{OT1}{pzc}{m}{it}

\begin{document}
 
 \title{Fix-lines and stability domain in the vicinity of the coupled
 third order resonance}
 
 \author{G.~Franchetti$^1$, F. Schmidt$^2$}
 \affiliation{$^1$GSI Darmstadt, Planckstrasse 1, 64291 Darmstadt, Germany}
 \affiliation{$^2$CERN CH-1211, Geneva 23, Switzerland}
 \date{\today}
 
 \begin{abstract}
 The single particle stability in a circular accelerator is of concern
 especially for operational regimes involving beam storage of hours.
 In the proximity to a resonance this stability domain shrinks, and
 the phase space fragments into a jungle of exotic objects like for
 instance ``fix-lines''. The concept of fix-points is easily
 understandable in a 2D phase space. It becomes quite challenging when
 the effect of resonances is considered in the 4D phase space, which
 leads then to the concept of fix-lines.  In this paper we investigate
 the fix-lines in the proximity of a coupled third order resonance and
 find the relation of these objects with the stability of motion. 
 \end{abstract}

 \maketitle

 %%%%%%%%%%%%%%%%%%%%%%%%%%%%%%%%%%%%%%%%%%%%%%%%%%%%%%%%%%%%%%%%%%
 \section{Introduction}
 The stability of particle motion in circular accelerators has always
 been a topic of heated discussion in the accelerator community. In
 fact, the problem of localized magnet non-linearities distributed
 around the accelerator has invoked its own field of nonlinear
 dynamics studies. The most complex consequences of these
 non-linearities are detuning with amplitude, excitation of
 resonances, chaotic motion and eventually particle
 loss~\cite{Chirikov79,lichtenberg}.

 The theory of resonances was developed by
 Hagedorn/Schoch~\cite{Hagedorn-I,Hagedorn-II,Schoch} in the 1950's
 and modified for CERN Accelerator School classes by Guignard in the
 1970's~\cite{guign1,guign2}. This theory treats the condition under
 which the motion of a particle is resonant. The treatment of motion
 is 2D, and the theory shows that the resonant phenomena can be
 described by a ``resonant driving term'', which is a quantity that
 takes into account the distribution of the multipolar errors along
 the ring and characterizes the resonance stop-band.

 The interplay of the stability domain and resonances is a very
 difficult subject and there seems to be no general theory that
 explains it completely. However, simulations show that in the
 proximity to a resonance the regime with stable motion typically
 shrinks, and the phase space fragments into regions with very
 different dynamical properties: from stable motion, to unstable, and
 chaotic motion~\cite{Chirikov79}.

 The difficulty of the dynamics in the vicinity of a resonance is
 remarkable even in regimes in which the motion do not exhibit
 chaoticity.  Concepts as fix-points easily understandable in a 2D
 phase space assume quite challenging aspects when the effect of
 resonances is considered in the 4D phase space.  For regular motion
 in 4D phase space, the fix-points become exotic objects that we call
 ``fix-lines'' \cite{Schmidt88}.  The structures created by the 2
 degree of freedom resonances have been subject of studies, which
 addressed the problem from the mapping approach
 \cite{Todesco94,VRAHATIS97}.

 Fix-lines become particularly relevant when the nonlinear dynamics is
 affected by space charge in bunched beams. The phenomena of periodic
 resonance crossing induced by synchrotron motion and space charge,
 create new and more complex dynamics which is very significant for
 present projects like the SIS100 of the FAIR project where bunches
 are stored for a long term in presence of significant space
 charge~\cite{FAIR}, or for the LIU project at CERN, where PSB, PS and
 SPS are upgraded in intensity and the same topology of high intensity
 problem is encountered~\cite{LIU}.  For 1D resonances the interplay
 of space charge with fix-points has been extensively studied,
 numerically and experimentally~\cite{sc_cern, sc_gsi}. On the other
 hand, studies on the interplay of space charge and 2D resonances has
 never be attempted because of the remarkable difficulty in
 characterizing stable resonances in 4D phase space (the fix-lines).

 In this paper we investigate the fix-lines in the proximity of a
 coupled third order resonance and find the relation of these objects
 with the stability properties of motion.  Hence we extend the theory
 as presented by Guignard \cite{guign2}.

 The plan of the paper is the following: In Sect. II we give an
 overview of the 2D resonances: this phenomenological analysis
 categorizes the resonances and discusses the main features of 1D and
 2D resonances. The concept of fix-line is introduced and examples are
 shown.  In Sect. III the phenomenology of the third order coupled
 resonances is being outlined. Fix-lines and tori cut are discussed
 using proper cuts in two dimensional planes.  Section IV describes
 the theory of particle motion in a constant focusing channel equipped
 with one sextupole, the theory is developed following a perturbative
 approach.  Section V deals with the problem of removing the time
 dependence from the Hamiltonian. We find that there are infinite
 canonical transformations able to remove the time dependence. This
 finding extends the theory of Schoch, and in fact it directly leads
 to a theory of the fix-lines (Sect. VI).  Section VII treats the
 issue of the stability of the motion around a fix-line, and presents
 exact formulas for the case of a single resonant term.  The analysis
 of stability allows to derive the secondary tunes of the motion
 around the fix-lines.  In Sect. VIII we discuss the stability of
 motion in the vicinity of the third order coupled resonance, and find
 with large generality a complete characterization of the stability of
 motion for the case of the dynamics dominated by one single resonant
 term.  Tests and comparison with tracking simulations are shown as
 well.

 In Sect. IX we extend the analysis to an AG structure equipped with
 many thin sextupoles. Following the corresponding arguments used for
 the constant focusing lattice, we directly retrieve a description of
 the dynamics in the proximity of a third order coupled resonance in
 terms of the resonance driving term as obtained from the theory of
 Schoch (for an arbitrary distribution of perturbative sextupolar
 errors). The comparison with multi-particle simulations is addressed
 as well. Lastly, in Section X we show that our theory allows to
 explain the results of the tori cut presented in Sect. II. Section XI
 is devoted to the conclusions and Sect. XII holds the
 acknowledgments.  In the appendix A (Sect. XIII) we discuss the
 perturbed solutions around the fix-lines, and in the appendix B
 (Sect. XIV) we elaborate in more details on mathematical aspects of
 the discussion in Sect. VIII.
 %

%%%%%%%%%%%%%%%%%%%%%%%%%%%%%%%%%%%%%%%%%%%%%%%%%%%%%%%%%%%%%%%%%%%%%%%%%%%%
 \section{Phenomenological analysis of resonance structures in 2
   degrees of freedom}
 %

%%%%%%%%%%%%%%%%%%%%%%%%%%%%%%%%%%%%%%%%%%%%%%%%%%%%%%%%%%%%%%%%%%%%%%%%%%%
 \subsection{Categories of resonances in the 4D phase space}
 Resonance structures in 1D of freedom have been studied in depth and
 will not be mentioned except for a schematic comparison with the 2D
 case.
 
\begin{figure}[H]
 \begin{center}
 \unitlength 0.9mm
 \begin{picture}(80,65)
 \put( -5, 0) {\epsfig{file=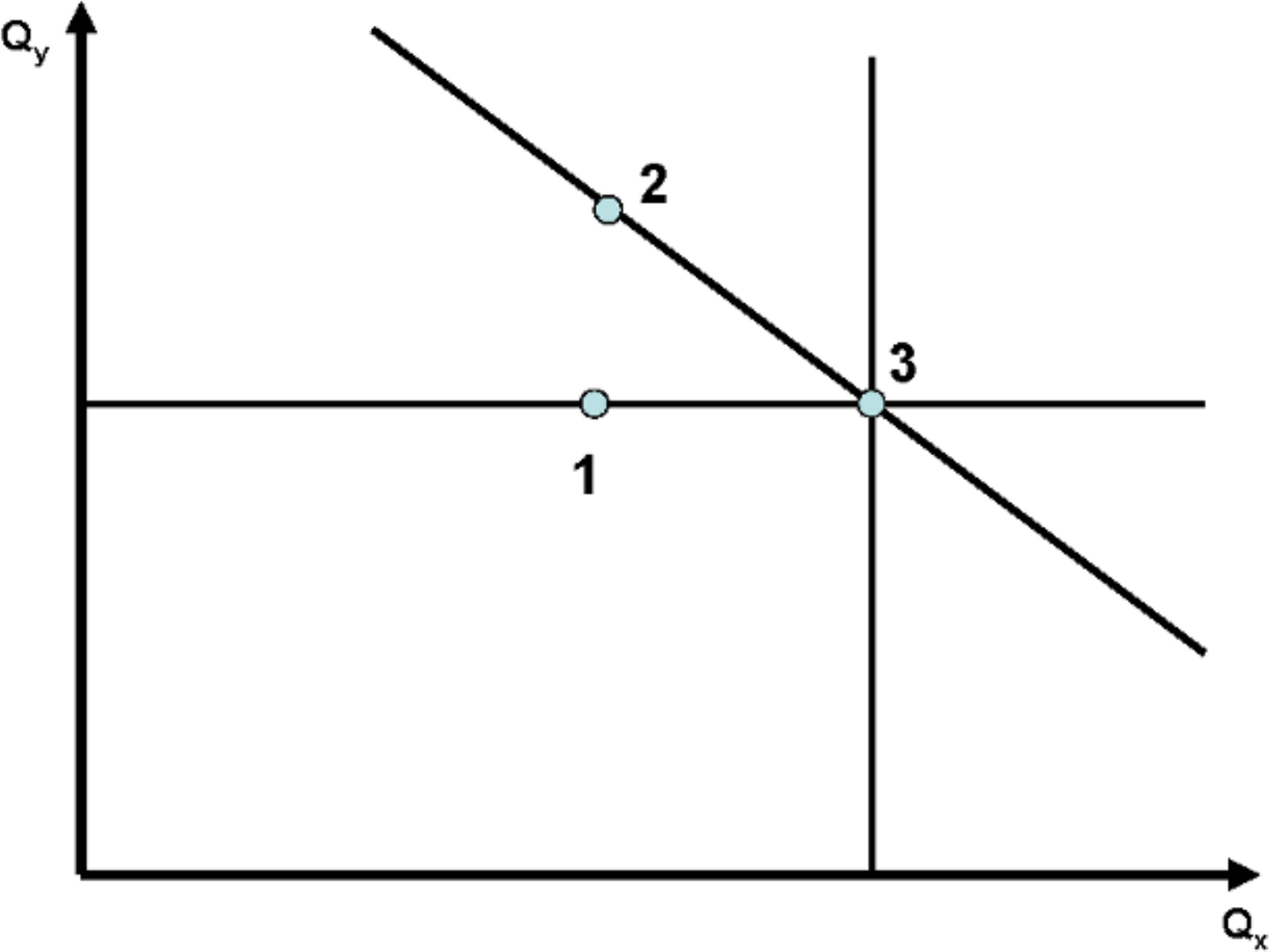,width=80mm}}
 \end{picture}
 \caption{Types of resonances in 4D phase space. 
 Resonance lines are typically presented in the $Q_x$-$Q_y$ diagram. 
 }
 \label{fig1:fig_1}
 \end{center}
 \end{figure}

 In 2 degrees of freedom we have basically 3 different types of
 resonances. In Fig.~\ref{fig1:fig_1} we find an example of a vertical
 1D resonance addressed with ``1''. In the vertical plane you do expect
 to see a resonance structure with distinct resonance islands, whilst
 in the horizontal plane the motion is non resonant. Therefore we will
 expect to find a number of resonances vertical islands that may be
 widened due to deformations in the horizontal plane. We will not
 follow this type of resonance any further since it is basically very
 similar to the classical and well-known 1D resonances.
 
 At the heart of this paper are the coupled resonance of type ``2'' in
 Fig.~\ref{fig1:fig_1}. The theoretical understanding of such
 coupled resonances including detuning with amplitude will be addressed
 in the following chapters. Here we want to show how they affect the
 motion in phase space.
 
 A peculiar type of resonance is depicted with ``3'' in
 Fig.~\ref{fig1:fig_1}. This is the 2D analogon of 1D fix-points,
 i.e. these fix-points are equivalent to the central fix-point at the
 origin of the 2D closed orbit if it is different from zero in any of
 the 4 coordinates. To some extent the motion in the vicinity of
 fix-points is better defined then close to a fix-line
 structure. The authors envisage a separate report just dedicated to
 those 2D fix-points.
 %
 %

%%%%%%%%%%%%%%%%%%%%%%%%%%%%%%%%%%%%%%%%%%%%%%%%%%%%%%%%%%%%%%%%%%%%%%%%%%%
 \subsection{Schematic description of 1 and 2 degree of freedom resonances}
\begin{figure}[H]
 \begin{center}
 \unitlength 0.9mm
 \begin{picture}(80,70)
 \put( -2, 0) {\epsfig{file=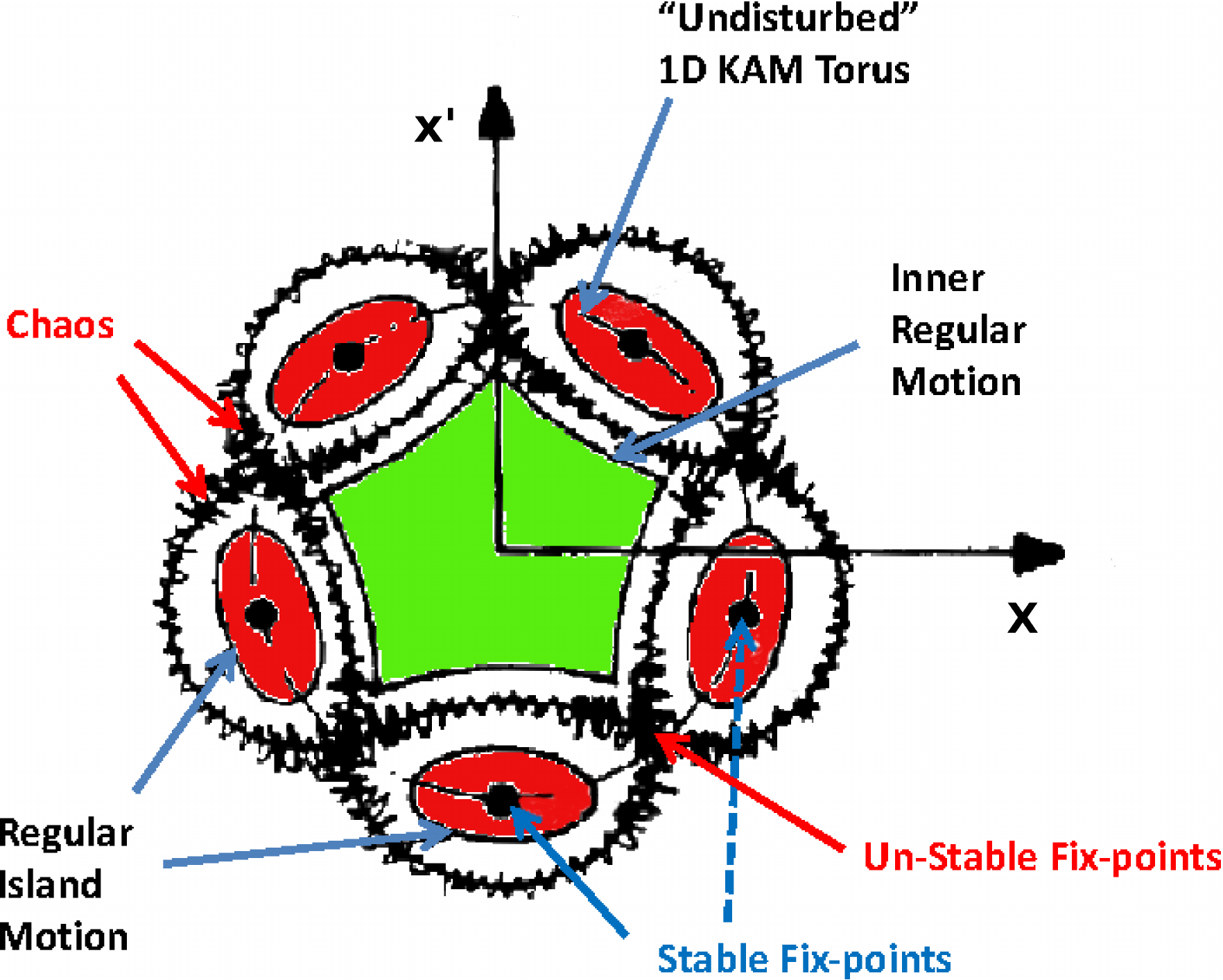,width=80mm}}
 \end{picture}
 \caption{Features of a 1D resonance stabilized through detuning with
   amplitude.}
 \label{fig1:fig_2}
 \end{center}
 \end{figure}

 Figure~\ref{fig1:fig_2} shows schematically how the 2D phase space
 looks for a stabilized resonance, i.e. the detuning with amplitude
 creates island instead of unbounded motion along hyperbolic lines to
 infinity. It is important to mention that the phase space in the plane
 of the resonance for the type ``1'' resonance in
 Fig.~\ref{fig1:fig_1} would look quite similar except that one will
 experience a widening of the projection of island motion that in the
 1D schematic depiction is a  measureless thin line around the
 fix-point.

 This classical island structure in 1D has mostly regular motion in
 the inner part at small amplitude to the central closed-orbit, which
 is the green area in the Fig.~\ref{fig1:fig_2}. For completeness
 one has to mention that careful inspection of the motion in that
 ``green'' area will reveal other fine island structures. Besides the
 regular motion around stable 1D fix-points, depicted as the red area in
 Fig.~\ref{fig1:fig_2} one finds deterministic chaotic motion in the
 vicinity of the unstable fix-points and the separatrix which separates
 the island motion from the ``green'' area at lower amplitude. At
 larger amplitude outside the island structure (for simplicity not
 shown in Fig.~\ref{fig1:fig_2}) one would expect regular motion up
 to a limit where motion becomes unbounded. This limit is called the
 dynamic aperture.
 
 Notice that there is a dashed circular line that connects the
 fix-points which is denoted by ``undisturbed'' KAM torus. By that we
 mean the undisturbed motion in absence of the non-linearities of the
 system.

 The last argument is a good link to resonances in 4D phase space, the
 real emphasis of this work. In 2 degrees of freedom the phase space
 is four dimensional. However, the motion of each individual particle
 is restricted to a two-dimensional manifold. In passing, we like to
 stress that the restriction to these two-dimensional manifolds is the
 very reason why there is a fundamental difference between motion in
 2D and 4D phase space respectively: in the case of 1 degree of
 freedom each regular particle motions is on a curved line that
 separates phase space into an inside and outside part so that no
 particle can cross-over from one part to the other. However, in 2
 degrees of freedom particles move on two-dimensional manifolds that
 cannot separate the 4D phase space into a  inside and outside part
 since that would require a three-dimensional manifold. In consequence
 this means that in 4D phase space all chaotic regions are linked
 through-out phase space such that there is no longer a strict concept
 of the dynamic aperture in systems larger than 1 degree of freedom. 
 
\begin{figure}[H]
 \begin{center}
 \unitlength 0.9mm
 \begin{picture}(80,180)
 \put( -10, 0) {\epsfig{file=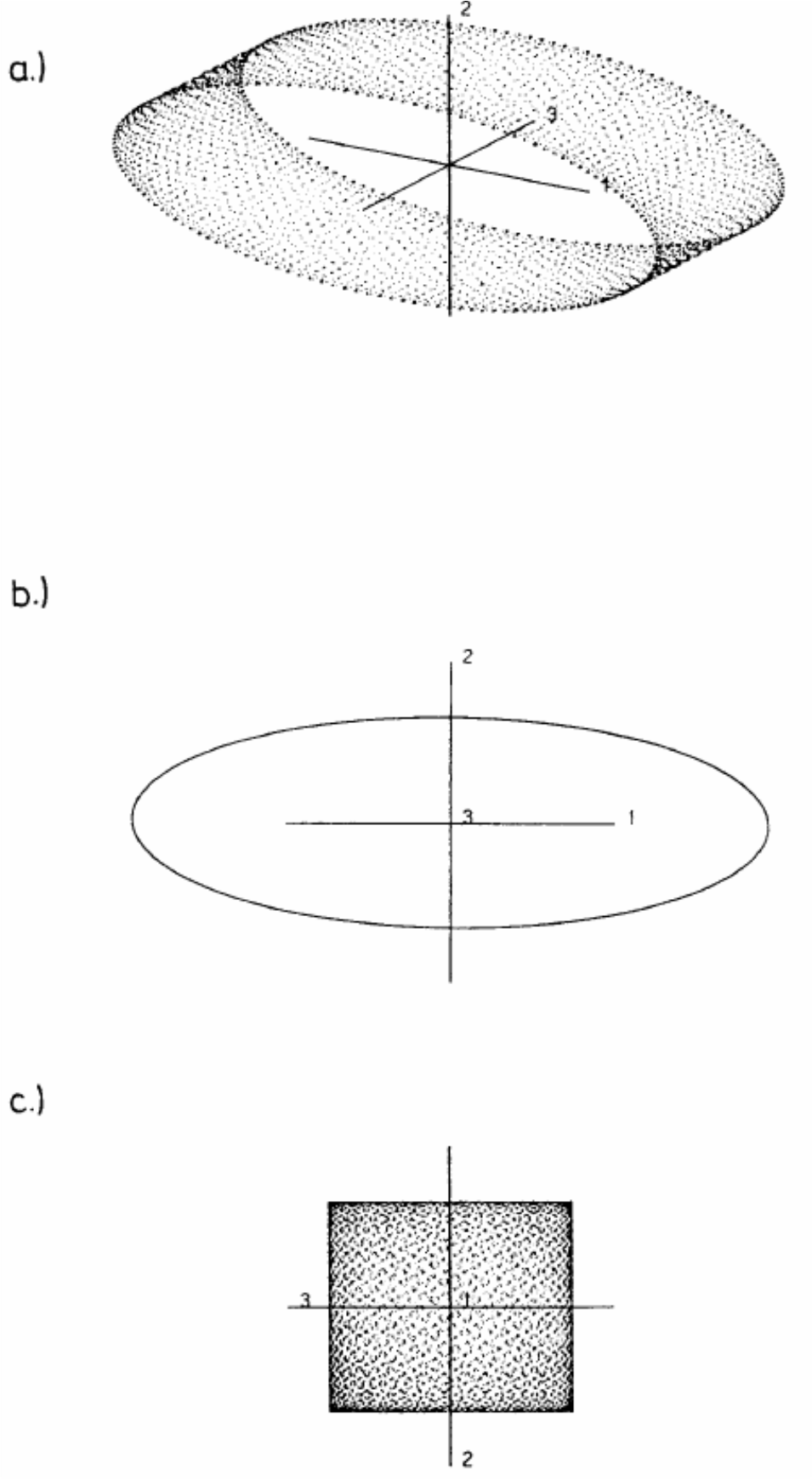,width=80mm}}
 \end{picture}
 \caption{Linear motion in 3D projections of the 4D phase space. The
   coordinate axis are denoted by 1, 2, 3 which stands for $x,x',y$ 
   respectively.}
 \label{fig1:fig_3}
 \end{center}
 \end{figure}

 While particle motion in 2D phase space is pretty much self-evident
 one needs some preparation to finds one's way in 4D phase space. To
 this end it is helpful to look at a 3D projection of linear and
 uncoupled motion in the 4D phase space. In Fig.~\ref{fig1:fig_3}
 the motion in phase space of $x, x', y$ is shown while $y'$ is still
 hidden. Part a) shows $x, x', y$ at some angle around the $x'$ axis: it
 becomes evident that the motion is on a torus. When looking straight
 from the horizontal viewpoint ($x, x'$ in part b) one finds a perfect
 ellipse without any thickness and for mixed coordinates ($y, x'$ in part
 c) one finds a rectangular shape. It goes without saying that the (
 $y,y'$) phase space is also an ellipse and that for all mixed coordinates
 one finds a rectangular shape.
 
\begin{figure}[H]
 \begin{center}
 \unitlength 0.9mm
 \begin{picture}(80,180)
 \put( -10, 0) {\epsfig{file=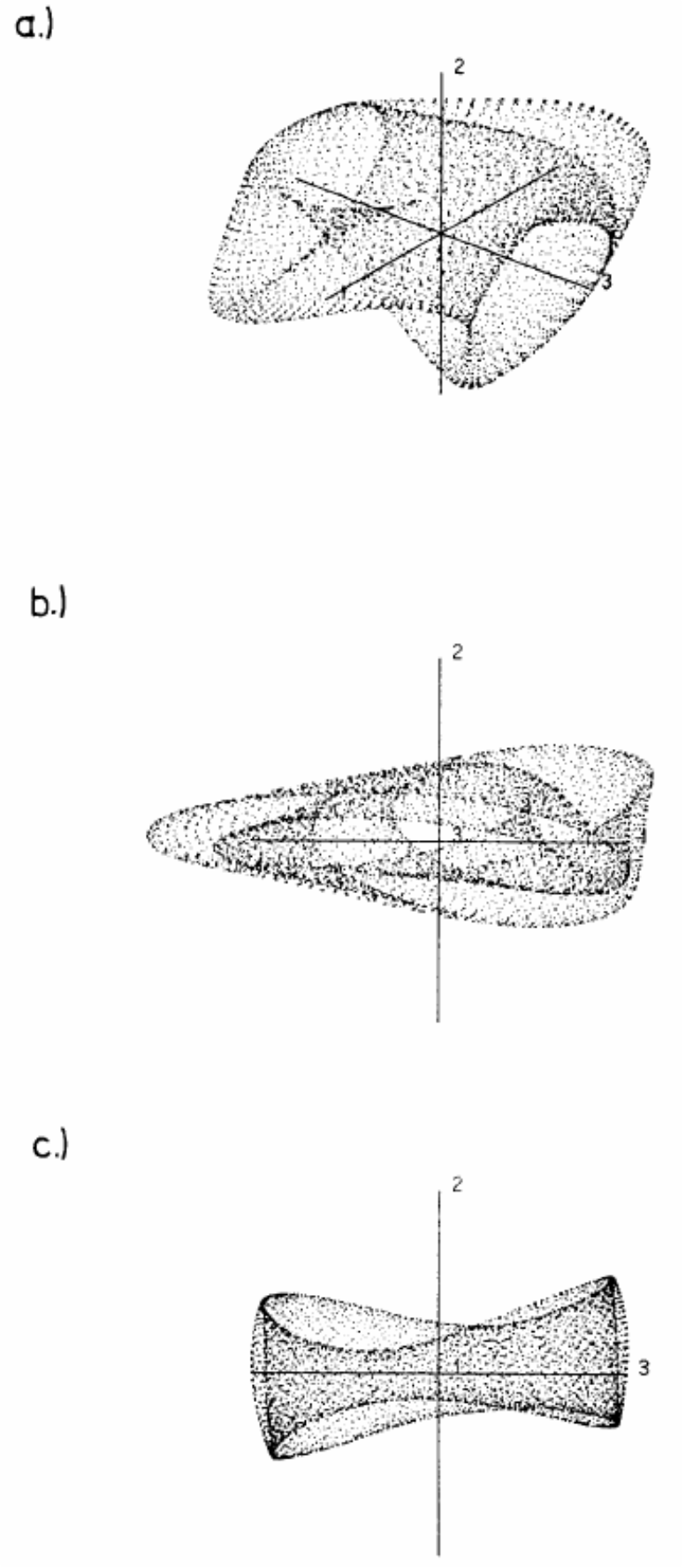,width=65mm}}
 \end{picture}
 \caption{Non-Linear and regular  motion in 3D projections of the 4D
   phase space. The coordinate axis are denoted by 1, 2, 3 which
   stands for $x,x',y$ respectively.}
 \label{fig1:fig_4}
 \end{center}
 \end{figure}

 It is important to realize that we are still lacking the $4^{th}$
 coordinate $y'$. A good way to describe this extra dimension is to
 understand that the particle moves on a double layered torus,
 i.e. two-dimensional manifolds in 4D. Introducing non-linearities
 tends to separate the layers of the torus so that very complex
 structures will be found in phase space. This becomes evident in
 Fig.~\ref{fig1:fig_4} which shows a case with strong
 non-linearities switched on: large amplitudes have been chosen but
 the motion remains regular and refined to a two-dimensional
 manifold. The rotation angle of part a) is chosen to make the double
 layer structure strikingly visible.
 
 Traditionally, there have been attempts to use other means E.G. a
 color code to allow for a full grasp of the dynamics in 4D. Set aside
 spectacular aesthetic pictures effectively this did little to
 substantially enhance our understanding of those dynamics. What
 really counts is that one can use cuts in phase space that take away
 overlapping pieces of the 2D manifold in the phase space
 projections.
 
\begin{figure}[H]
 \begin{center}
 \unitlength 0.9mm
 \begin{picture}(80,67)
 \put( -10, 0) {\epsfig{file=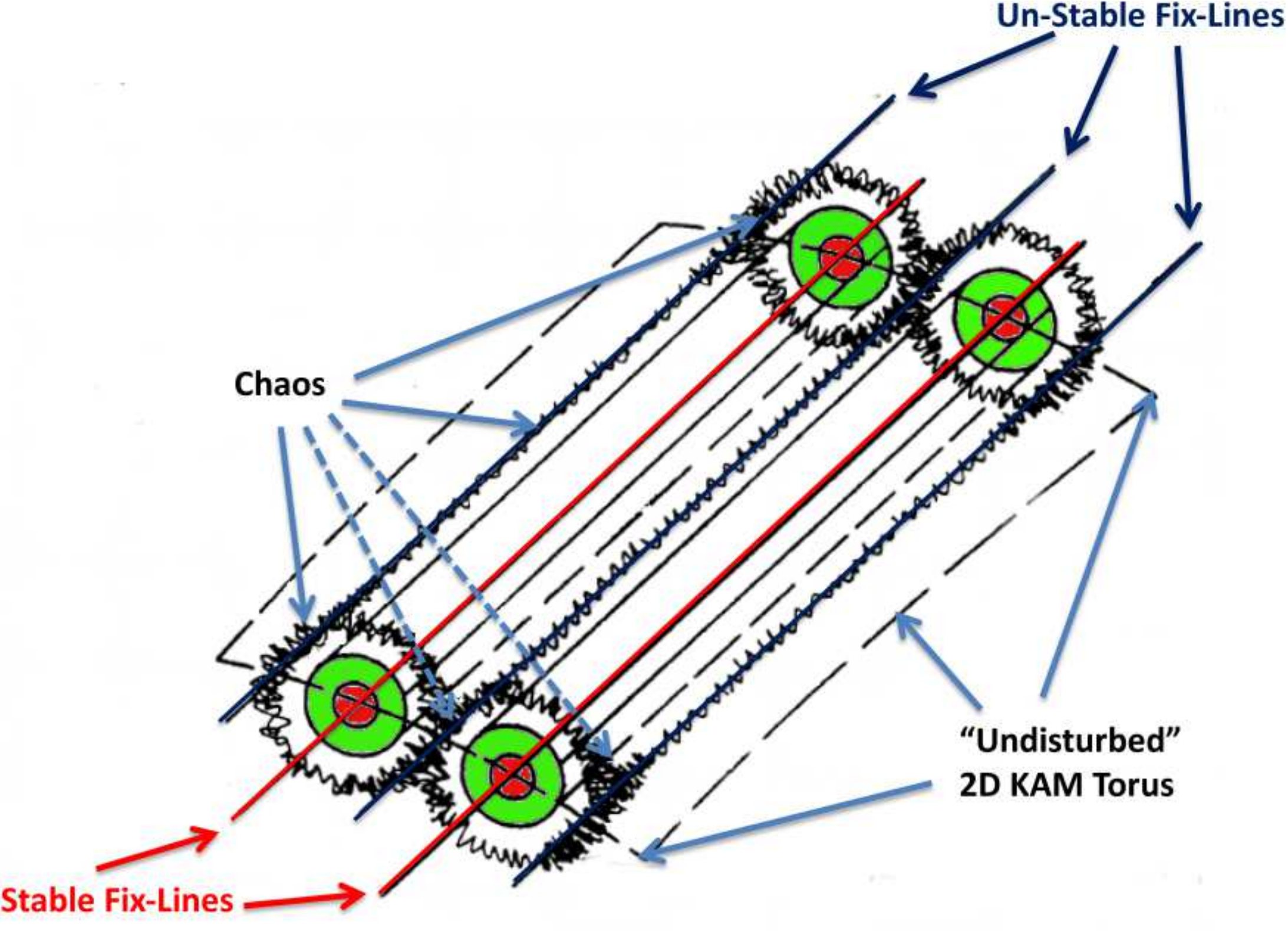,width=85mm}}
 \end{picture}
 \caption{Features of a 2D fix-line resonance.}
 \label{fig1:fig_5}
 \end{center}
 \end{figure}

 Schematically one can have a look at how the ``undisturbed'' 2D KAM
 torus in case of linear motion is broken up into a fix-line
 structures as shown in Fig.~\ref{fig1:fig_5}. One finds that of this
 torus we are left with a set of alternating stable and unstable
 fix-lines: in the vicinity of the stable fix-lines the particles move
 on tori around the fix-line (again a 2D manifold), while the motion
 at the unstable fix-lines and the limiting ``separatrix torus'' is
 of chaotic nature.
 
 The comparison of Fig.~\ref{fig1:fig_2} and
 Fig.~\ref{fig1:fig_5} shows how island structures in 2D are a
 generalization of those in 1D, albeit in 4D we have additional
 complexity. A more complete analogon to the 1D case is to be expected
 around 2D fix-points as shown below.
  
\begin{figure}[H]
 \begin{center}
   \unitlength 0.9mm
 \begin{picture}(80,50)
 \put( -5, 0) {\epsfig{file=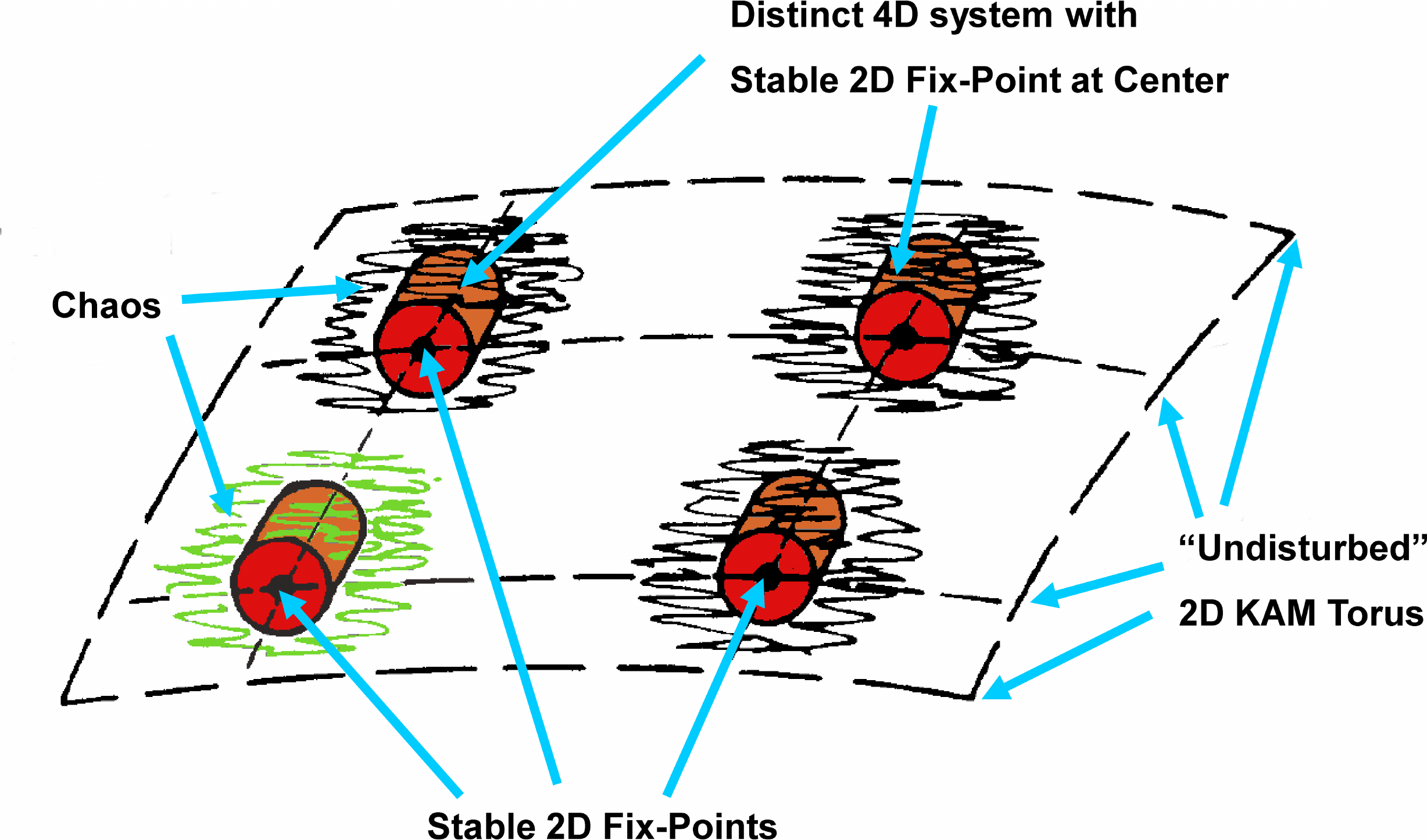,width=80mm}}
 \end{picture}
 \caption{Features of a 2D fix-point resonance.}
 \label{fig1:fig_6}
 \end{center}
 \end{figure}

 In case of a fix-line there is one condition fixed between the tunes
 therefore the stable 2D manifolds break up into 1D objects. In case
 there is a second condition fulfilled between the tunes these
 fix-lines are further broken up so that only fix-points, stable or
 unstable, remain. In Fig.~\ref{fig1:fig_6} it is schematically
 shown how the particle motion evolves around stable fix-points. In
 fact, one expects structures that resemble the motion around the
 central fix-point around the closed-orbit, i.e. regular motion in
 small distance to the fix-point and the set-in of chaos at larger
 amplitudes. In the figure the lower left system the evolution of the
 chaotic particles is colored in green for better visibility.

 %
 %\newpage
 %
 %
 %

%%%%%%%%%%%%%%%%%%%%%%%%%%%%%%%%%%%%%%%%%%%%%%%%%%%%%%%%%%%%%%%%%%%%%%%%%%%
 \subsection{Simulation examples of 2D fix-line and fix-point resonances}
\begin{figure}[H]
 \begin{center}
 \unitlength 0.7mm
 \begin{picture}(80,240)
 \put(-6,0){
 \put( -10, 160) {\epsfig{file=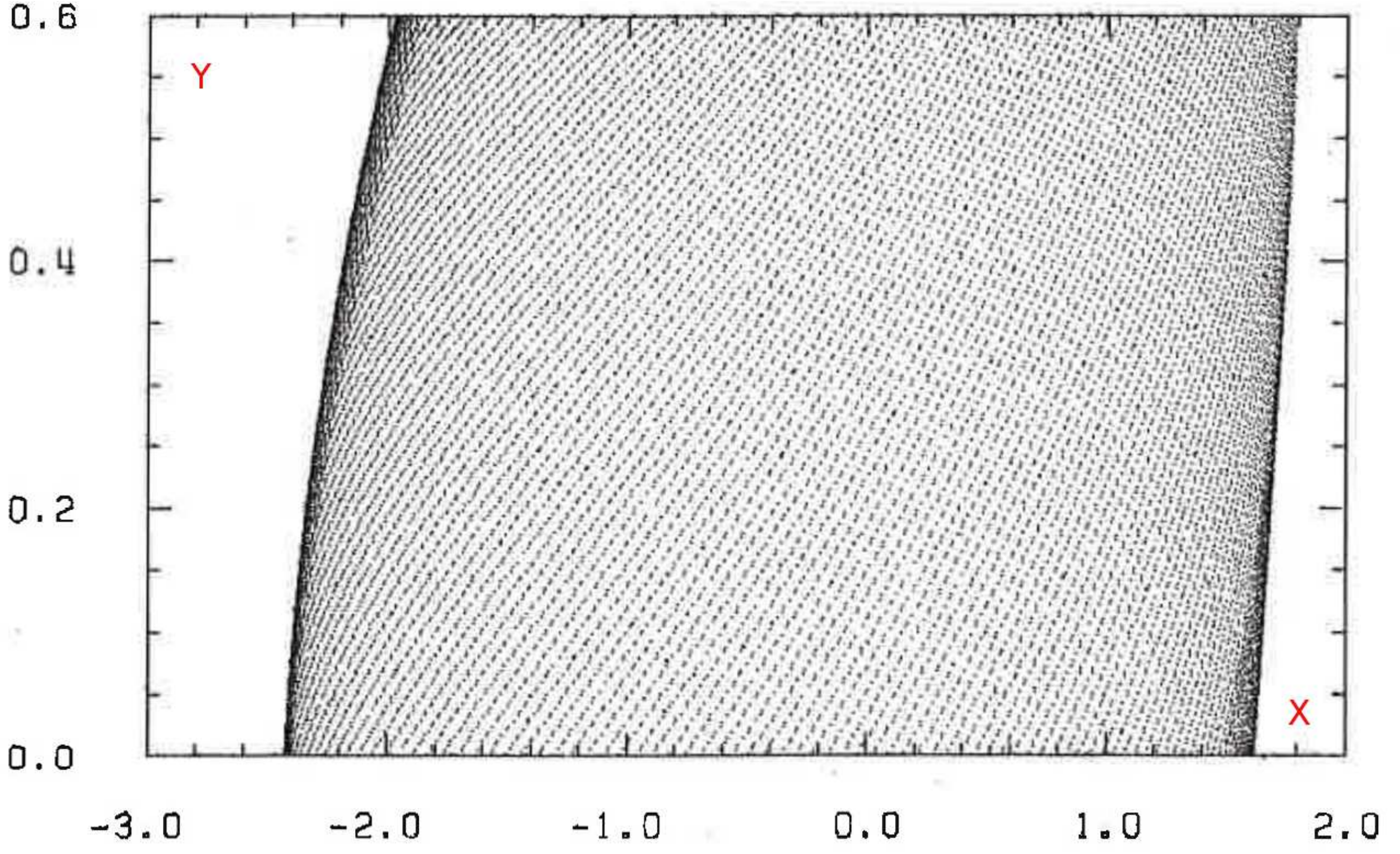,width=80mm}}
 \put( -10, 80)  {\epsfig{file=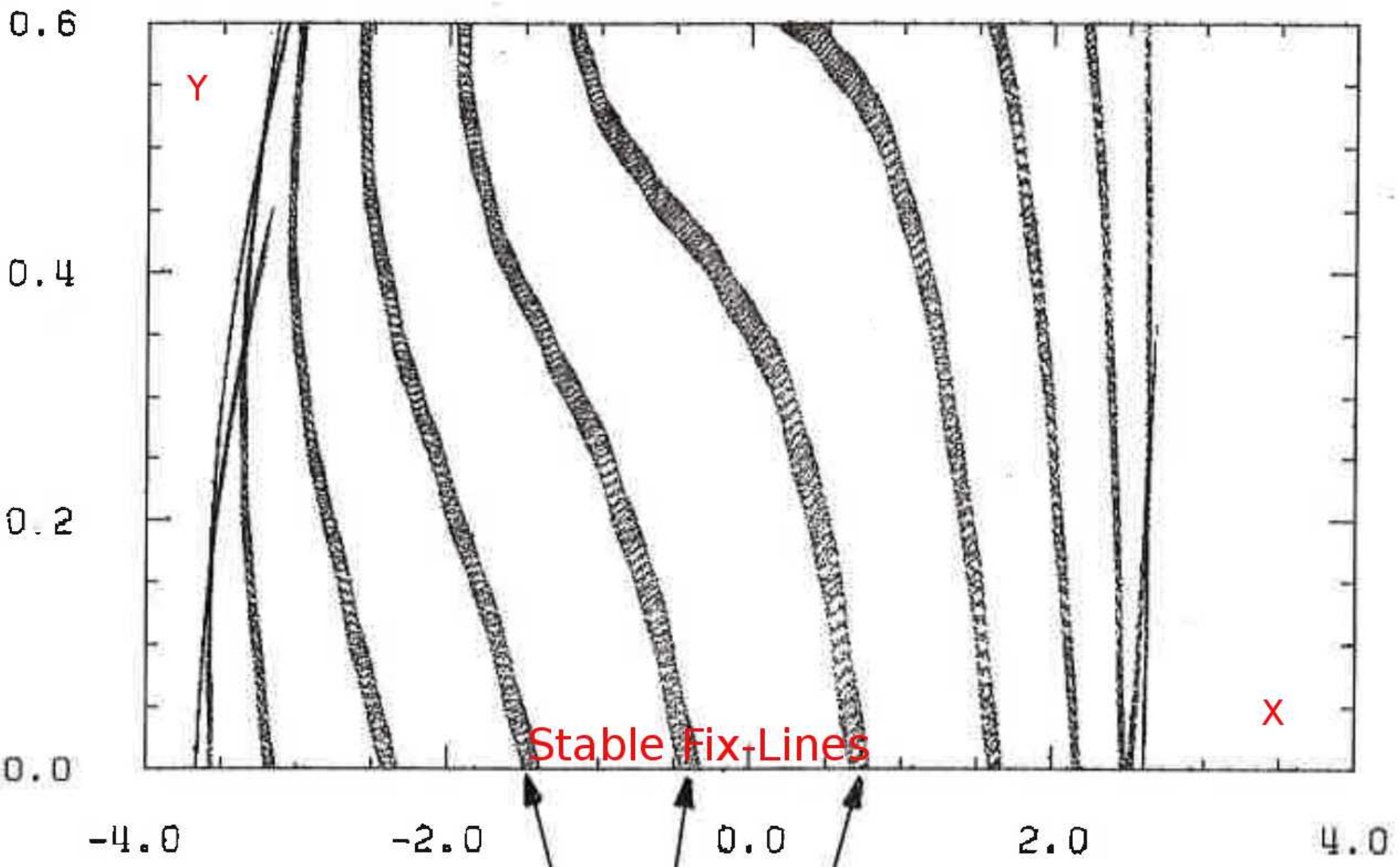,width=80mm}}
 \put( -10,  0)  {\epsfig{file=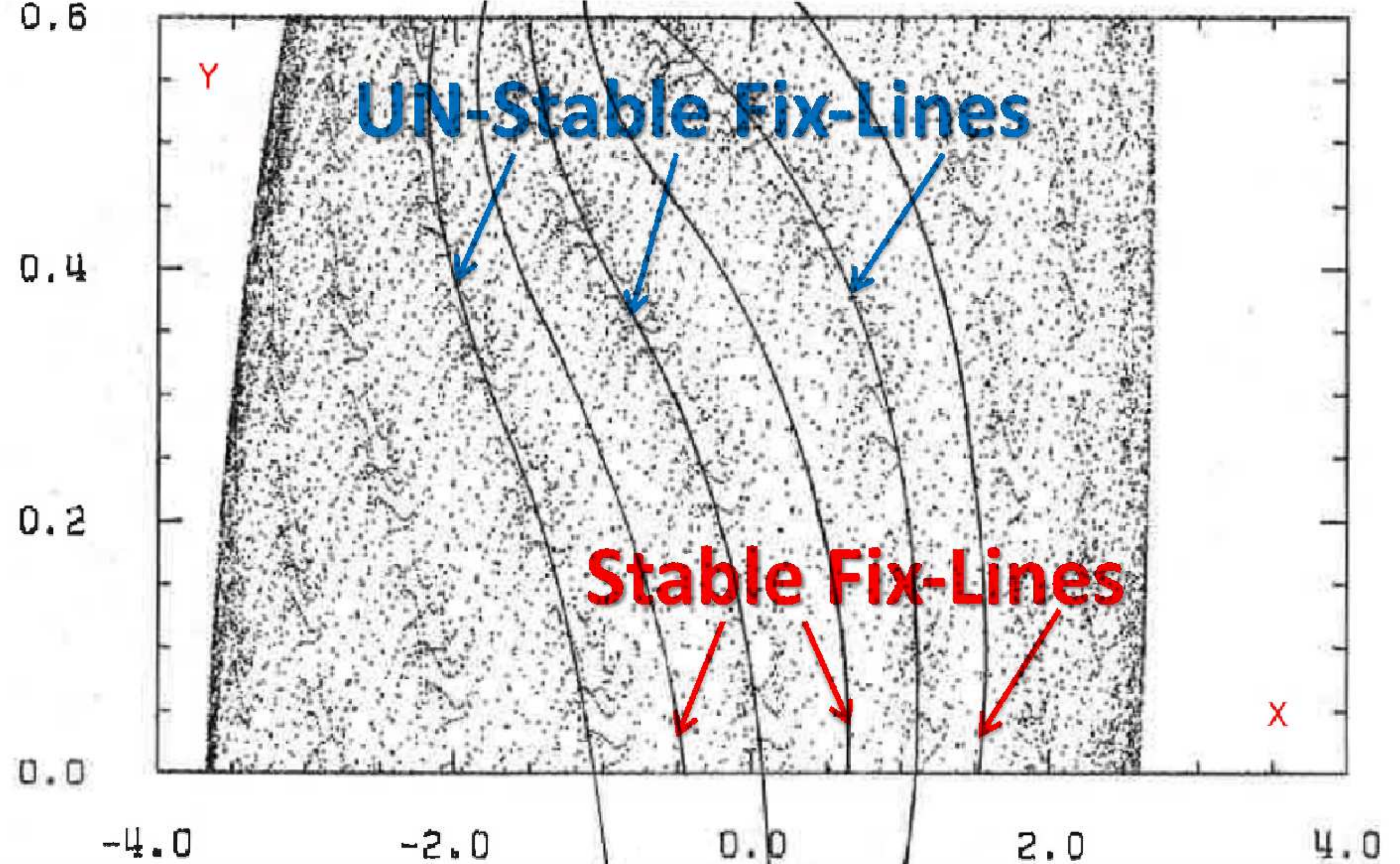,width=80mm}}
 \put( -10,235) {a)}
 \put( -10,155) {b)}
 \put( -10,75)  {c)}
           }
 \end{picture}
 \caption{2D fix-line phase space cut-out.}
 \label{fig1:fig_7}
 \end{center}
 \end{figure}
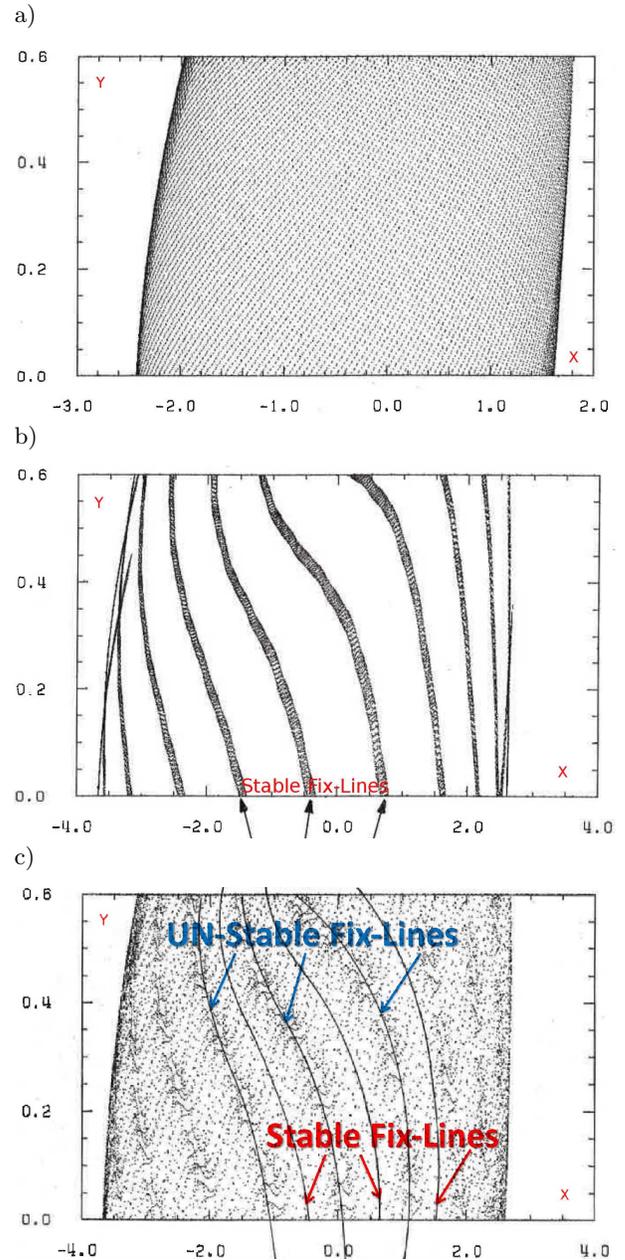

 In Fig.\ref{fig1:fig_3} and Fig.\ref{fig1:fig_4} we have
 already shown simulation examples for linear and non-linear motion. As
 explained above we are applying cuts in phase space such that one can
 inspect the two-dimensional manifold unobstructed by projections
 of other pieces of this manifold. Fig.~\ref{fig1:fig_7} part
 a) depicts a regular and dense coverage of such a manifold.

 In part b) the amplitudes have been varied in such a way that the
 tunes are on resonance and the motion is restricted to a torus around
 and with a small distance to the stable fix-lines. When the distance
 to the stable fix-lines is increased the motion becomes chaotic and
 the unstable fix-lines become visible. In Fig.~\ref{fig1:fig_7}
 part c) both the stable and unstable fix-lines are indicated with
 red and blue arrows respectively. The two types of fix-lines are
 interleaved and the motion near the unstable fix-lines exhibits the
 typical large chaotic variations. It should be mentioned that the motion of
 part c) encloses that of part b).

 Lastly, we are also presenting an example of particle motion in the
 vicinity of 2D fix-points. In Fig.~\ref{fig1:fig_8} part a) close to
 the fix-points the motion type is regular. When the distance to the
 stable 2D fix-points is increased, the motion will exhibit chaotic
 behavior which is the case in Fig.~\ref{fig1:fig_8} part b) where one
 finds fuzzy images due to the chaoticity. In part a) the motion
 around one of the fixed point is circled, while the motion in part b)
 is around the same fix-point. It is interesting to demonstrate that
 chaotic motion of part b) encloses the stable motion of part a). To
 this end these two trajectories are drawn on top of each other and
 turned in the three dimensional phase space projections x, x’, y so
 that one can see that both motion types evolve around the same
 fix-point (see Fig.~\ref{fig1:fig_8} part c) ): one can nicely find
 regular motion close to the fix-point and fuzzy, i.e. chaotic motion
 at larger amplitude. It has to be mentioned that this holds true in
 all 3D projections, albeit without a visual demonstration shown here.
 %

%%%%%%%%%%%%%%%%%%%%%%%%%%%%%%%%%%%%%%%%%%%%%%%%%%%%%%%%%%%%%%%%%%%%%%%%%%
 \subsection{Locking to a 2D fix-line structure}
 In the analysis of motion in the vicinity of resonances it is
 important to consider a balance of resonance driving terms and the
 detuning with amplitudes. Often, too simple models do not really
 describe what will be found in well designed accelerators. Designers
 go to great pains to minimize driving terms and choose tune working
 points with optimal stability, i.e. large dynamic aperture.
 
 To this end we are using an example with a sizable number of
 FODO structures each with sextupoles with an average non-zero value 
 and some random component. This set-up can be seen as a reasonable good
 approximation for a realistic accelerator.
 
 To study the effect of the sextupole coupling resonance $Q_x + 2Q_y$
 the bare tunes are set close to that resonance. We have tested 11 runs
 incremented by a fixed 4D coordinate vector.
 
 In Fig.~\ref{fig1:fig_9} part a) the phase advance per
 turn is shown over several thousand turns. Despite the apparent large
 variations of the phase advance one also finds the tunes of each case
 and a line connecting the tunes. Due to the specific detuning with
 amplitude the resonance line $Q_x + 2Q_y$ (shown in the figure as well)
 is being approached until at step ``8'' when the motion is locked to
 the resonance. With larger amplitudes in this 2D case we are finding a
 phenomenon not known in 1D: for the next 3 steps the tunes are
 continuing to move but remaining locked to the resonance. It is
 important to understand that at all amplitudes in between case ``8''
 and ``11'' the motions remains locked to this resonance. Each point
 though on the resonance line is a fix-line system in itself,  as
 described in the previous sections: a fix-line at the center and
 motion around it while keeping identical horizontal and vertical
 tunes.
 In this study we have encountered rather accidentally particle
 ``10'' which fulfills almost perfectly the condition for a fix-line.

 Figure~\ref{fig1:fig_9} part b) shows a blow-up of this fix-line
 motion in the plane of the phases. It might be surprising at first
 sight that the particle is never actually on the resonance. The
 averaged phase advance, i.e. the tunes are indicated by the blue star
 in the figure. The tunes clearly never coincides with the individual
 phase advance values per turn (shown as red dots).

 \begin{figure}[H]
 \begin{center}
 \unitlength 0.7mm
 \begin{picture}(80,315)
 \put( -10, 230) {\epsfig{file=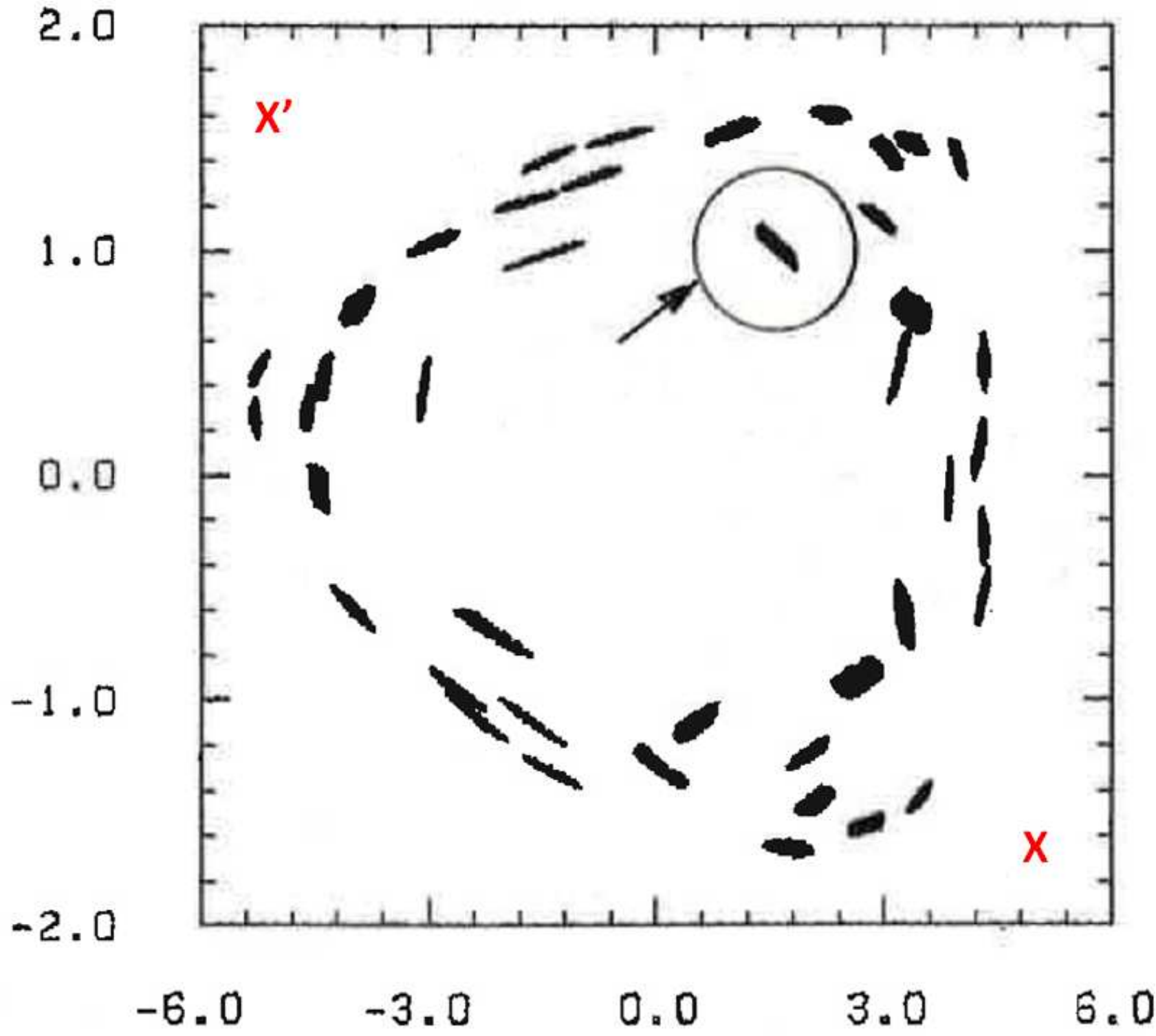,width=65mm}}
 \put( -10, 140) {\epsfig{file=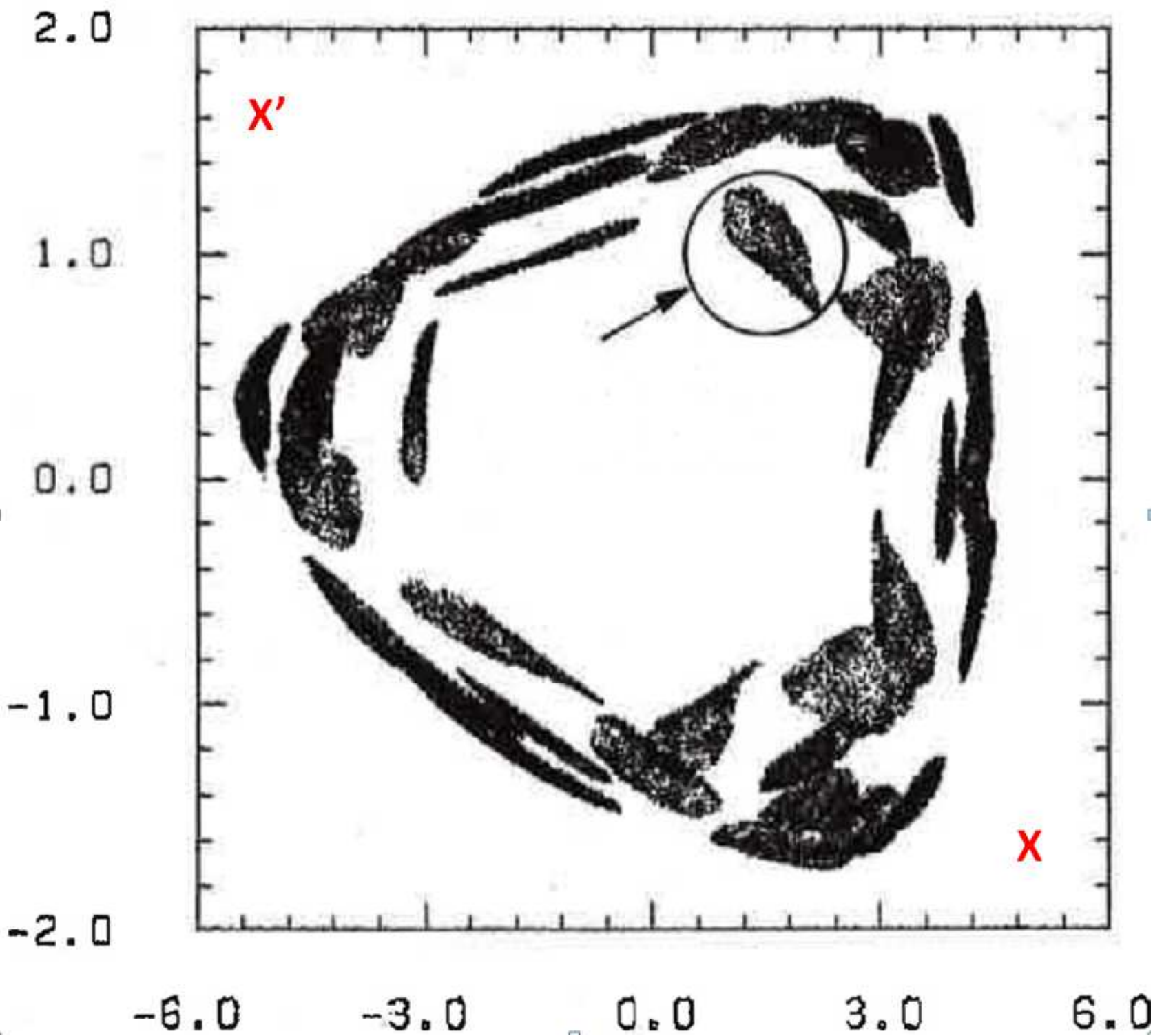,width=65mm}}
 \put( -10,  0)  {\epsfig{file=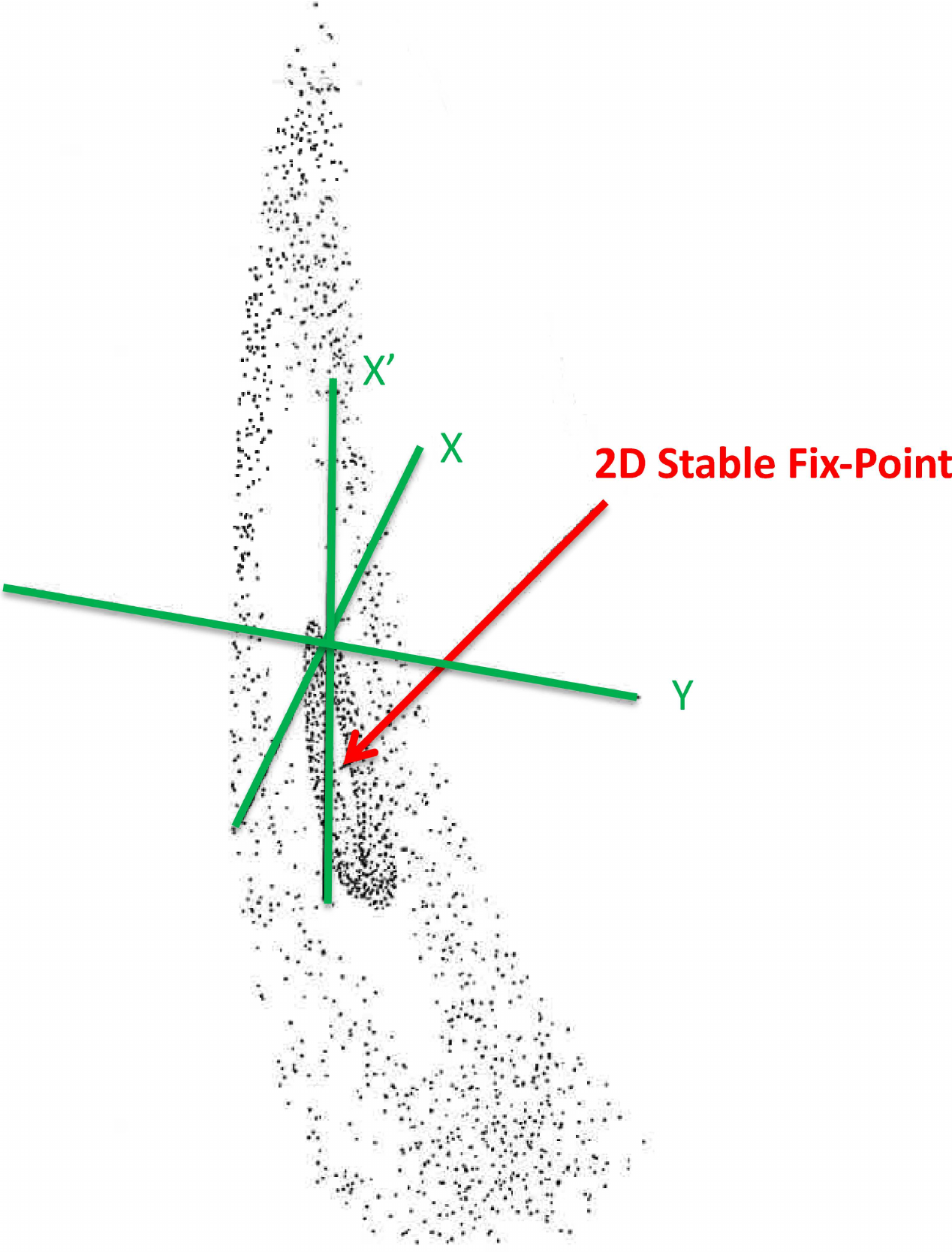,width=65mm}}
 \put( -10,315) {a)}
 \put( -10,225) {b)}
 \put( -10,130) {c)}

 \end{picture}
 \caption{2D motion in the vicinity of a stable 2D fix-point.}
 \label{fig1:fig_8}
 \end{center}
 \end{figure}
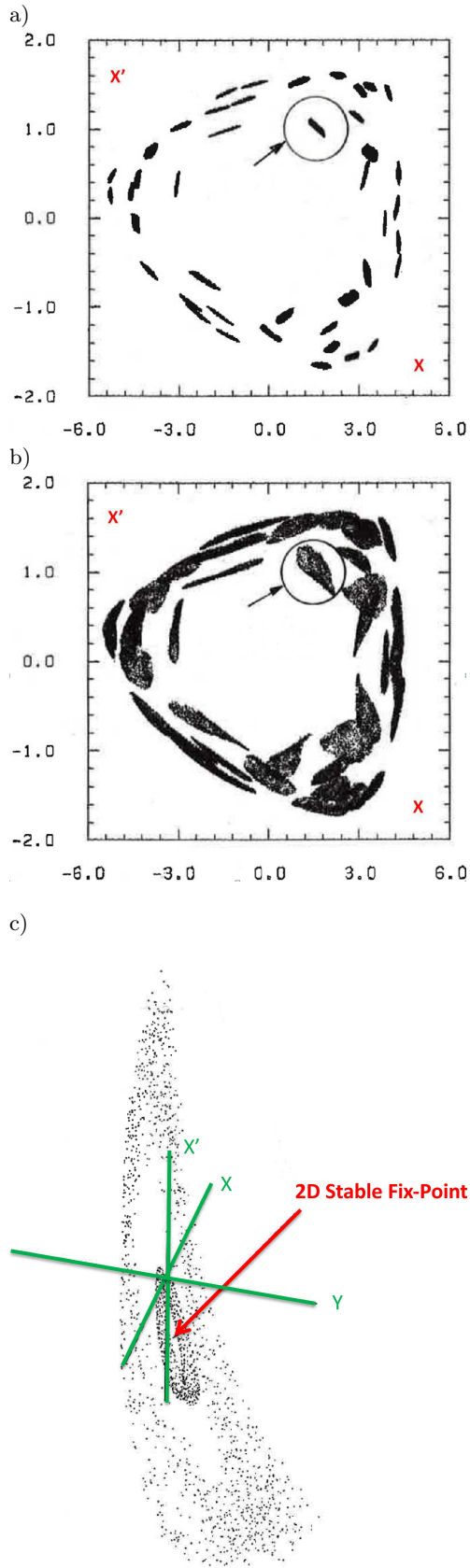
 \begin{figure}[H]
 \begin{center}
 \unitlength 0.7mm
 \begin{picture}(80,170)
 \put( -20, 90) {\epsfig{file=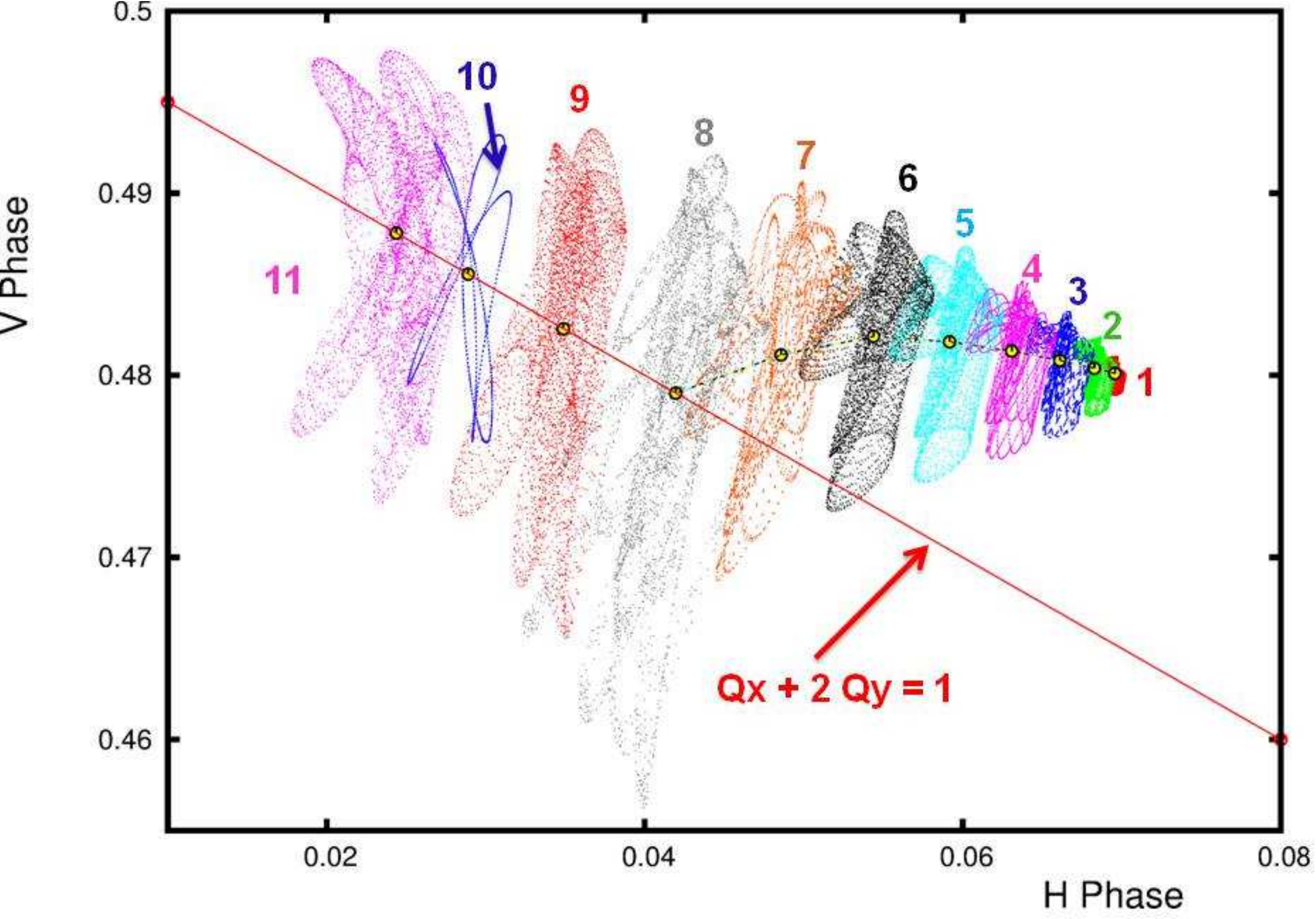,width=80mm}}
 \put( -20,  0) {\epsfig{file=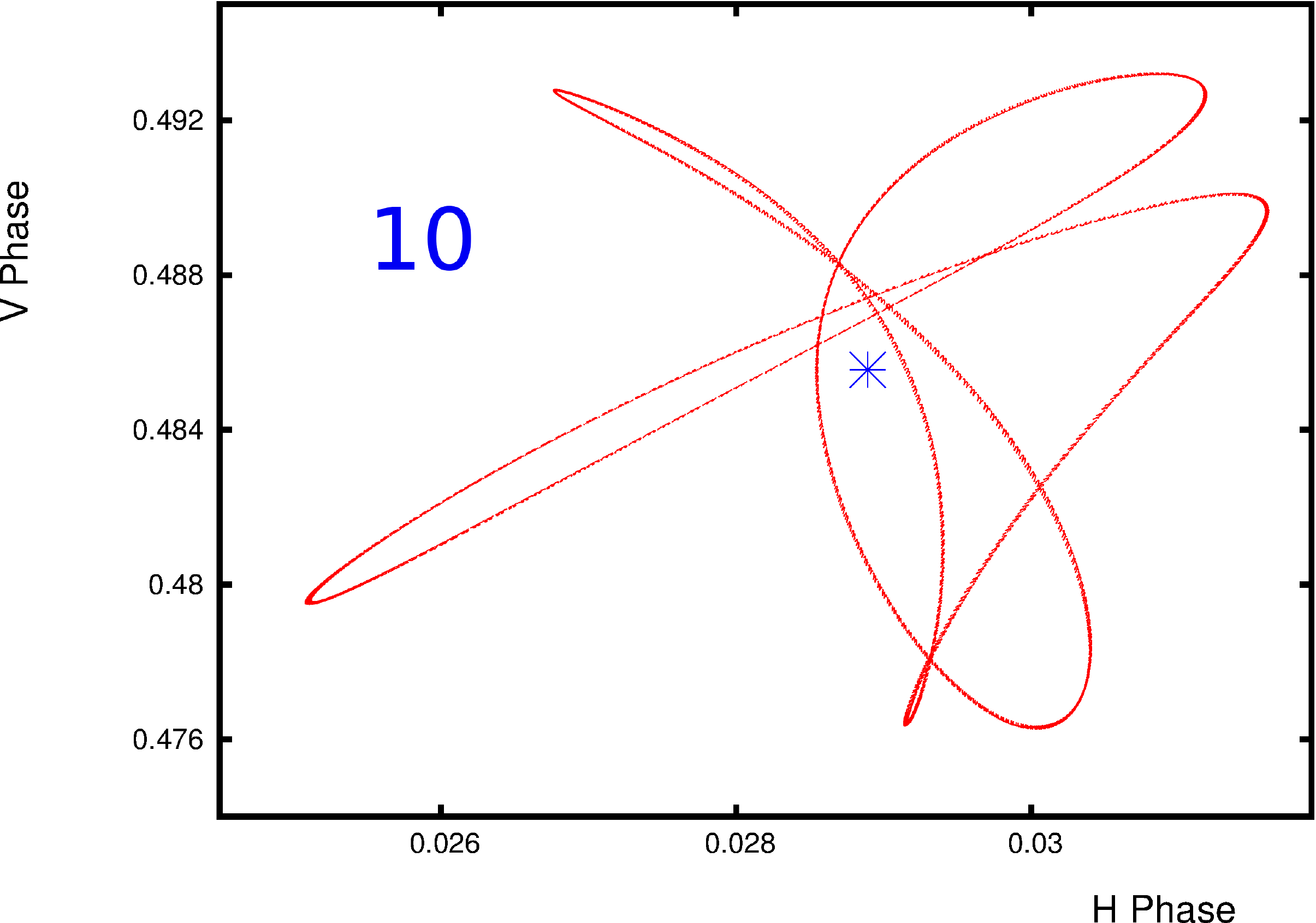,width=80mm}}
 \put( -20,175) {a)}
 \put( -20,85)  {b)}

 \end{picture}
 \caption{Phase-advance per turn; Tunes ($\equiv$ averaged phase
   advance); Orbiting around the 2D fix-line from case 8 through case
   11. Part b) is a blow-up of case 10 that is an example of a
   fix-line in the plane of the phases. Interesting enough the phases
   never really respect the resonance condition instead the tunes do.}
 \label{fig1:fig_9}
 \end{center}
 \end{figure}

%%%%%%%%%%%%%%%%%%%%%%%%%%%%%%%%%%%%%%%%%%%%%%%%%%%%%%%%%%%%%%%%%%%%%%
 %
 \section{Linking the phenomenological phase space depiction with the
   theoretical derivation}
 
 The theory of fix-lines resonance structures will be derived in the
 following chapters. We would like to use a particular theoretical
 prediction to demonstrate how the particle motion of a large scale
 $Q_x + 2\times Q_y$ fix-line structure due to a single sextupole in a
 constant focusing lattice can be viewed in the 4D phase space.
 
\begin{figure}[H]
 \begin{center}
 \unitlength 0.7mm
 \begin{picture}(80,95)
   \put( -15, 0) {\epsfig{file=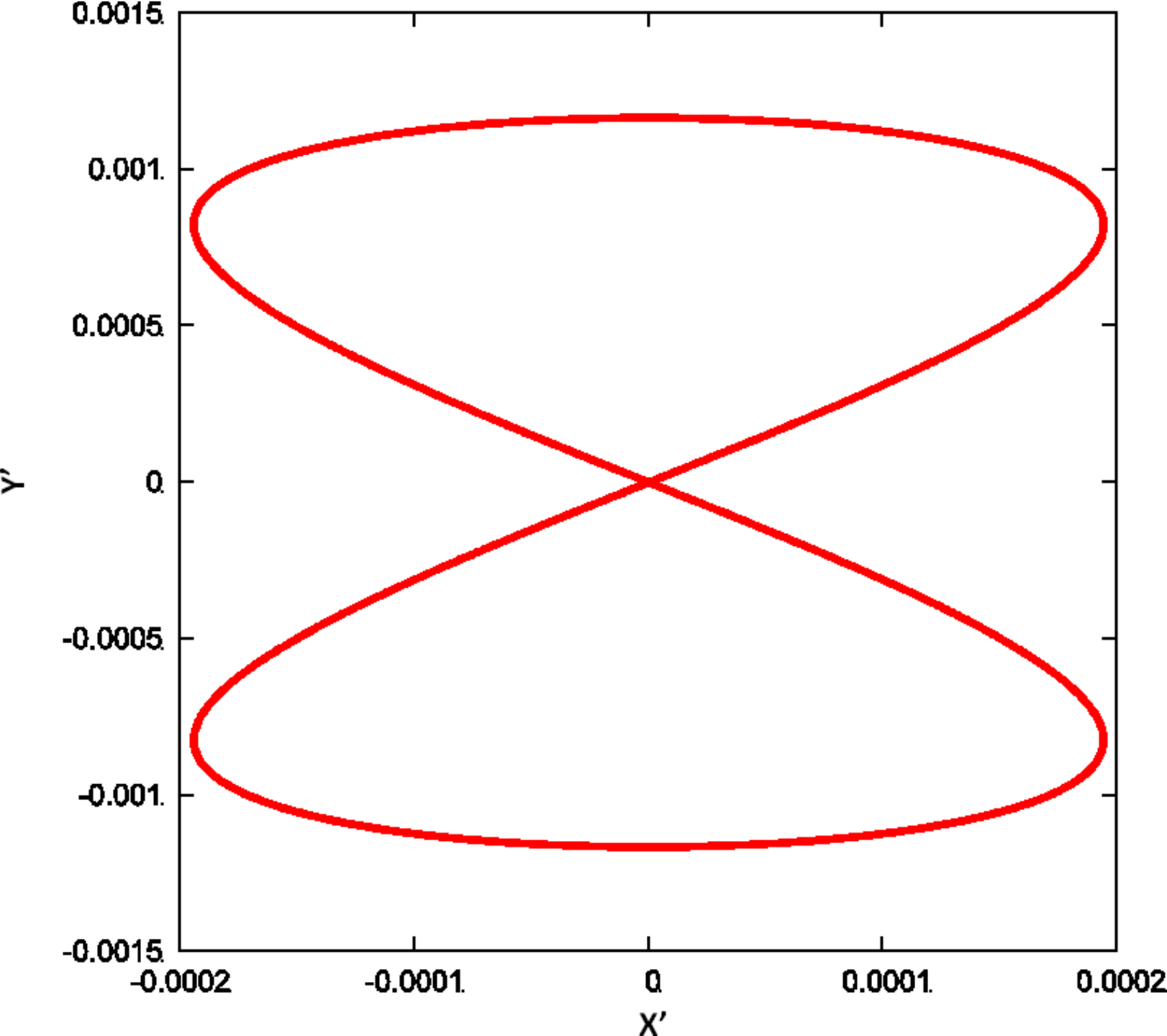,width=75mm}}
 \end{picture}
 \caption{The fix-line is seen in the $y'$ versus $x'$ plane as a figure 8
   like structure.}
 \label{fig1:fig_10}
 \end{center}
 \end{figure}

 It will be shown that in the mixed phase space plane
 \mbox{\bf $y'$ versus $x'$} we expect a figure 8 like structure (see
 Fig.~\ref{fig1:fig_10}) for the particle motion on the fix-line.

\begin{figure}[H]
 \begin{center}
 \unitlength 0.7mm
 \begin{picture}(80,105)
 \put( -20, 0) {\epsfig{file=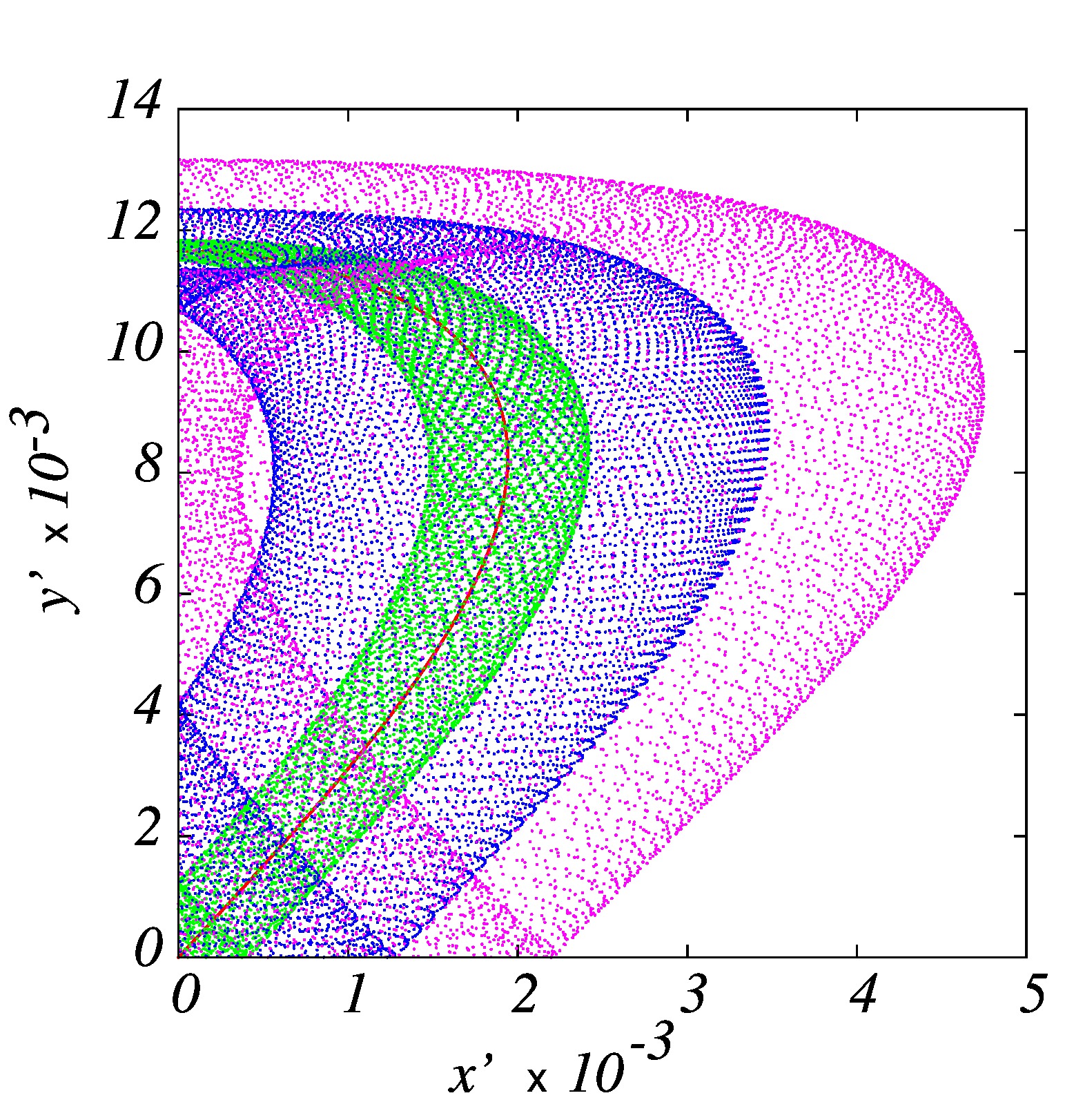,width=80mm}}
 \end{picture}
 \caption{Single passage of the fix-line structure in the $y'$ versus $x'$
   phase space projection.}
 \label{fig1:fig_11}
 \end{center}
 \end{figure}

\begin{figure}[H]
 \begin{center}
 \unitlength 0.7mm
 \begin{picture}(80,111)
 \put( -20, 0) {\epsfig{file=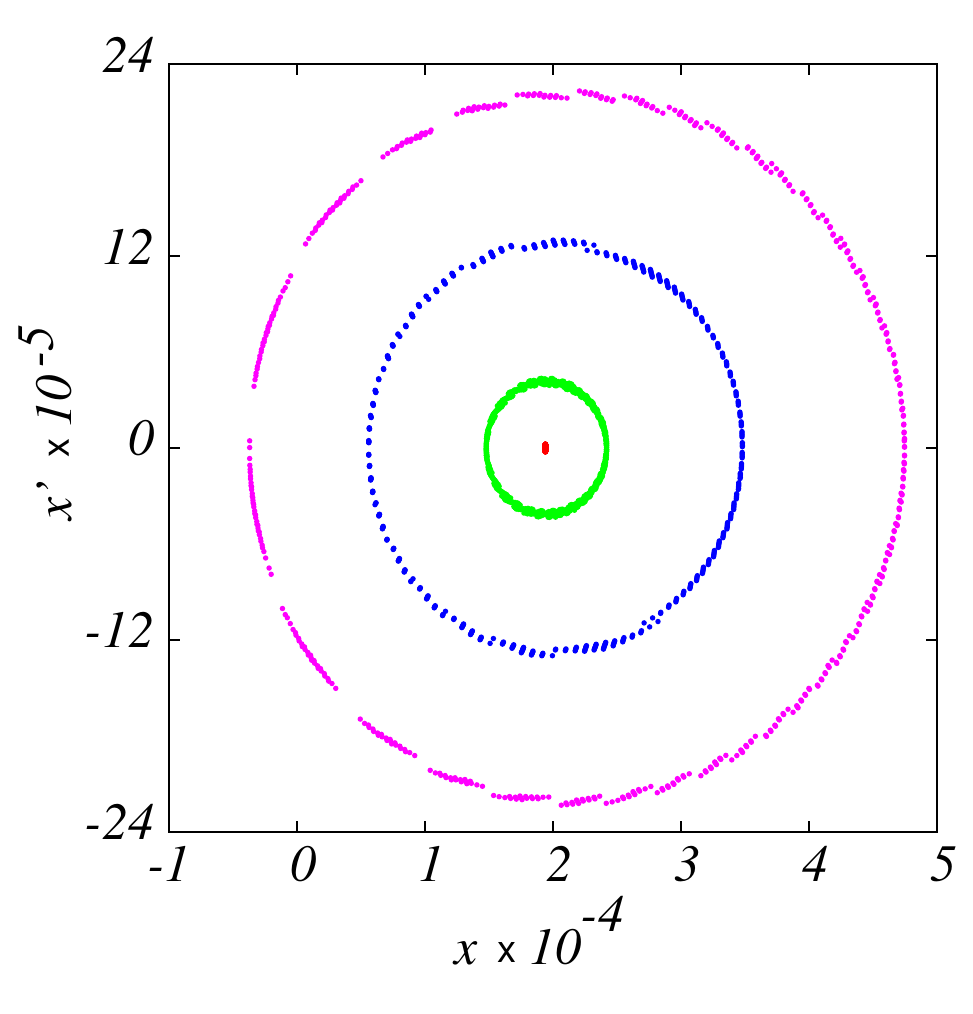,width=80mm}}
 \end{picture}
 \caption{Horizontal phase space projection after restricting the
   vertical motion to a small angle.}
 \label{fig1:fig_12}
 \end{center}
 \end{figure}

\begin{figure}[H]
 \begin{center}
 \unitlength 0.7mm
 \begin{picture}(80,112)
 \put( -20, 0) {\epsfig{file=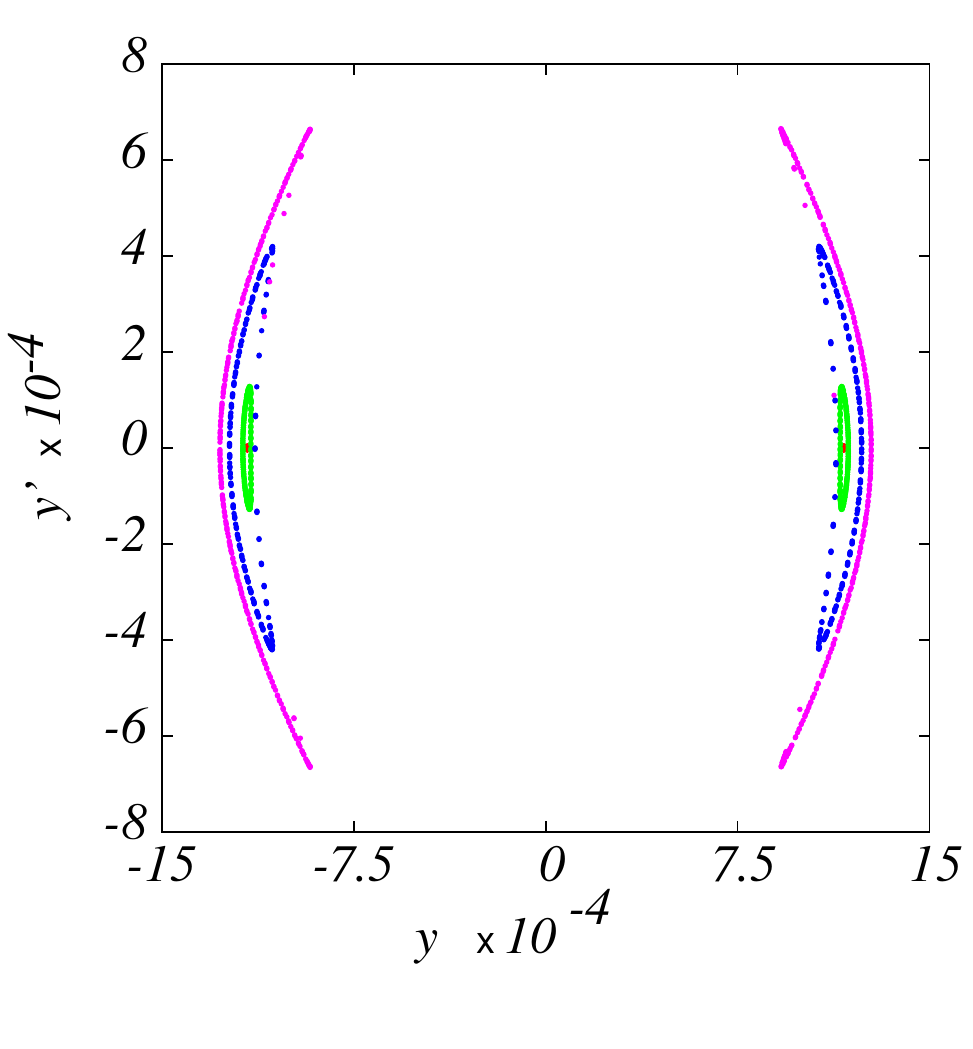,width=80mm}}
 \end{picture}
 \caption{Vertical phase space projection after restricting the
   horizontal motion to a small angle.}
 \label{fig1:fig_13}
 \end{center}
 \end{figure}

 The particle motion on the stable fix-line and on the tori around it
 creates a fix-line structure that is distinct from regular motion
 around the closed orbit. To visualize this fix-line structure we show
 the stable fix-line and 3 tori around it in a restricted phase space
 as described in the earlier chapters. We have to do so to view only a
 single passage of the fix-line structure in the \mbox{\bf $y'$ versus
 $x'$} phase space projection. To this end we restrict {\bf $x$} and {\bf
   $y$} to positive values. Figure~\ref{fig1:fig_11} gives a
 first impression of this fix-line structure. However, it remains
 difficult to figure out graphically how the tori are stacked
 around the fix-line. Therefore, we apply a more vigorous cut in the
 phase space by restricting the particle motion to a very small angle
 in either the horizontal or vertical plane and depicting the resulting
 phase space projection in the other plane.
 Figure~\ref{fig1:fig_12} shows that in the horizontal plane the
 fix-line (red dots) lies indeed in the center with the particle
 motion located on the three tori with increasing distance around the
 fix-line. The same is true in the vertical plane as seen in
 Fig.~\ref{fig1:fig_13}, except now we find two such structures which
 reflects the fact that we are investigating the \mbox{$Q_x + 2\times
   Q_y$} resonance.
 \begin{figure}[H]
 \begin{center}
 \unitlength 0.7mm
 \begin{picture}(80,110)
 \put( -20, 0) {\epsfig{file=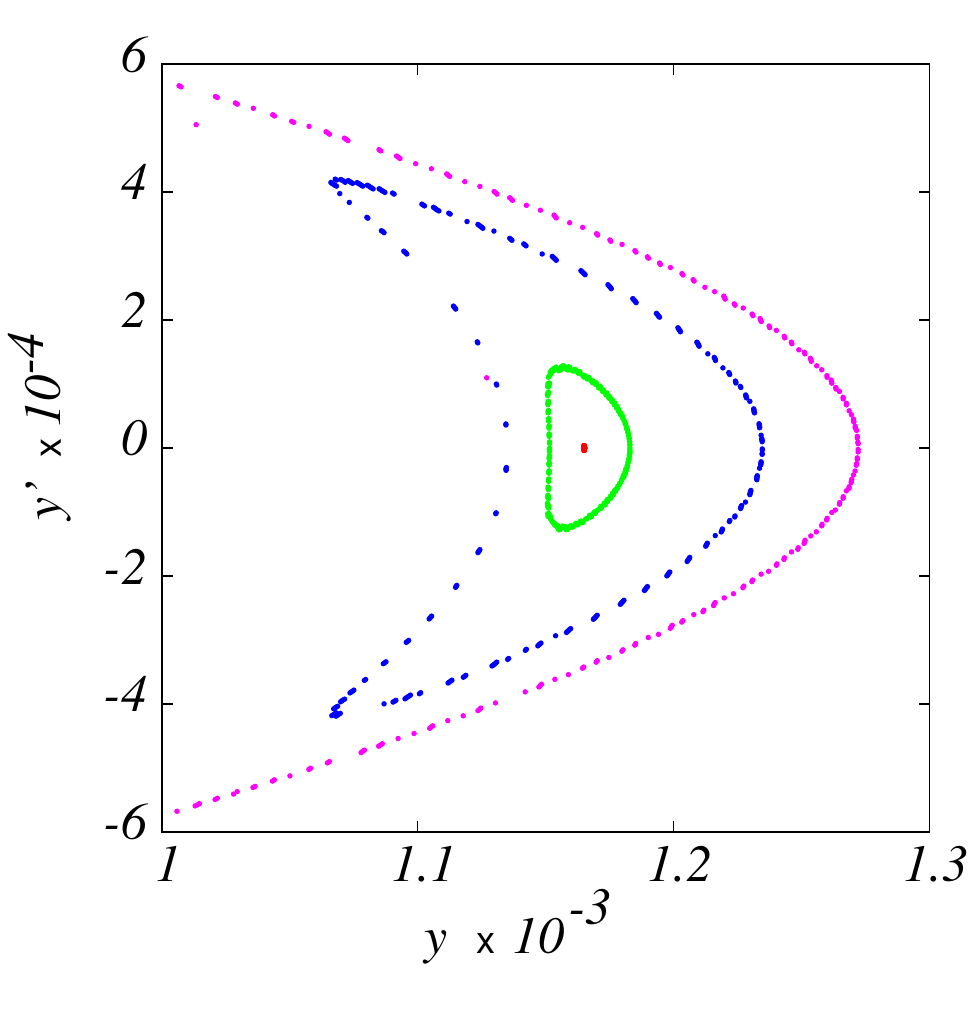,width=80mm}}

 \end{picture}
 \caption{Right part (see Fig.~\ref{fig1:fig_13}) of vertical
   phase space projection after restricting the horizontal motion to a
   small angle.}
 \label{fig1:fig_14}
 \end{center}
 \end{figure}
 We have added a close-up (Fig.~\ref{fig1:fig_14}) for positive {\bf
   $y$} values so that details with the tori and the fix-line at the
 center become more apparent.

 In the following chapters we derive a theory of the particle motion
 in the vicinity of the third order coupling resonance. We start from
 a constant focusing model, which allows a simpler description of the
 dynamics.  Starting from the description of the linear motion, we
 derive the motion of a particle by varying the constants (emittance
 $a_x,a_y$ and phases $\varphi_x,\varphi_y$). We obtain a system of
 differential equations in $a_x,a_y,\varphi_x,\varphi_y$ which is
 ``time dependent''. Then we demonstrate that there are infinite
 canonical transformations capable of removing the time dependence in
 the Hamiltonian of the system.  Once the time dependence is removed,
 the simplest solution is obtained when the new variables are
 constant. This particular solution is proven to be the fix-line, and
 we will obtain analytically the result of Fig.~\ref{fig1:fig_10}. It
 turns out that the parametrization of the canonical transformation is
 the parametrization of the fix-lines, hence we find that there are
 infinite fix-lines contrary to the 1D case with one set of only 3
 (unstable) fix-points.  Perturbations from the solutions allow to
 discuss the stability of the fix-line and to derive secondary
 tunes, i.e. counting the oscillations around the fix-lines. Moreover,
 we find that sextupoles by themselves are capable to create stable
 fix-lines, a feature not know in the 1D case.
 
 The particle dynamics become better described in a special
 combination of the ``varying constants''
 $a_x,a_y,\varphi_x,\varphi_y$: we find that the motion in this
 special combination has another invariant that allows to discuss the
 stability.  Consequently, we are able to describe the stability
 domain in the vicinity of the third order coupled resonance. As may
 have been expected we find that the unstable fix-lines define the
 border of the stability regime. We derive the properties of the
 stability domain in terms of scaled quantities.  The scaling factor
 depends on the distance from the resonance and on the driving term.
 
 Lastly, we apply our approach to the AG machines. The conclusion
 remains the same, and the scaling parameters is incorporating the
 complexity of the AG lattice along with the distribution of
 sextupolar errors.
 
 %%%%%%%%%%%%%%%%%%%%%%%%%%%%%%%%%%%%%%%%%%%%%%%%%%%%%%%%%%%%%%%%%%
 \section{Equation of motion for lattice with constant focusing}
 Let's start with the equations of motion
 describing the dynamics of a particle in a circular accelerator
 of radius $R$.
 \begin{equation}
 \begin{aligned}
   \frac{d^2 x}{ds^2} + k_x x &= f_x(s,x,y), \\
   \frac{d^2 y}{ds^2} + k_y y &= f_y(s,x,y), \\
 \end{aligned}
 \label{eq:1}
 \end{equation}
 here $k_x,k_y$ are constants.
 The tunes of the machine $Q_x,Q_y$ are defined as
 $
   \sqrt{k_x}  R = Q_x, \quad
   \sqrt{k_y}  R = Q_y.
 $
 In our discussion $f_x,f_y$ are created by a magnetic multipole, then
 \begin{equation}
   -f_x(s) + i f_y(s) = \frac{k_n(s)}{n!}(x+iy)^n.
 \label{eq:2}
 \end{equation}
 With this notation for $n=1$ we get
 $
   -f_x(s) = k_1(s) x, \quad
    f_y(s) = k_1(s) y,
 $
 hence, a positive $k_1$ yields a focusing force in the $x$ plane.
 
 Let's make the change of the variable $s$ to the new variable
 $s \rightarrow \theta$ so that
 $
   \frac{d x}{ds} =   \frac{d x}{d\theta}\frac{1}{R}.
 $
 We use the notation
 $
   x' =   \frac{d x}{d\theta}.
 $
 With this change of variable the equations of motion become
 \begin{equation}
 \begin{aligned}
   x'' + Q^2_x x &= R^2 f_x(s(\theta),x,y), \\
   y'' + Q^2_y y &= R^2 f_y(s(\theta),x,y),  \\
 \end{aligned}
 \label{eq:3}
 \end{equation}
 and the length of the one turn is $\Delta\theta = 2\pi$.
 The integrated strength of $k_2(\theta)$ in the $\theta$ coordinate
 is linked to the integrated strength in the $s$ coordinate via the
 relation
 \begin{equation}
   \int_{\theta_1}^{\theta_2} k_2(\theta)d\theta  =
   \int_{s_1}^{s_2} k_2(s)\frac{d\theta}{ds}ds  = \frac{K_2}{R}
 \label{eq:4}
 \end{equation}
 where $K_2$ is the integrated strength in the $s$ coordinate.

 %%%%%%%%%%%%%%%%%%%%%%%%%%%%%%%%%%%%%%%%%%%%%%%%%%%%%%%%%%%%%%%%%
 \subsection{Perturbative approach: the dynamics with one sextupole}
 We will consider next the accelerator with a single sextupole for
 exciting sextupolar resonances.
 The Hamiltonian of this system is $H=H_0+H_1$ with
 \begin{equation}
   H_0 = \frac{1}{2}x^{'2} + \frac{Q_x^2}{2}x^2 +
       \frac{1}{2}y^{'2} + \frac{Q_y^2}{2}y^2
 \label{eq:5}
 \end{equation}
 and
 \begin{equation}
   H_1 = R^2 k_2(\theta) \left(\frac{x^3}{6} - \frac{xy^2}{2}\right). 
 \label{eq:6}
 \end{equation}
 Therefore for a localized sextupole of integrated strength $K_2$ located
 at $\theta=0$ in a circular machine, using Eq.~\ref{eq:4}
 we write
 \begin{equation}
   k_2(\theta) = \sum_{m=-\infty}^{\infty}\frac{K_2}{R}\delta(\theta+2\pi m).
 \label{eq:7}
 \end{equation} 
 where the sum over infinite delta functions $\delta()$ is used for
 modeling the periodicity of one sextupolar error in the machine.  
 The Hamiltonian $H_1$ becomes then
 \begin{equation}
   H_1 = R K_2\sum_{m=-\infty}^{\infty}\delta(\theta + 2\pi m)
   \left(\frac{x^3}{6} - \frac{xy^2}{2}\right). 
 \label{eq:8}
 \end{equation}
% where $K_2$ is the integrated strength of the sextupole in the
% $s$ coordinate.
 
 The solution of the unperturbed system with Hamiltonian $H_0$ is
 the solution of the canonical equations
 \begin{equation}
 \begin{aligned}
 \frac{dx}{d\theta}  =&
  \frac{\partial H_0}{\partial x'},  & \frac{dx'}{d\theta}
   =& -\frac{\partial H_0}{\partial x},  \\
 \frac{dy}{d\theta}  =&
 \frac{\partial H_0}{\partial y'}, & \frac{dy'}{d\theta}
 =& -\frac{\partial H_0}{\partial y}, \\
 \end{aligned}
 \label{eq:9}
 \end{equation}
 which yields the equations of motion, that are
 \begin{equation}
   x'' + Q_x^2 x = 0, \quad   y'' + Q_y^2 y = 0.
 \label{eq:10}
 \end{equation}
 The solutions are 
 \begin{equation}
 \begin{aligned}
   x = \sqrt{\beta_x a_x}\cos(Q_x\theta+\varphi_x), \\
   y = \sqrt{\beta_y a_y}\cos(Q_y\theta+\varphi_y),  \\
 \end{aligned}
 \label{eq:11}
 \end{equation}
 with $\beta_x = 1/Q_x, \beta_y = 1/Q_y$. The parameters $a_x,a_y$
 are the single particle invariant, i.e. the usual emittances expressed
 in the reference frame with $\theta$ coordinate. Hence they are not the
 emittances in the laboratory frame.
 
 Next we use the canonical theory of perturbation, in which we add
 $H_1$ and compute the evolution in the system $H_0+H_1$
 of the otherwise constants $a_x,\varphi_x,a_y,\varphi_y$.
 The equations of motion of $a_x,\varphi_x,a_y,\varphi_y$ are
 \begin{equation}
 \begin{aligned}
 a_x'       &= -&2 \frac{\partial H_1}{\partial \varphi_x}, & \quad \varphi_x' &=  &2 \frac{\partial H_1}{\partial a_x},       \\
 a_y'       &= -&2 \frac{\partial H_1}{\partial \varphi_y}, & \quad \varphi_y' &=  &2 \frac{\partial H_1}{\partial a_y}. 
 \end{aligned}
 \label{eq:12}
 \end{equation}
 Note that this is a canonical system with Hamiltonian $2 H_1$.
 Next we substitute in $H_1$ the Eqs.~\ref{eq:11} re-written in the form
 \begin{equation}
 \begin{aligned}
     x = \sqrt{\beta_x a_x}\frac{e^{i A_x} +e^{-i A_x}}{2} \\
     y = \sqrt{\beta_y a_y}\frac{e^{i A_y} +e^{-i A_y}}{2} \\
 \end{aligned}
 \label{eq:13}
 \end{equation}
 with $A_x = Q_x\theta+\varphi_x$, and $A_y = Q_y\theta+\varphi_y$ defined
 just for convenience.
 We also expand the periodic $\delta$ as
 \begin{equation}
   \sum_{m=-\infty}^{\infty}\delta(\theta + 2\pi m) =
   \frac{1}{2\pi} \sum_{l=-\infty}^{\infty} e^{-il\theta}
 \label{eq:14}
 \end{equation}
 and substitute it in $H_1$.
 After some algebraic calculation, $H_1$ becomes
 \begin{equation}
 \begin{split}
 H_1 = &  R K_2 \frac{1}{48\pi} (\beta_x a_x)^{3/2} \times \frac{1}{2} \times \\
       &  \sum_{j=0}^{3}\sum_{l=-\infty}^{\infty} \binom{3}{j}
          \cos[((2j-3)Q_x-ç∂l)\theta + (2j-3)\varphi_x] - \\
       & - R K_2 \frac{1}{16\pi} \frac{1}{2}
           \sqrt{\beta_x a_x}\beta_ya_y \left\{
       \sum_{l=-\infty}^{\infty}\sum_{j=0}^{2}\sum_{j'=0}^{1}
       \binom{1}{j'}\binom{2}{j}\right.\\
       &
       \left.
       \cos[[(2j'-1)Q_x + (2j-2)Q_y-l)]\theta + \right.\\
       &\left. + (2j'-1)\varphi_x+(2j-2)\varphi_y]
       \right\}.  \\
 \end{split}
 \label{eq:15}
 \end{equation}
 At this point we keep the slow varying terms only, that is the terms
 close to the resonance $Q_x+2Q_y=N$. Here $N$ is an integer used
 throughout this article with the same meaning: it defines the
 location of the resonance.  Let's define
 \begin{equation}
   \Delta_r = -N+Q_x+2Q_y,
 \label{eq:16}
 \end{equation}
 then the slower varying terms are obtained in the second term for
 $l=N,j'=1,j=2$, and $l=-N,j'=0,j=0$.
 Hence the slower harmonics Hamiltonian $H_{s1}$, which mainly contributes
 to the dynamics is
 \begin{equation}
   H_{s1} = -R K_2 \frac{1}{16\pi} \sqrt{\beta_xa_x}\beta_ya_y
   \cos[\Delta_r \theta + \varphi_x+2\varphi_y].
 \label{eq:17}
 \end{equation}
 We define for convenience
 \begin{equation}
   \Lambda = -R K_2 \frac{1}{16\pi} \sqrt{\beta_x}\beta_y.
 \label{eq:18}
 \end{equation}
 If instead we consider a continuous sextupolar component with strength
 $k_2(\theta)={\cal K}_2\cos(N\theta)$, the expression for $\Lambda$ is
 \begin{equation}
   \Lambda = -R^2 {\cal K}_2 \frac{1}{16}\sqrt{\beta_x}\beta_y. 
 \label{eq:19}
 \end{equation}
 Note that ${\cal K}_2$ in this experssion is the strength of a 
 distributed sxtupole not to be confused with the integrated strength 
 $K_2$ in Eq.~\ref{eq:18}. 
 We therefore find that the general form of the slowly varying Hamiltonian 
 is 
 \begin{equation}
   H_{s1} = \Lambda \sqrt{a_x}a_y
   \cos[\Delta_r \theta + \varphi_x+2\varphi_y]
 \label{eq:20}
 \end{equation} 
 and that it is time dependent as
 \begin{equation}
   \frac{\partial H_{s1}}{\partial\theta} =
 - \Lambda \sqrt{a_x}a_y
   \sin[\Delta_r \theta + \varphi_x+2\varphi_y]  \Delta_r \ne 0.
 \label{eq:21}
 \end{equation}

 %%%%%%%%%%%%%%%%%%%%%%%%%%%%%%%%%%%%%%%%%%%%%%%%%%%%%%%%%%%%%%%%%%%%%%5
 \section{Removing the time dependency}
 We redefine the variables $\varphi_x,\varphi_y$ so that they incorporate
 the term $\Delta_r\theta$.
 Let's define
 \begin{equation}
 \begin{aligned}
   \tilde\varphi_x = \varphi_x + t_x \theta, \\
   \tilde\varphi_y = \varphi_y + t_y \theta, \\
 \end{aligned}
 \label{eq:22}
 \end{equation}
 with $t_x,t_y$ some constants to be defined later.
 Then the argument of the cosine in Eq.~\ref{eq:20} becomes
 \begin{equation}
   \Delta_r \theta +  \varphi_x +
                     2\varphi_y =
   \tilde\varphi_x  + 2\tilde\varphi_y
    + [\Delta_r - (t_x + 2t_y)] \theta,
 \label{eq:23}
 \end{equation}
 therefore if $\Delta_r - (t_x + 2t_y) = 0$ the time dependence disappears.
 We then consider the coordinate transformation
 \begin{equation}
 \begin{aligned}
   \tax            &= a_x, \\
   \tilde\varphi_x &= \varphi_x + t_x \theta, \\
   \tay            &= a_y, \\
   \tilde\varphi_y &= \varphi_y + t_y \theta. \\
 \end{aligned}
 \label{eq:24}
 \end{equation}
 If we define
 \begin{equation}
   \tilde H_{s1}(\tax,\tilde\varphi_x,\tay,\tilde\varphi_y,\theta) =
   H_{s1}\left(
   \tax, \tilde\varphi_x  - t_x \theta,
   \tay, \tilde\varphi_y  - t_y \theta,\theta
                 \right), 
 \label{eq:25}
 \end{equation}
 then
 \begin{equation}
 \begin{aligned}
 2\frac{\partial \tilde H_{s1}}{\partial \tpx} &= 2\frac{\partial H_{s1}}{\partial \varphi_x}  = - a_x' = - \tax', \\
 2\frac{\partial \tilde H_{s1}}{\partial \tax} &= 2\frac{\partial H_{s1}}{\partial a_x}  =  \varphi_x' = \tpx' - t_x,\\
 2\frac{\partial \tilde H_{s1}}{\partial \tpy} &= 2\frac{\partial H_{s1}}{\partial \varphi_y}  = - a_y' = - \tay', \\
 2\frac{\partial \tilde H_{s1}}{\partial \tay} &= 2\frac{\partial H_{s1}}{\partial a_y}  =  \varphi_y' = \tpy' - t_y.\\
 \end{aligned}
 \label{eq:26}
 \end{equation}
 Therefore we have constructed a function (Eq.~\ref{eq:25})
 that is time independent, but if we try to derive the equations of motion from it,
 we find Eqs.~\ref{eq:26}.
 These equations are clearly not the canonical equations,
 that means that the function
 Eq.~\ref{eq:25} is not an Hamiltonian.
 This happens because in
 \begin{equation}
   2\frac{\partial \tilde H_{s1}}{\partial \tax}
   = \tpx' - t_x
 \label{eq:27}
 \end{equation}
 there is  the constant term  $-t_x$ that should not be there.
 We then define the new function
 \begin{equation}
 \begin{split}
   \tilde H_{s1}(\tax,\tilde\varphi_x,\tay,\tilde\varphi_y,\theta) =
   &H_{s1}\left(
   \tax, \tilde\varphi_x  - t_x \theta,
   \tay, \tilde\varphi_y  - t_y \theta,\theta
                 \right) + \\
   & + \tax \frac{t_x}{2} + \tay \frac{t_y}{2} \\
 \end{split}
 \label{eq:28}
 \end{equation}
 which is also time independent.
 Next we check if one can get the canonical equations
 \begin{equation}
 \begin{aligned}
 2\frac{\partial \tilde H_{s1}}{\partial \tpx} &= 2\frac{\partial H_{s1}}{\partial \varphi_x} = - \tax', \\
 2\frac{\partial \tilde H_{s1}}{\partial \tax} &= 2\frac{\partial H_{s1}}{\partial a_x} + t_x = \varphi_x' + t_x = \tpx',\\
 2\frac{\partial \tilde H_{s1}}{\partial \tpy} &= 2\frac{\partial H_{s1}}{\partial \varphi_y} = - \tay', \\
 2\frac{\partial \tilde H_{s1}}{\partial \tay} &= 2\frac{\partial H_{s1}}{\partial a_y} + t_y = \varphi_y' + t_y = \tpy'.\\
 \end{aligned}
 \label{eq:29}
 \end{equation}
 Therefore  we obtain
 \begin{equation}
 \begin{aligned}
 2\frac{\partial \tilde H_{s1}}{\partial \tpx} &= - \tax' \\
 2\frac{\partial \tilde H_{s1}}{\partial \tax} &=   \tpx'\\
 2\frac{\partial \tilde H_{s1}}{\partial \tpy} &= - \tay' \\
 2\frac{\partial \tilde H_{s1}}{\partial \tay} &=   \tpy'\\
 \end{aligned}
 \label{eq:30}
 \end{equation}
 which are the canonical equations  with the 
 Hamiltonian $2\tilde H_{s1}$. 
 By construction the Hamiltonian $\tilde H_{s1}$ is time independent.
 In fact
 \begin{equation}
 \begin{aligned}
   \tilde H_{s1}(\tax,\tilde\varphi_x,\tay,\tilde\varphi_y,\theta) =&
   \Lambda \sqrt{\tax}\tay
   \cos[\tilde\varphi_x+2\tilde\varphi_y] + \\
   &+ \tax \frac{t_x}{2} + \tay \frac{t_y}{2} \\ 
 \end{aligned} 
 \label{eq:31}
 \end{equation}
 does not depend on $\theta$. 
 Note that the procedure described here is equivalent to transform the
 Hamiltonian $2 H_{s1}$  (from Eq.~\ref{eq:20}) to the Hamiltonian 
 $2 \tilde H_{s1}$ (from Eq.~\ref{eq:31}) 
 through the canonical transformation Eqs.~\ref{eq:24} 
 constructed by using the generating function 
 \begin{equation}
   F_2(\varphi_x,\varphi_y,\tax,\tay) = 
   \varphi_x \tax + \varphi_y \tay + t_x\tax\theta + t_y\tay\theta
 \label{eq:31b}
 \end{equation}
 with $t_x+2t_y=\Delta_r$.

 %%%%%%%%%%%%%%%%%%%%%%%%%%%%%%%%%%%%%%%%%%%%%%%%%%%%%%%%%%%%%%%%%%%%%%%%%%%
 \subsection{Summary}
 We can summarize the situation as follows.
 There are infinite canonical transformations of the form
 \begin{equation}
 \begin{aligned}
   \varphi_x = \tpx - t_x \theta, \qquad a_x = \tax, \\
   \varphi_y = \tpy - t_y \theta, \qquad a_y = \tay, \\
 \end{aligned}
 \label{eq:32}
 \end{equation}
 with $t_x+2 t_y = \Delta_r$ that allow to describe the time evolution
 as solution of canonical equations of the time independent Hamiltonian
 \begin{equation}
 \begin{split}
   \tilde H_{s1}(\tax,\tilde\varphi_x,\tay,\tilde\varphi_y) &=
   \Lambda \sqrt{\tax}\tay
   \cos[\tilde\varphi_x+2\tilde\varphi_y] + \\
   &+ \tax \frac{t_x}{2} + \tay \frac{t_y}{2} + {\cal F}(\tax,\tay). \\
 \end{split}
 \label{eq:33}
 \end{equation}
 The function ${\cal F}(\tax,\tay)$ collects all the nonlinear terms 
 arising from the theory when other multipolar components (non resonant) 
 are included in the Hamiltonian $H_1$. 
 Following the procedure previously outlined, the slowly varying terms will
 be found in the zero-th order harmonics, that will create an additional
 term in the slowly varying Hamilton $H_{s1}$.
 Space charge can be treated in the same way.
 The explicit form of the canonical equations in the coordinates
 $\tax,\tilde\varphi_x,\tay,\tilde\varphi_y$ is
 \begin{equation}
 \begin{aligned}
 -\tax'           &=   -2\Lambda \sqrt{\tax}\tay  \sin[\tilde\varphi_x+2\tilde\varphi_y]\\
 \tilde\varphi_x' &=    2\Lambda \frac{1}{2\sqrt{\tax}}\tay \cos[\tilde\varphi_x+2\tilde\varphi_y] + t_x + 2{\cal F}_x\\
 -\tay'           &=   -4\Lambda \sqrt{\tax}\tay  \sin[\tilde\varphi_x+2\tilde\varphi_y]\\
 \tilde\varphi_y' &=    2\Lambda \sqrt{\tax} \cos[\tilde\varphi_x+2\tilde\varphi_y] + t_y  + 2{\cal F}_y\\
 \end{aligned}
 \label{eq:34}
 \end{equation}
 where we used the short notation
 ${\cal F}_x = \partial {\cal F} / \partial \tax$, and
 ${\cal F}_y = \partial {\cal F} / \partial \tay$.
 From these equations we find immediately with 
 $
    2\tax' - \tay' = 0:
 $
 \begin{equation}
    2\tax - \tay = C
 \label{eq:35}
 \end{equation}
 with $C$ a constant.
 As the constants $t_x,t_y$ must satisfy the condition $t_x+2t_y=\Delta_r$,
 the solution
 $\tax,\tilde\varphi_x,\tay,\tilde\varphi_y$ depends
 on two free parameters, a choice could be $t_x,C$.

 %%%%%%%%%%%%%%%%%%%%%%%%%%%%%%%%%%%%%%%%%%%%%%%%%%%%%%%%%%%%%%%%%%%%%%%
 \section{Selecting the canonical transformation}
 
 In this section we search in the space
 $(\tax,\tilde\varphi_x,\tay,\tilde\varphi_y)$ the solution of 
 $\tax'=0,\tilde\varphi_x'=0,\tay'=0,\tilde\varphi_y'=0$.
 The canonical equations then read
 \begin{equation}
 \begin{aligned}
 0 &=\frac{\partial \tilde H_{s1}}{\partial \tpx} =    -\Lambda \sqrt{\tax}\tay              \sin[\tilde\varphi_x + 2\tilde\varphi_y],\\
 0 &=\frac{\partial \tilde H_{s1}}{\partial \tax} =     \Lambda \frac{1}{2\sqrt{\tax}}\tay   \cos[\tilde\varphi_x + 2\tilde\varphi_y] + \frac{t_x}{2} + {\cal F}_x,\\
 0 &=\frac{\partial \tilde H_{s1}}{\partial \tpy} =  -2 \Lambda \sqrt{\tax}\tay              \sin[\tilde\varphi_x + 2\tilde\varphi_y],\\
 0 &=\frac{\partial \tilde H_{s1}}{\partial \tay} =     \Lambda \sqrt{\tax}                  \cos[\tilde\varphi_x + 2\tilde\varphi_y] + \frac{t_y}{2}  + {\cal F}_y.\\
 \end{aligned}
 \label{eq:36}
 \end{equation}
 The equations~\ref{eq:36}-1,\ref{eq:36}-3 are satisfied for
 \begin{equation}
   \sin[\tilde\varphi_x+2\tilde\varphi_y] = 0, 
 \label{eq:37}
 \end{equation}
 that is for
 $
    \tilde\varphi_x+2\tilde\varphi_y = \pi M,
 $
 with $M=0,1$.
 The other 2 equations become
 \begin{equation}
 \begin{aligned}
 0 &=\Lambda  \frac{1}{2\sqrt{\tax}}\tay (-1)^M + \frac{t_x}{2} + {\cal F}_x, \\
 0 &= \Lambda \sqrt{\tax}  (-1)^M + \frac{t_y}{2}  + {\cal F}_y. \\
 \end{aligned}
 \label{eq:38}
 \end{equation}
 From the condition $t_x+2t_y=\Delta_r$ we can eliminate $t_x$
 and we get the system
 \begin{equation}
 \left\{
 \begin{aligned}
 0 &=\Lambda \frac{1}{2\sqrt{\tax}}\tay (-1)^M
     + \frac{\Delta_r}{2} - t_y + {\cal F}_x\\
 0 &= \Lambda \sqrt{\tax}  (-1)^M + \frac{t_y}{2}  + {\cal F}_y.\\
 \end{aligned}
 \right.
 \label{eq:39}
 \end{equation}
 This system can be solved in $\tax,\tay$, from which we find also 
 $C$ through Eq.~\ref{eq:35}.  
 We have therefore established the correspondence
 \begin{equation}
   t_y \longrightarrow (\tax,\tay,t_x,C),
 \label{eq:40}
 \end{equation}
 which means that
 all the solutions of Eqs.~\ref{eq:38} are on a 1-dimensional curve
 in the plane $(\tax,\tay)$.
 This 1-dimensional curve can be expressed as a function of $\tax,\tay$,
 without any reference to $t_x,t_y,C$,
 and together with the phase relation one gets

 \begin{equation}
 \left\{
 \begin{aligned}
 &0 =\Lambda (-1)^M
     \left[\frac{\tay}{2\sqrt{\tax}} +
          2\sqrt{\tax}
     \right]
     + \frac{\Delta_r}{2} + {\cal F}_x  + 2{\cal F}_y\\
 &\tilde\varphi_x + 2\tilde\varphi_y = \pi M. \\
 \end{aligned}
 \right.
 \label{eq:41}
 \end{equation}
 
 The significance of this equation is the following: for any pair
 $(\tax,\tay)$ that satisfies Eq.~\ref{eq:41} (top) there exists an
 associated canonical transformation characterized by the pair
 $(t_x,t_y)$ obtained from Eqs.~\ref{eq:38}. In the new coordinates
 identified by $(t_x,t_y)$ the pair $(\tax,\tay)$ joined with
 $\tilde\varphi_x + 2\tilde\varphi_y = \pi M$ solves Eq.~\ref{eq:36},
 i.e. in the system of coordinates identified by $(t_x,t_y)$,
 $(\tax,\tay)$ are constant.  Therefore all the $(\tax,\tay)$
 satisfying Eq.~\ref{eq:41} are all possible stationary solutions of
 Eq.~\ref{eq:36} (for the correspondent canonical
 transformations). The set of these solutions $(\tax,\tay)$ is a 1
 dimensional curve (or a collection of 1 dimensional curves for
 complicated ${\cal F}_x,{\cal F}_y$ dependencies).

 %%%%%%%%%%%%%%%%%%%%%%%%%%%%%%%%%%%%%%%%%%%%%%%%%%%%%%%%%%%%%%%%%%%%%%%%%%%
 \subsection{The fix-line}
 Next we discuss the meaning of the solution
 $\tax'=0,\tilde\varphi_x'=0,\tay'=0,\tilde\varphi_y'=0$.
 As $\tilde\varphi_x'=\tilde\varphi_y'=0$ we get
 \begin{equation}
 \begin{aligned}
   \tilde\varphi_x = \tilde\varphi_{x,0}, \\
   \tilde\varphi_y = \tilde\varphi_{y,0}, \\
 \end{aligned}
 \label{eq:42}
 \end{equation}
 and returning back to the coordinates of the Hamiltonian $H_{s1}$
 (Eq.~\ref{eq:20}, there written without ${\cal F}$)
 we find
 \begin{equation}
 \begin{aligned}
   \varphi_x = \tilde{\varphi}_{x,0} - t_x \theta, \\
   \varphi_y = \tilde{\varphi}_{y,0} - t_y \theta, \\
 \end{aligned}
 \label{eq:43}
 \end{equation}
 with $\tilde{\varphi}_{x,0}+2\tilde{\varphi}_{y,0}=\pi M$,
 resulting for the coordinates of Eqs.~\ref{eq:11}
 \begin{equation}
 \begin{aligned}
   x  &= \sqrt{\beta_xa_x}\cos[(Q_x- t_x)\theta + \tilde{\varphi}_{x,0} ], \\
   x' &= -(Q_x- t_x)\sqrt{\beta_xa_x}\sin[(Q_x- t_x)\theta + \tilde{\varphi}_{x,0} ], \\
   y  &= \sqrt{\beta_ya_y}\cos[(Q_y- t_y)\theta + \tilde{\varphi}_{y,0} ], \\
   y' &= -(Q_y- t_y)\sqrt{\beta_ya_y}\sin[(Q_y- t_y)\theta + \tilde{\varphi}_{y,0} ]. \\
 \end{aligned}
 \label{eq:44}
 \end{equation}
 We will next prove that the points of this trajectory at a given 
 longitudinal position (for example at $\theta=0$) lie on 
 a one dimensional closed curve that we call fix-line. 
 The proof of this follows:
 
 By using the conditions $\tpx+2\tpy=\pi M$, $t_x+2t_y=\Delta_r$, and 
 Eq.~\ref{eq:16}, it is straightforward to find that
 \begin{equation}
   (Q_x- t_x)\theta + \tilde{\varphi}_{x,0}  + 2[(Q_y- t_y)\theta + \tilde{\varphi}_{y,0}] =
   N \theta + \pi M. 
 \label{eq:45}
 \end{equation}
 The Poincar\'e surface of section is identified by the condition
 $\theta = 2\pi {\cal N}$
 with ${\cal N}$ an integer correspondent to the ${\cal N}$-th turn. 
 Next we limit the discussion to the $x,y$ coordinates without losing 
 generality. 
 By using Eq.~\ref{eq:44}, we find that on the Poincar\'e 
 surface of section the coordinates $x,y$ at the turn ${\cal N}$ become 
 \begin{equation}
 \begin{aligned}
   x_{\cal N} = \sqrt{\beta_xa_x}\cos[(Q_x- t_x)2\pi {\cal N} + \tilde{\varphi}_{x,0} ], \\
   y_{\cal N} = \sqrt{\beta_ya_y}\cos[(Q_y- t_y)2\pi {\cal N} + \tilde{\varphi}_{y,0} ], \\
 \end{aligned}
 \label{eq:46}
 \end{equation} 
 and using Eq.~\ref{eq:45} we find
 \begin{equation}
 \begin{aligned}
   x_{{\cal N}} =& \sqrt{\beta_xa_x}\cos[-2(Q_y - t_y)2\pi {\cal N} \\
   &-2\tilde{\varphi}_{y,0}+\pi M + N 2\pi {\cal N}], \\
   y_{{\cal N}} =& \sqrt{\beta_ya_y}\cos[(Q_y- t_y)2\pi {\cal N} 
              + \tilde{\varphi}_{y,0} ], \\
 \end{aligned}
 \label{eq:47}
 \end{equation}
 with $\tilde\varphi_{y,0}$ as an initial phase that sets the starting
 point. 
 Therefore we reach the result that the coordinates $x,y$ of Eq~\ref{eq:44}, 
 on a surface of the Poincar\'e section,  
 can be parameterized (ignoring multiples of $2\pi$) as
 \begin{equation}
 \begin{aligned}
   x_t &= \sqrt{\beta_xa_x}\cos(-2 t +\pi M ), \\
   y_t &= \sqrt{\beta_ya_y}\cos(   t ), \\
 \end{aligned}
 \label{eq:48}
 \end{equation}
 \begin{figure}[H]
 \begin{center}
 \unitlength 0.74mm
 \begin{picture}(80,115)
 \put(3,0){
 \put(-25, 55)   {\epsfig{file=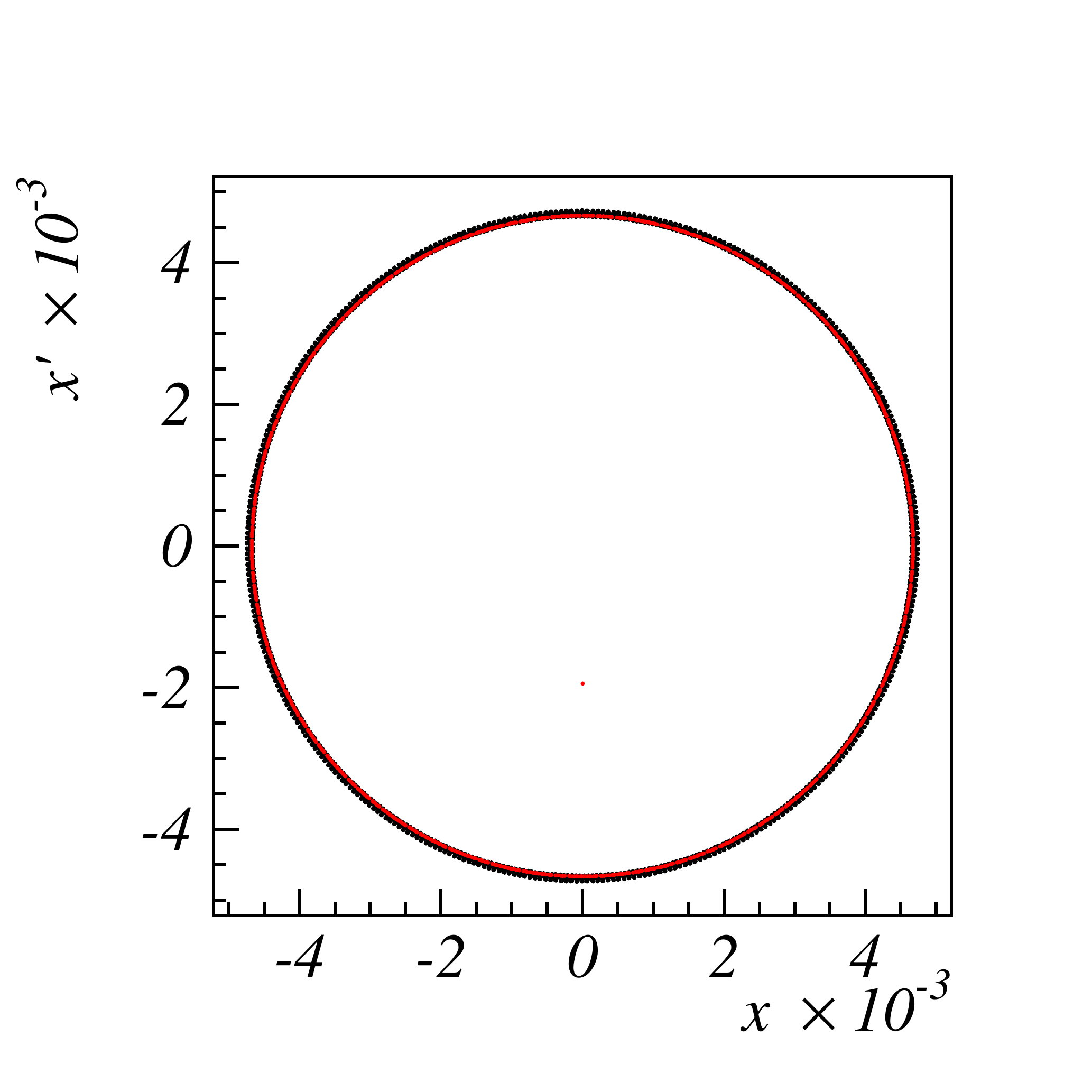,width=49mm}}
 \put( 35, 55)   {\epsfig{file=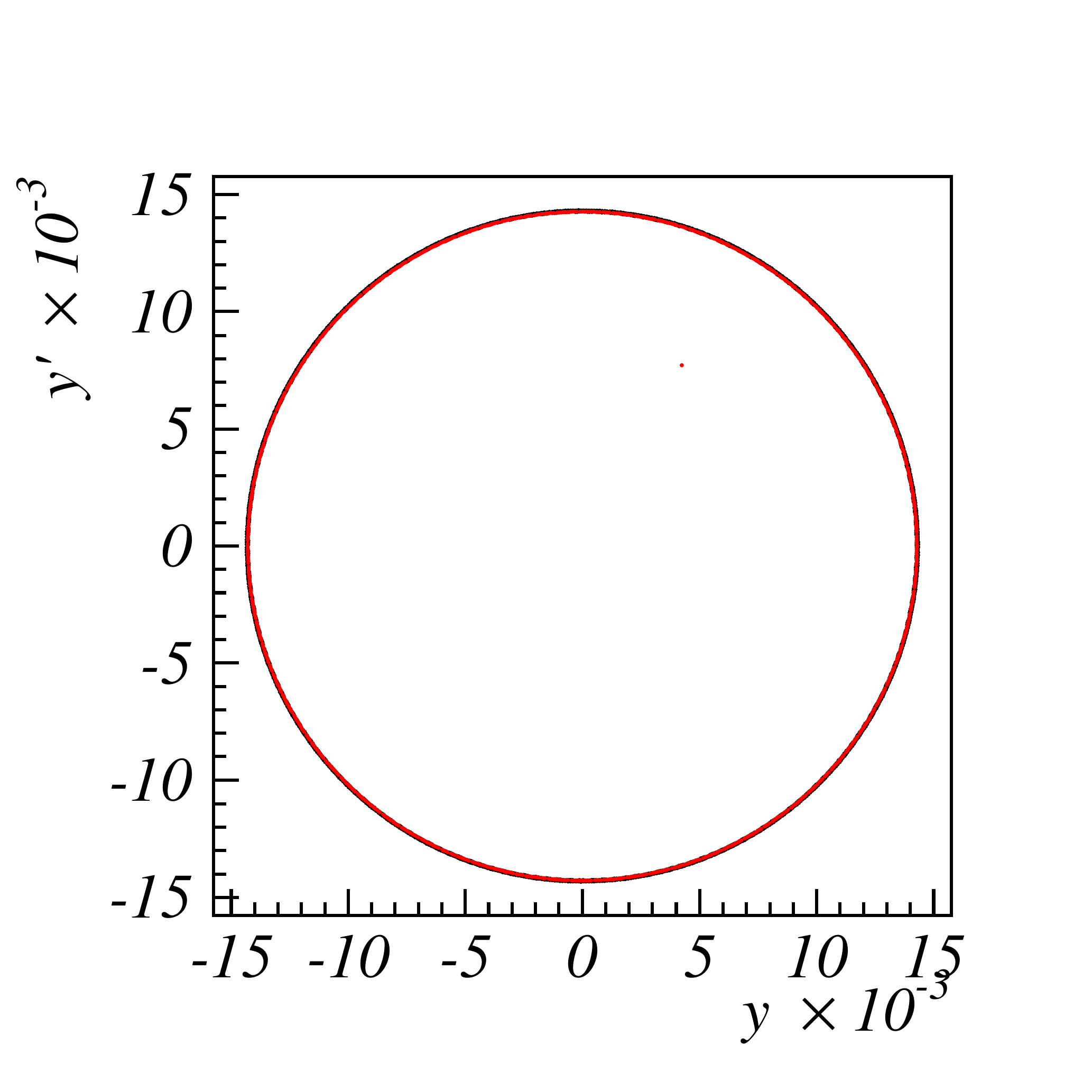,width=49mm}}
 \put(-25,   -3) {\epsfig{file=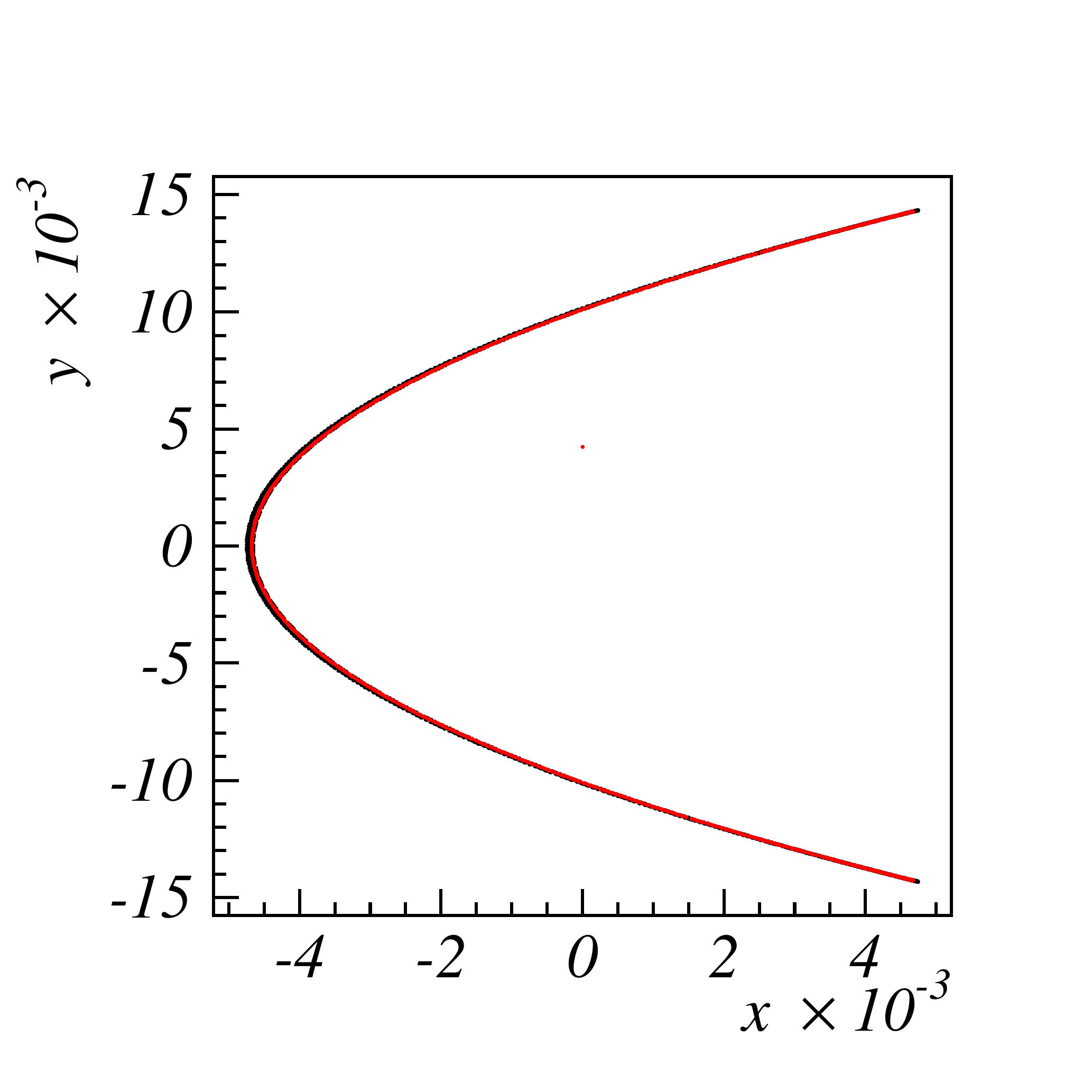,width=49mm}}
 \put( 35,   -3) {\epsfig{file=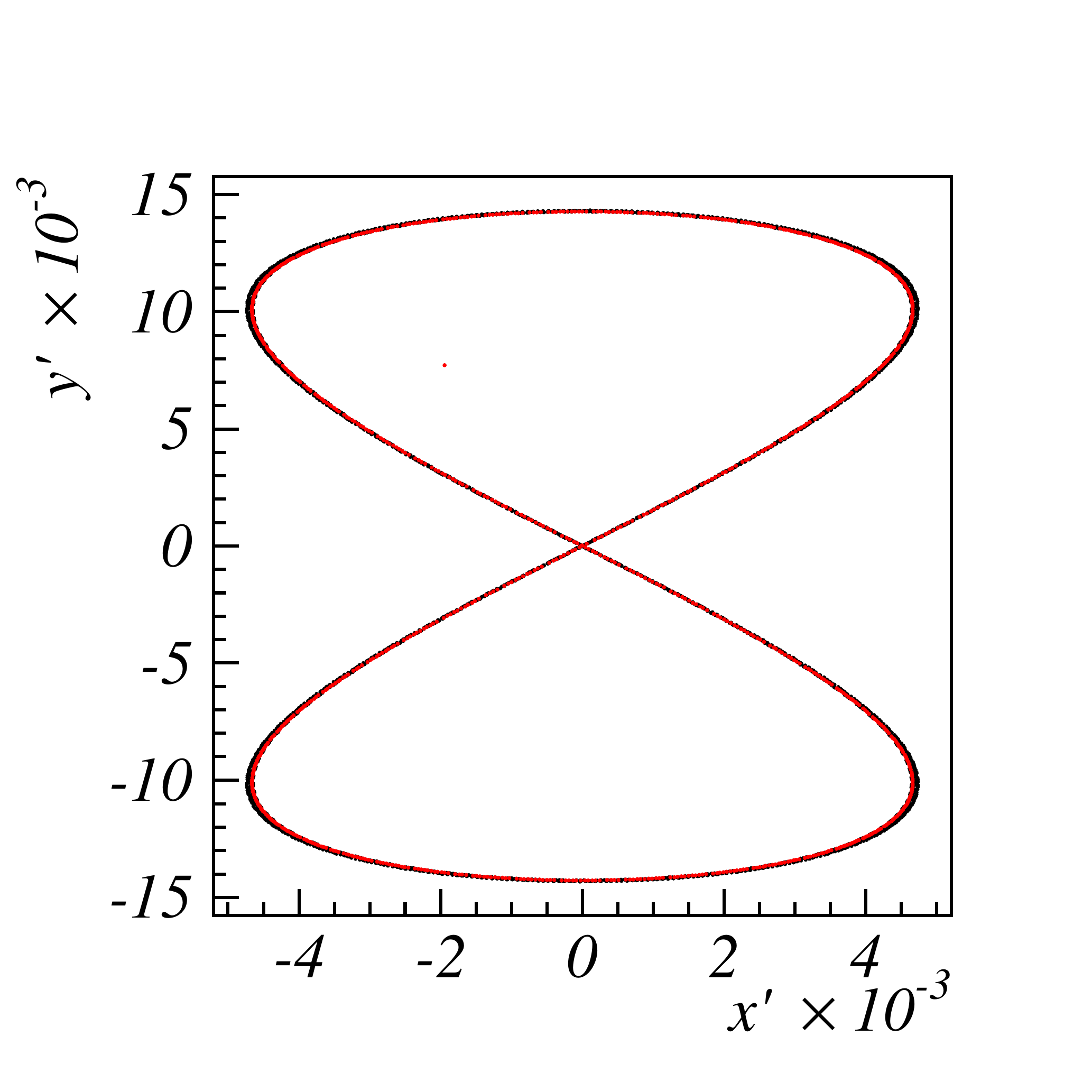,width=49mm}}
 \put(-15,114)   {a)}
 \put( 45,114)   {b)}
 \put(-15,56)    {c)}
 \put( 45,56)    {d)}
          }
 \end{picture}
 \caption{
 The fix-line as predicted by the theory (red dots)
 and by simulations (black dots). 
 Pictures a), b) circular shape shows that $\tax,\tay$ 
 are constant. 
 }
 \label{fig1:fig_15}
 \end{center}
 \end{figure}
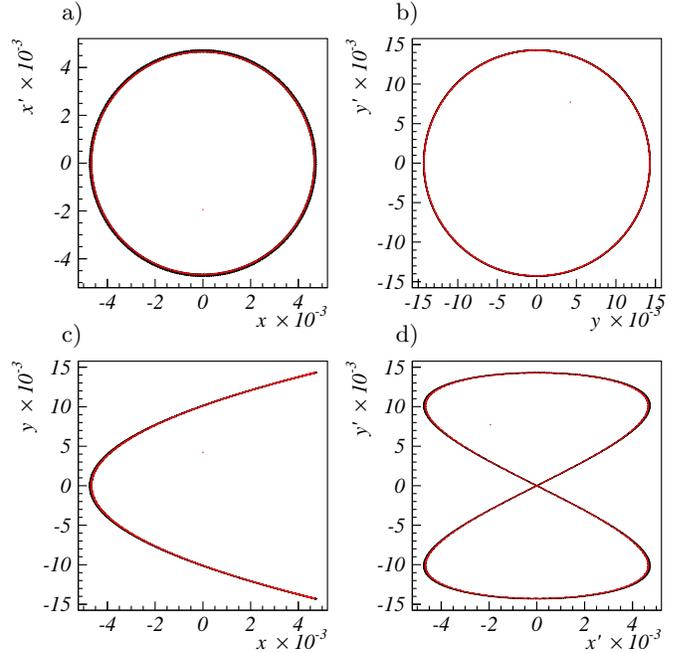
 where $a_x,a_y$ must satisfy Eq.~\ref{eq:41} (top), 
 and $0<t<2\pi$ is here a variable that parameterizes the curve. 
 Therefore the coordinates $x,y$ of the solution in Eq.~\ref{eq:44} are 
 always found on the curve expressed by Eq.~\ref{eq:48}. 
 The same argument applies to any other pairs of coordinates of Eq~\ref{eq:44}. 
 We therefore find that at a given longitudinal position the trajectory  
 Eq~\ref{eq:44} lies on a closed curve. 
 This curve is invariant in its shape, and 
 each point in this curve is mapped after one turn again on the same curve, 
 for this reason this curve is called fix-line, which is a generalization of 
 the fix-point for 1D stable resonances.  
 In Fig.~\ref{fig1:fig_15} we compare the fix-line as predicted by the theory
 (red dots) with simulation results (black dots). 
 The figure shows 4 pictures for the planes, $x-x', y-y',x-y$, and $x'-y'$. 
 It is clearly visible that for the variables $x,y$, and $x'-y'$ the 
 shape is that one predicted from the theory. 
 However, the planes $x-x'$, $y-y'$ exhibits a circular shape 
 showing that $\tax,\tay$ are constant!  
 The simulations are obtained by exciting a single
 harmonics, and $Q_x = 1.324, Q_y = 1.84, \Delta_r = 4.64\times 10^{-3}$,
 and $\Lambda = -7.444\times 10^{-3}$ m$^{-1}$.

 %%%%%%%%%%%%%%%%%%%%%%%%%%%%%%%%%%%%%%%%%%%%%%%%%%%%%%%%%%%%%%%%%%%%%%%%%%%
 \subsection{Stability}
 We consider here a particle slightly displaced 
 from the stationary solution 
 of Eqs.~\ref{eq:36} (this solution is the fix-line) of the 
 amount  $\delta\tax,\delta\tay,\delta\tpx,\delta\tpy$. 
 In order to obtain the equation of motion of the perturbation 
 we expand equations~\ref{eq:34} in 
 $\delta\tax,\delta\tay,\delta\tpx,\delta\tpy$, 
 and keep the first order only. 
 The canonical equations of the perturbation 
 take the form (properly re-arranged) 
 \begin{widetext}
 \begin{equation}
 \begin{aligned}
 \delta\tax' &=
 -2\frac{\partial^2 \tilde H_{s1}}
 {\partial \tpx\partial \tax} \delta\tax
 -2\frac{\partial^2 \tilde H_{s1}}
 {\partial \tpx\partial \tay} \delta\tay
 -2\frac{\partial^2 \tilde H_{s1}}
 {\partial \tpx^2} \delta\tpx
 -2\frac{\partial^2 \tilde H_{s1}}
 {\partial \tpx\partial \tpy} \delta\tpy,
 \\
 \delta\tay' &=
 -2\frac{\partial^2 \tilde H_{s1}}
 {\partial \tpy\partial \tax} \delta\tax
 -2\frac{\partial^2 \tilde H_{s1}}
 {\partial \tpy\partial \tay} \delta\tay
 -2\frac{\partial^2 \tilde H_{s1}}
 {\partial \tpy\partial \tpx} \delta\tpx
 -2\frac{\partial^2 \tilde H_{s1}}
 {\partial \tpy^2} \delta\tpy,
 \\
 \delta\tpx' &=
 2\frac{\partial^2 \tilde H_{s1}}
 {\partial \tax^2} \delta\tax
 +2\frac{\partial^2 \tilde H_{s1}}
 {\partial \tax\partial \tay} \delta\tay
 +2\frac{\partial^2 \tilde H_{s1}}
 {\partial \tax\partial\tilde\varphi_x} \delta\tpx
 +2\frac{\partial^2 \tilde H_{s1}}
 {\partial \tax\partial \tpy} \delta\tpy,
 \\
 \delta\tpy' &=
 2\frac{\partial^2 \tilde H_{s1}}
 {\partial \tay\partial \tax} \delta\tax
 +2\frac{\partial^2 \tilde H_{s1}}
 {\partial \tay^2} \delta\tay
 +2\frac{\partial^2 \tilde H_{s1}}
 {\partial \tay\partial\tilde\varphi_x} \delta\tpx
 +2\frac{\partial^2 \tilde H_{s1}}
 {\partial \tay\partial \tpy} \delta\tpy,
 \\
 \end{aligned}
 \label{eq:49}
 \end{equation}
 \end{widetext}

 where the slowly varying Hamiltonian is given by Eq.~\ref{eq:33}.
 Any term that has only one derivative with respect to
 $\tilde\varphi_x,\tilde\varphi_y$ yields zero because
 that term is proportional to
% \begin{equation}
$   \sin[\tilde\varphi_x+2\tilde\varphi_y]$, 
% \label{eq:50}
% \end{equation}
 which is zero on the fix-line.
 Therefore we get
 \begin{equation}
 \begin{aligned}
 \delta\tax' &=
 -2\frac{\partial^2 \tilde H_{s1}}
 {\partial \tpx^2} \delta\tpx
 -2\frac{\partial^2 \tilde H_{s1}}
 {\partial \tpx\partial \tpy} \delta\tpy,
 \\
 \delta\tay' &=
 -2\frac{\partial^2 \tilde H_{s1}}
 {\partial \tpy\partial \tpx} \delta\tpx
 -2\frac{\partial^2 \tilde H_{s1}}
 {\partial \tpy^2} \delta\tpy,
 \\
 \delta\tpx' &=
 2\frac{\partial^2 \tilde H_{s1}}
 {\partial \tax^2} \delta\tax
 +2\frac{\partial^2 \tilde H_{s1}}
 {\partial \tax\partial \tay} \delta\tay,
 \\
 \delta\tpy' &=
 2\frac{\partial^2 \tilde H_{s1}}
 {\partial \tay\partial \tax} \delta\tax
 +2\frac{\partial^2 \tilde H_{s1}}
 {\partial \tay^2} \delta\tay.
 \\
 \end{aligned}
 \label{eq:51}
 \end{equation}
 We define for convenience
 \begin{equation}
 \begin{split}
 &\lambda =
 2\frac{\partial^2 \tilde H_{s1}}{\partial \tpx^2} =
 2\frac{1}{2}\frac{\partial^2 \tilde H_{s1}}{\partial \tpx\partial \tpy} =
 2\frac{1}{4}\frac{\partial^2 \tilde H_{s1}}{\partial \tpy^2}, \\
 &A_{xx} = 2\frac{\partial^2 \tilde H_{s1}}
 {\partial \tax^2},
 \quad
 A_{xy} = 2\frac{\partial^2 \tilde H_{s1}}{\partial \tax\partial \tay}, \\
 \quad
 &A_{yx} = 2\frac{\partial^2 \tilde H_{s1}}{\partial \tay\partial \tax}, 
 \quad
 A_{yy} = 2\frac{\partial^2 \tilde H_{s1}}{\partial \tay^2}. \\
 \end{split}
 \label{eq:52}
 \end{equation}
 Therefore we always find that the motion of the perturbation is given
 by the following system of equations
 \begin{equation}
 \begin{aligned}
 \delta\tax' &= - \lambda\delta\tpx -2\lambda \delta\tpy,
 \\
 \delta\tay' &= -2\lambda\delta\tpx -4\lambda\delta\tpy,
 \\
 \delta\tpx' &= A_{xx} \delta\tax + A_{xy} \delta\tay,
 \\
 \delta\tpy' &= A_{yx} \delta\tax + A_{yy} \delta\tay,
 \\
 \end{aligned}
 \label{eq:53}
 \end{equation}
 which after some algebra becomes
 \begin{equation}
 \begin{aligned}
 \delta\tax'' =& -\omega^2\delta\tax + \lambda(A_{xy} + 2 A_{yy}) C_p,  \\
 \delta\tay'' =& -\omega^2\delta\tay - \lambda(A_{xx} + 2 A_{yx}) C_p,  \\
 \delta\tpx'' =& -\lambda(A_{xx} + 2 A_{xy})(\delta\tpx +2\delta\tpy),  \\
 \delta\tpy'' =& -\lambda(A_{xy} + 2 A_{yy})(\delta\tpx +2\delta\tpy),  \\
 \end{aligned}
 \label{eq:54}
 \end{equation}
 where we defined
 \begin{equation}
   \omega^2=\lambda(A_{xx} + 2 A_{yx} + 2A_{xy} + 4 A_{yy}).
 \label{eq:55}
 \end{equation}
 The coefficient $C_p$ is a constant obtained from integrating 
 Eqs.~\ref{eq:53}, we call it $C_p$ with index $p$ as we refer to 
 the perturbation around the fix-line. 

The value of $C_p$ depends on the initial condition of the perturbation 
$\delta\tax,\delta\tay,\delta\tpx,\delta\tpy$. 
From Eq.~\ref{eq:53} we find
 \begin{equation}
   2\delta\tax = \delta\tay + C_p.
 \label{eq:56}
 \end{equation}
 We also find
 \begin{equation}
   (\delta\tpx+2\delta\tpy)'' = -\omega^2(\delta\tpx+2\delta\tpy).
 \label{eq:57}
 \end{equation}
 We conclude that for
 \begin{equation}
   \omega^2 > 0
 \label{eq:58}
 \end{equation}
 the motion of $\delta\tax,\delta\tay$, and $\delta\tpx+2\delta\tpy$
 is stable.
 The evolution of $\delta\tpx,\delta\tpy$ 
 instead is not bounded if $C_p \ne 0$ as a term linear in $\theta$ 
 appears in the solution of the equation of motion. 
 The derivation of the solution is reported in Appendix A. 
 An interesting consequence in the case of $C_p \ne 0$ is that there 
 exists a neighbouring fix-line identified by 
 $\tax^c, \tay^c$, and a point on it 
 $\tax^c, \tay^c, \tpx^c, \tpy^c$, 
 so that the initial perturbation assumes values 
 $(\delta\tax)_1, (\delta\tay)_1, (\delta\tpx)_1, (\delta\tpy)_1$, and 
 $C'_p = 2(\delta\tax)_1 - (\delta\tay)_1 = 0$. 
 That means that the initial perturbation oscillates now around 
 the new fix-line point. 
 The proof of this result is shown in the Appendix A.

 %%%%%%%%%%%%%%%%%%%%%%%%%%%%%%%%%%%%%%%%%%%%%%%%%%%%%%%%%%%%%%%%%%%%%%%%%%%%%%5
 \section{Equation of motion with the resonant term only}
 We discuss the fix-lines for the simplified case of ${\cal F} =0$,
 i.e. in absence of detuning induced by other non-linearities like
 space charge, or strong nonlinear components due to magnets.
 The equation of stationary solutions reads 
 (see Eq.~\ref{eq:39} with ${\cal F}_x={\cal F}_y=0$) 
 \begin{equation}
 \left\{
 \begin{aligned}
 0 &=\Lambda  \frac{1}{2\sqrt{\tax}}\tay (-1)^M
     + \frac{\Delta_r}{2} -t_y\\
 0 &= \Lambda \sqrt{\tax} (-1)^M + \frac{t_y}{2}.\\
 \end{aligned}
 \right.
 \label{eq:62}
 \end{equation}
 We remind here for convenience that $\Delta_r = Q_x + 2Q_y$. 
 Adding the second equation multiplied by 2 to the first one $t_y$
 can be eliminated and we obtain
 \begin{equation}
 0 =\Delta_r \Lambda (-1)^M
     \left[\frac{1}{2\sqrt{\tax}}\tay +
          2\sqrt{\tax}
     \right] + \frac{\Delta_r^2}{2}.
 \label{eq:63}
 \end{equation}
 Therefore if the solutions $\tax,\tay$ exists, it also holds
 \begin{equation}
   \Delta_r \Lambda (-1)^M \le 0, 
 \label{eq:64}
 \end{equation}
 which puts a constrains on $t_x,t_y$ as follows
 \begin{equation}
 \left\{
 \begin{aligned}
 0 &= \Delta_r \Lambda \frac{1}{2\sqrt{\tax}}\tay (-1)^M  + \Delta_r \frac{t_x}{2}\\
 0 &= \Delta_r \Lambda \sqrt{\tax}  (-1)^M + \Delta_r \frac{t_y}{2}\\
 \end{aligned}
 \right.
 \rightarrow
 \left\{
 \begin{aligned}
 &\Delta_r t_x \ge 0\\
 &\Delta_r t_y \ge 0\\
 \end{aligned}
 \right.
 \label{eq:65}
 \end{equation}
 As a consequence if the solution exists, then it must be
 \begin{equation}
   \Delta_r^2 \ge \Delta_r t_x \ge 0.
 \label{eq:66}
 \end{equation}
 In that case we can parameterize $(t_x,t_y)$ as follows
 \begin{equation}
   t_x = \tau \Delta_r, \qquad t_y = \frac{\Delta_r}{2}(1-\tau), 
 \label{eq:67}
 \end{equation}
 with $0\le \tau \le 1$ and the solutions for the fix-lines take the
 form
 \begin{equation}
 \left\{
 \begin{aligned}
 \tax &= \frac{\Delta_r^2}{16\Lambda^2 }(1-\tau)^2\\
 \tay &= \frac{\Delta_r^2}{ 4\Lambda^2 }\tau(1-\tau)\\
 \end{aligned}
 \right.
 \label{eq:68}
 \end{equation}
 At this point, as a last step, it should be checked if the solution
 really exists.  Substituting the previous equations into
 Eq.~\ref{eq:63} we find that the solution exists as long as
 \begin{equation}
   \frac{\Lambda}{|\Lambda|}\frac{\Delta_r}{|\Delta_r|}(-1)^M + 1 = 0, 
 \label{eq:69}
 \end{equation}
 which can be satisfied only if $\Lambda \Delta_r (-1)^M < 0$.
 Our analysis shows that in proximity of the coupled 3rd order resonance
 there exist an infinite number of fix-lines.

 %%%%%%%%%%%%%%%%%%%%%%%%%%%%%%%%%%%%%%%%%%%%%%%%%%%%%%%%%%%%%%%%%%%%%%%%%%%
 \subsection{Analysis of the stability around the fix-line}
 By using Eqs.~\ref{eq:65}, the stability coefficients of Eqs.~\ref{eq:52}
 are
 {\small
 \begin{equation}
 \begin{aligned}
   A_{xx}   &= -2 \Lambda \frac{1}{4} \tax^{-3/2}\tay (-1)^M = \frac{t_x}{2\tax}, \\
   A_{xy}   &=  A_{yx} = 2 \Lambda \frac{1}{2}  \tax^{-1/2} (-1)^M = - \frac{t_x}{\tay}, \\
   A_{yy}   &=  0, \\
   \lambda &= -2 \Lambda \sqrt{\tax}\tay (-1)^M = \tay t_y, \\
 \end{aligned}
 \label{eq:70}
 \end{equation}
 }
 where we have used the relation Eqs.~\ref{eq:36} on the fix-line for
 ${\cal F}=0$.
 After some algebra we find $\omega^2 = t_x^2 - 4t_xt_y$,
 which, in terms of the parametrization of the fix-line, becomes
 \begin{equation}
   \omega^2 = \Delta_r^2\tau(3\tau-2). 
 \label{eq:71}
 \end{equation}
This equation shows that with $\tau > 2/3$,   $\omega^2$ is positive, hence 
Eqs.~\ref{eq:54} yields stable motion, whereas for $\tau < 2/3$, 
$\omega^2$ is negative which means that the motion is unstable. 

 Therefore this analysis shows that with $\tau>2/3$ a fix-line is stable.

\begin{figure}[H]
 \begin{center}
 \unitlength 0.7mm
 \begin{picture}(80,225)
 \put(-27, 110) {\epsfig{file=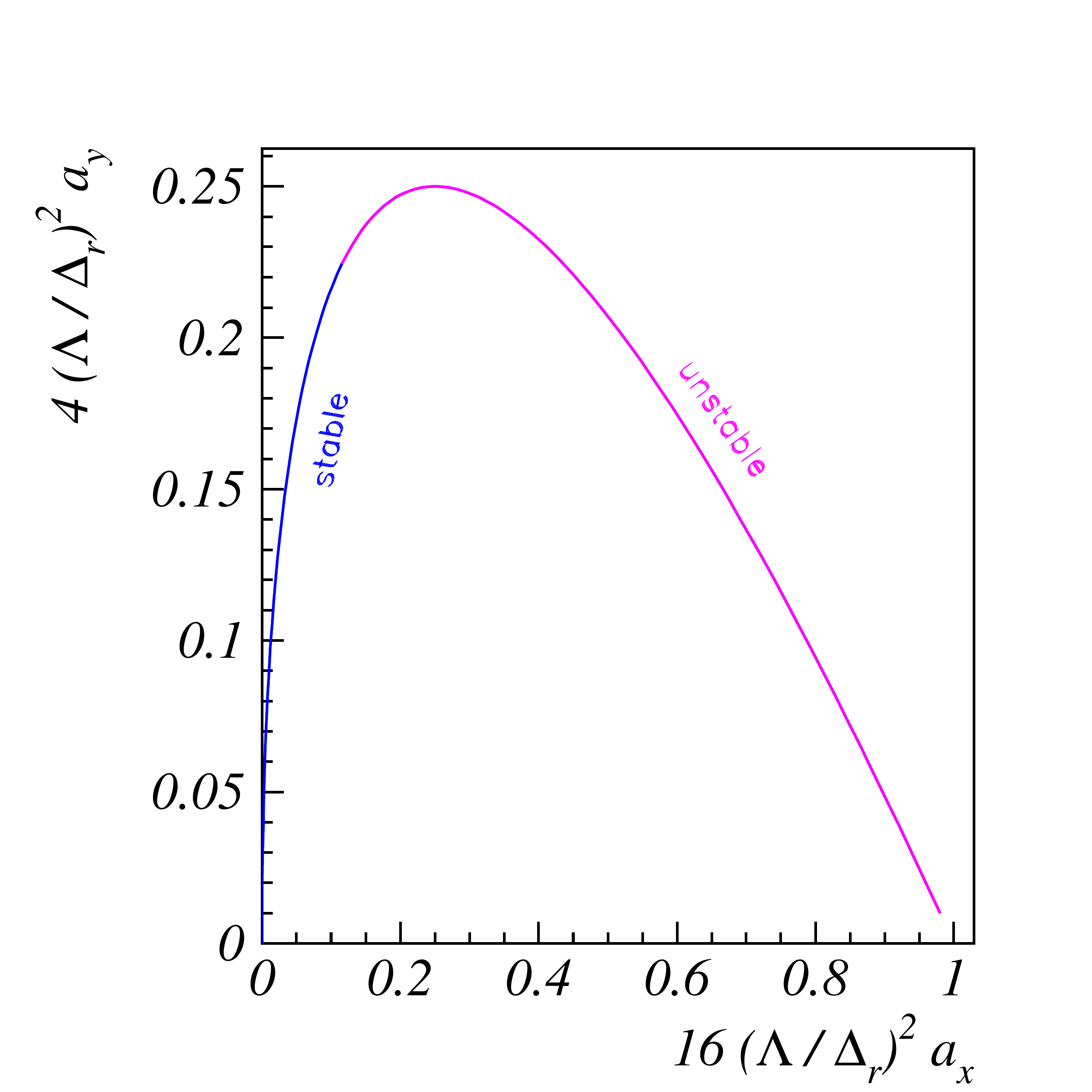,width=89mm}}
 \put(-20,   0) {\epsfig{file=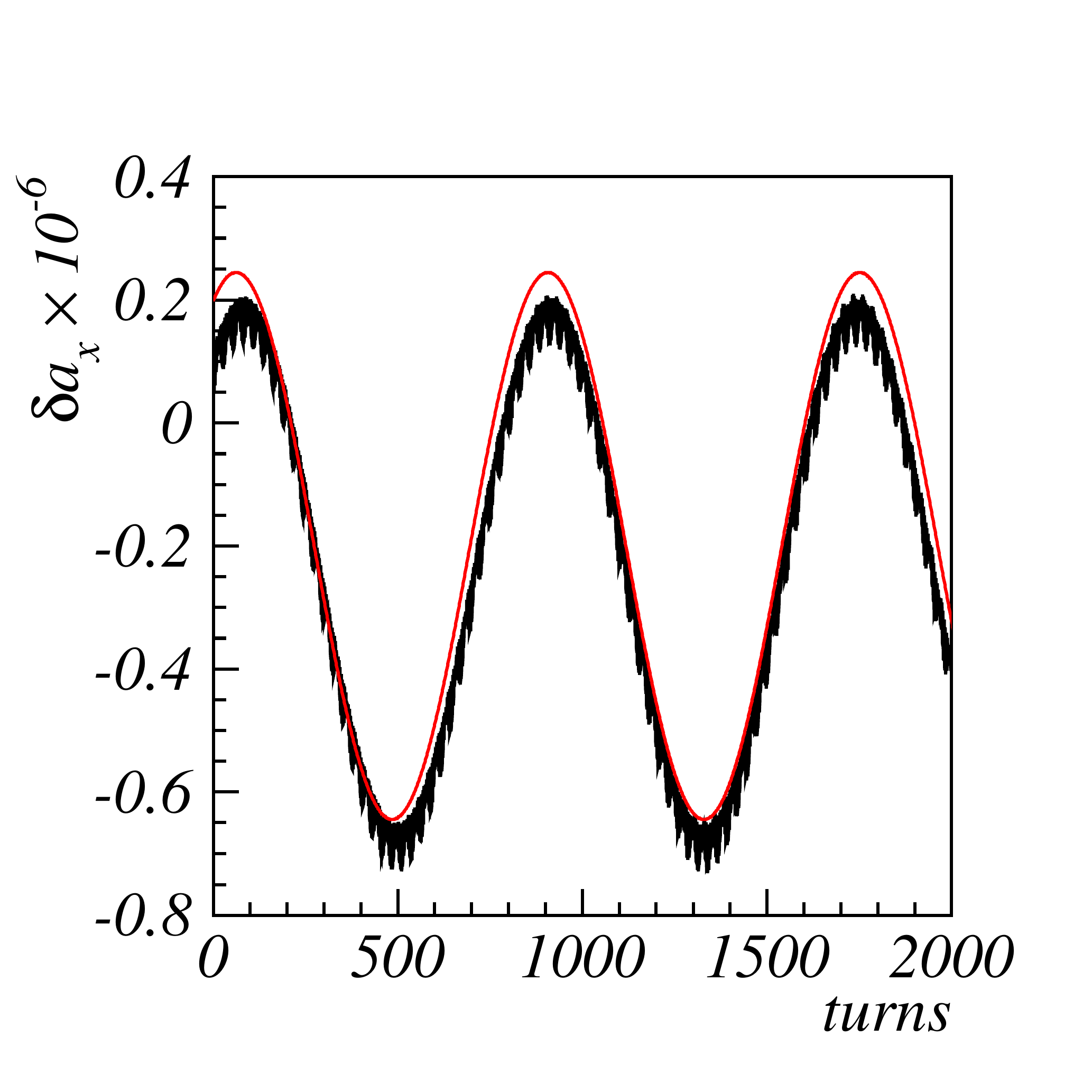,width=85mm}}
 \put(-10,223) {a)}
 \put(-10,110) {b)}
 \end{picture}
 \caption{
 On the top are shown the set of stable and unstable fix-lines.  The
 bottom picture shows the oscillations of $\tax$ around the fix-line
 value.  In this example $C_p\ne 0$, hence the oscillation of the
 perturbation $\delta\tax$, is not centered around zero, i.e. on the
 unperturbed fix-line.  The center of oscillation $\delta\tax^c$ is
 located in this example at $-0.2\times 10^{-6}$.  
 The black curve is obtained from computer simulation of the 
 accelerator model,  while the red curve is obtained from the theory. 
 The wavelength of the red curve is obtained from
 Eq.~\ref{eq:71}.
}
 \label{fig1:fig_16}
 \end{center}
 \end{figure}

 The situation is summarized in Fig.~\ref{fig1:fig_16} top, where the
 collection of all fix-lines is represented. The point of merging of
 the stable with the unstable fix-line has $\omega^2=0$ and that
 corresponds to $\tau= 2/3$.

 Figure~\ref{fig1:fig_16} bottom shows the comparison of the 
 oscillation of an initial condition slightly off a fix-line.
 It is evident that the center of oscillation is not $(0,0)$ because
 $2 \delta\tax \ne \delta\tay$ (hence $C_p \ne 0$). 
 The red curve is drawn from the theoretical model
 by fitting the amplitude, however, the wavelength is given by $\omega$.
 The parameters of the simulations are the same as for Fig.~\ref{fig1:fig_15}, 
 namely,  $\Delta_r=4.64\times 10^{-3}$, 
 $\Lambda = -7.444\times 10^{-3}$ m$^{-1}$. 
 The fix-line is taken for $\tau=0.7$, which implies 
 $t_x=3.248\times 10^{-3}$, and $t_y=6.960\times 10^{-4}$. 
 %
 %
 %

 %%%%%%%%%%%%%%%%%%%%%%%%%%%%%%%%%%%%%%%%%%%%%%%%%%%%%%%%%%%%%%%%%%%%%%
 \section{Stability of Motion}
 We make use of the properties of the equations of motion
 Eqs.~\ref{eq:34}.  In particular of the invariant $2\tax - \tay = C$
 (Eq.~\ref{eq:35}).  By using it, we can drop one of the two equations
 of Eqs.~\ref{eq:34}, and summing the other two we get
 {\small
 \begin{equation}
 \begin{aligned}
 -\tax'                                 &=       -2\Lambda \sqrt{\tax}(2\tax-C)  \sin[\tilde\varphi_x+2\tilde\varphi_y],\\
 \tilde\varphi_x' + 2 \tilde\varphi_y'  &= \quad  2\Lambda \frac{1}{2\sqrt{\tax}}(2\tax-C) \cos[\tilde\varphi_x+2\tilde\varphi_y] + \\
                                        & \quad  +4\Lambda\sqrt{\tax} \cos[\tilde\varphi_x+2\tilde\varphi_y] + \Delta_r + 2{\cal F}_x+4{\cal F}_y.\\
 \end{aligned}
 \label{eq:72}
 \end{equation}
 }

 These equations show that the natural variable for the system is
 $\Omega=\tilde\varphi_x + 2 \tilde\varphi_y$, and the equations become
 \begin{equation}
 \begin{aligned}
 -\tax'  =&  -2\Lambda \sqrt{\tax}(2\tax-C)  \sin[\Omega],\\
 \Omega' =&    \Lambda\left[ \frac{1}{\sqrt{\tax}}(2\tax - C) + 
 4\sqrt{\tax}\right]\cos[\Omega] + \\
 &+ \Delta_r + 2{\cal F}_x+4{\cal F}_y.\\
 \end{aligned}
 \label{eq:73}
 \end{equation}
 For the case in which we take only the resonant term we have
 ${\cal F}_x = {\cal F}_y=0$, and the equations take the form
 \begin{equation}
 \begin{aligned}
 \tax'   &=   2\Lambda \sqrt{\tax}(2\tax-C)  \sin[\Omega],\\
 \Omega' &=    \Lambda\left[ \frac{1}{\sqrt{\tax}}(2\tax - C) + 4\sqrt{\tax}\right]\cos[\Omega] + \Delta_r.\\
 \end{aligned}
 \label{eq:74}
 \end{equation}
 Now consider the function
 \begin{equation}
   I(\tax,\Omega) = 2\Lambda \sqrt{\tax}(2\tax-C)  \cos[\Omega] + \tax\Delta_r, 
 \label{eq:75}
 \end{equation}
 and note that
 \begin{equation}
 \begin{aligned}
   \frac{\partial I}{\partial \tax} =&
  \Lambda \left[\frac{1}{\sqrt{\tax}}(2\tax-C) + 4\sqrt{\tax} \right]
  \cos[\Omega] + \Delta_r = \Omega', \\
   \frac{\partial I}{\partial \Omega} =&
   -2\Lambda \sqrt{\tax}(2\tax-C)  \sin[\Omega] = - \tax'. 
 \end{aligned}
 \label{eq:76}
 \end{equation}
 Therefore we find
 \begin{equation}
   \frac{dI}{d\theta} = \frac{\partial I}{\partial \tax}\tax' + \frac{\partial I}{\partial \Omega}\Omega'
   = \Omega'\tax' +(-\tax')\Omega' = 0. 
 \label{eq:77}
 \end{equation}
 That means that $I(\tax,\Omega)$ is an invariant of motion and the
 trajectories of particles are set by the value of the invariant
 (that we call $I_0$).  By using $I(\tax,\Omega)$ it is possible to
 discuss the stability of the motion of particles.  In fact given the
 initial conditions of $\tax,\tay,\Omega$ we find $C$ from the
 relation $2\tax-\tay = C$, hence we find the value of the invariant
 $I_0=I(\tax,\Omega)$.  The trajectory of that particle is then given
 by the set $\Gamma_{I_0} = \{(\tax,\Omega): I(\tax,\Omega) = I_0\}$.
 We observe that if the trajectory extends up to infinity, it means
 that $\tax\rightarrow\infty$, hence $\tay\rightarrow\infty$, while
 $0\le\Omega\le 2\pi$.  In this situation the first term of
 Eq.~\ref{eq:75} diverges as $ 4\Lambda \tax^{3/2} \cos[\Omega]$, and
 this divergence is faster than the one of the second term, which
 diverges as $\tax\Delta_r$.  That means that $\Omega$ must change
 accordingly for keeping unchanged the value of the invariant,
 therefore for $\tax\rightarrow\infty$ we find $\Omega \rightarrow
 \pi/2+n\pi$, with $n$ an integer.
 
 We observe that the function $I=I(\tax,\Omega)$ defines a function
 $\tax=\tax(\Omega)$, which has the following property: if $\Omega =
 2\pi n$ ($n$ is an integer) and $\tax$ is not a fix-line, then
 $\frac{d\tax}{d\Omega}=0$.  In fact, as $I$ is an invariant,
 $dI/d\Omega=0$, but
 \begin{equation}
   \frac{dI}{d\Omega} =
   \frac{\partial I}{\partial \tax} \frac{d\tax}{d\Omega} +
   \frac{\partial I}{\partial \Omega}
 \label{eq:78}
 \end{equation}
 that is
 \begin{equation}
   \frac{dI}{d\Omega}=\Omega' \frac{d\tax}{d\Omega} -\tax' = 0.
 \label{eq:79}
 \end{equation}
 Therefore for $\Omega = \pi n$ it holds $\sin[\Omega]=0$, hence 
 from Eqs.~\ref{eq:73}, we find $\tax'=0$. 
 On the other hand, if $\tax,\Omega$ is not a fix-line then 
 $\Omega'\ne 0$ (because $\tpx'=\tpy'=0$ is the condition for 
 finding a fix-line).  
 Therefore if $\Omega = 2\pi n$, and $\tax$ is not a fix-line then 
 \begin{equation}
   \frac{d\tax}{d\Omega} = 0.
 \label{eq:80}
 \end{equation}
 This property tells us that the implicit functions $\tax=\tax(\Omega)$
 defined by $I=I(\tax,\Omega)$
 are periodic in $\Omega$ and at the periodicity they merge smoothly 
 (that is the first order derivative in $\Omega$ is zero at $\Omega = 2\pi n$).

 As the fix-lines are described by Eqs.~\ref{eq:68} we will consider
 the following scaling
 \begin{equation}
 \left\{
 \begin{aligned}
 \tax &= \frac{\Delta_r^2}{16\Lambda^2 }\hax\\
 \tay &= \frac{\Delta_r^2}{16\Lambda^2 }\hay\\
 \end{aligned}
 \right.
 \label{eq:81}
 \end{equation}
 The relation $2\tax=\tay + C$ becomes
 \begin{equation}
   2\hax = \hay + \frac{16\Lambda^2 }{\Delta_r^2} C. 
 \label{eq:82}
 \end{equation}
 It is convenient to define the parameter 
 \begin{equation}
   \xi = \frac{16\Lambda^2 }{\Delta_r^2} \frac{C}{2} 
 \label{eq:83}
 \end{equation}
 so that we obtain the relation $2\hax-\hay = 2 \xi$. 
 The invariant $I$ defined in Eq.~\ref{eq:75} rescaled in the
 ``normalized coordinates'' $\hax,\hay$ reads
 \begin{equation}
   \hat I = \frac{I 16 \Lambda^2}{\Delta_r^3} 
   = \mu\sqrt{\hax} ( \hax - \xi)\cos\Omega + \hax, 
 \label{eq:84}
 \end{equation}
 with
 \begin{equation}
   \mu =
    \frac{\Lambda}{|\Lambda|}
    \frac{|\Delta_r|}{\Delta_r}. 
 \label{eq:85}
 \end{equation}
 This relation satisfies the relation $\mu^2=1$. 
 Note that, as $2\hax-\hay = 2 \xi$ it follows that $\hax$ must
 simultaneously satisfy the following conditions
 \begin{equation}
 \left\{
 \begin{aligned}
 \hax & \ge 0\\
 \hax & \ge \xi\\
 \end{aligned}
 \right.
 \label{eq:86}
 \end{equation}
 which express simply the condition $\hax \ge 0, \hay \ge 0$ 
 intrinsic in the definition of $\hax,\hay$. 
 In addition we observe that all the possible curves
 $\hat I(\hax,\Omega)=const.$ never intercept.
 
 The trajectories can be found from $\Omega$ as function of $\hax$ 
 inverting the Eq.~\ref{eq:84}, obtaining
 \begin{equation}
   \cos[\Omega] = \mu\frac{1}{\sqrt{\hax}}\frac{\hat I - \hax}{\hax-\xi}, 
 \label{eq:87}
 \end{equation}
 where we used the property $\mu^2=1$ (from Eq.~\ref{eq:85}). To
 study this equation we make use of an auxiliary function defined as
 \begin{equation}
   \ya(\hax) = -\mu\cos[\Omega] =
   -\frac{1}{\sqrt{\hax}}\frac{\hat I - \hax}{\hax-\xi}.
 \label{eq:88}
 \end{equation} 
 This function is defined in the plane $\hax,\ya$, which is the plane
 about which we will discuss next. Note that $\xi$ is here a constant
 of motion defined by the initial conditions. Hence by varying $\hat
 I$ we identify a specific level line of the invariant. If $-1 \le \ya
 \le 1$ then $\ya=-\mu\cos[\Omega]$ can be solved and we find
 $\Omega$; if $\ya<-1$ or $1<\ya$ then $\ya=-\mu\cos[\Omega]$ cannot
 be solved so that these values of $\ya$ are not acceptable.

%%%%%%%%%%%%%%%%%%%%%%%%%%%%%%%%%%%%%%%%%%%%%%%%%%%%%%%%%%%%%%%%%%%%%%%5
 \subsection{Fix-lines and invariants}
 Next we prove the following property: 
 consider a level line identified by $\hat I$, of the form 
 $\ya=\ya(\hax,\hat I,\xi)$ defined by Eq.~\ref{eq:88}. 
 If this curve is tangent to the line $\ya=1$ (in the plane $\hax,\ya$) 
 then the tangent point $(\hax,\ya)$ identifies a fix-line. 

 Proof: 
 Firstly, we observe that Eq.~\ref{eq:84} is simply a rewrite of the
 Hamiltonian, from which we can obtain the equations of motion as 
 \begin{equation}
 -\hax'   
  = \Delta_r\frac{\partial \hat I}{\partial \Omega}, \qquad
  \Omega' 
  = \Delta_r\frac{\partial \hat I}{\partial \hax}. 
 \label{eq:91}
 \end{equation}

 The function  $\ya=\ya(\hax)$ is an implicit function of the 
 form $\hat I = \hat I(\hax,\ya(\hax),\xi)$ with $\hat I=\textrm{const.}$. 
 Therefore 
 \begin{equation}
    \frac{d}{d\hax}\hat I(\hax,\ya(\hax),\xi) = 
   \frac{\partial\hat I}{\partial \hax} +
   \frac{\partial\hat I}{\partial \ya} \frac{d\ya}{d\hax} 
   = 0. 
 \label{eq:89}
 \end{equation} 
 If for $\hax$ the level line $\ya=\ya(\hax)$ is tangent to a
 horizontal line, then we have $d\ya(\hax)/d\hax=0$.  This applied in
 Eq.~\ref{eq:89} implies that on the tangent point one finds
 \begin{equation}
   \frac{\partial\hat I}{\partial\hax}(\hax,\ya(\hax),\xi) = 0, 
 \label{eq:90}
 \end{equation}
 which via the second part of the Eqs.~\ref{eq:91} yields $\Omega'=0$
 (or $\ya'=0$).  Secondly, we use the condition that $\ya=\ya(\hax)$
 is tangent to $\ya=1$; that means that $\cos[\Omega]=-\mu$,
 i.e. $\sin[\Omega]=0$, which yields $\partial\hat I/\partial\Omega =
 0$.  Therefore via the first part of Eqs.~\ref{eq:91} we find
 $\hax'=0$.  If at any time $\hax'=\Omega'=0$ then it follows that
 $\hax = \textrm{const.}, \Omega = \textrm{const.}$, because
 $\hax,\Omega$ are the canonical variables of the Hamiltonian
 Eq.~\ref{eq:84}.

 The condition of $\ya=\ya(\hax)$ being tangent to 
 $\ya=1$ implies $\cos[\Omega]=-\mu$, which we can write 
 as $\cos[\Omega]= (-1)^M$, with $M$ an integer. 
 By using this we find that Eq.~\ref{eq:90} is just a reformulation 
 of Eq.~\ref{eq:63}, which is the equation defining $\hax,\hay$ of 
 a fix-line. 
 This equation admits always a solution, in fact the condition $\ya=1$, 
 which reads 
 \begin{equation}
   -\frac{|\Lambda|}{\Lambda}
    \frac{\Delta_r}{|\Delta_r|}
    (-1)^M = 1, 
 \label{eq:94}
 \end{equation}
 is just the condition of existence of a fix-line Eq.~\ref{eq:69}. 

 Therefore the point $\hax$ where $\ya=\ya(\hax)$ is tangent to the 
 horizontal line $\ya=1$ identifies a fix-line. 
 As we have already seen this fix-line is parameterized by $\tau$, 
 and the relation between $\tau$ and the fix-line $\hax,\hay$  
 (see Eq.~\ref{eq:68}) reads
 \begin{equation}
   \hax = (1-\tau)^2, \qquad \hay = 4\tau(1-\tau). 
 \label{eq:92}
 \end{equation}
 For these two values of $\hax,\hay$ we find that $\tpx'=\tpy'=0$ 
 (Eqs.~\ref{eq:62}), and the angles $\tpx,\tpy$ have to satisfy the condition 
 $\tpx+2\tpy=M\pi$. 
 The relation of $\xi$ to $\tau$ is 
 \begin{equation}
   \xi = 3\tau^2-4\tau+1. 
 \label{eq:93}
 \end{equation}

 %%%%%%%%%%%%%%%%%%%%%%%%%%%%%%%%%%%%%%%%%%%%%%%%%%%%%%%%%%%%%%%%%%%%%%%5
 \subsection{Stability of motion}
 \label{SOM}
 
 The stability of a trajectory is characterized by the level lines 
 as follows:
 
 {\it A particle motion defined by the initial condition 
 $((\hax)_0,\ya_0)$ is stable if $-1 < \ya_0 < 1$ and 
 its level line $(\hax,\ya(\hax))$ crosses both the lines 
 $\ya=1$ and $\ya=-1$ or twice the same line  $\ya=1$ or $\ya=-1$. 
 Only in this case the motion is bounded by the periodicity of the
 coordinate $\Omega$.
 }
 
 We next discuss the properties of stability as a function of $\xi$.

 %%%%%%%%%%%%%%%%%%%%%%%%%%%%%%%%%%%%%%%%%%%%%%%%%%%%%%%%%%%%%%%%%%%%%%%%%%
 \subsubsection{Region $1 < \xi$}
 Observing 
 %Figs.~\ref{fig1:fig_26} 
 part a) of Fig.~\ref{fig1:fig_17} 
 the red line $\hat I = \xi$ separates two classes of level curves, 
 the curves below the 
 red line go to $-\infty$ for $\hax$ approaching $\xi$; 
 Appendix B discusses the geometrical properties of these level lines. 
 It is clear that if
 $1/\sqrt{\xi} < 1$, i.e. for $\xi > 1$, the red curve that separates the
 two classes of level lines is located between $\ya=1$ and $\ya=0$ so that
 no particle can be stable.
 Any pair $(\hax,\ya)$ chosen in $\xi < \hax$ and $-1<\ya<1$ defines a
 level line that goes to $\hax \rightarrow \infty, \ya\rightarrow 0$ 
 without crossing $\ya = \pm 1$.

 %%%%%%%%%%%%%%%%%%%%%%%%%%%%%%%%%%%%%%%%%%%%%%%%%%%%%%%%%%%%%%%%%%%%%%%%%%%
 \subsubsection{Region $0 < \xi \le 1$}
 This case is also illustrated in part a) of Fig.~\ref{fig1:fig_17}. 
 The level line tangent to $\ya=1$ 
 determines a region dashed in pink in which 
 every initial $((\hax)_0,\ya_0)$ evolves on a level line that
 is bounded. Specifically as the particle trajectory 
 $(\hax,\ya)$ cannot exceed
 $-1 < \ya < 1$, it follows that the particle bounces back and forth on the
 lines $\ya=-1$, and $\ya=1$ keeping the invariant constant. 
 Particles on the right of the level line tangent $\ya=1$ cannot be stable.
 It appears evident that the maximum region of stability, marked by 
 a region of thin red lines is determined by the level line tangent 
 to $\ya=1$. 
 As discussed in the previous section, the point tangent to the line 
 $\ya=1$ is a fix-line. 
 This point is indicated as $(\hax^*)_+$ in part a) of 
 Fig.~\ref{fig1:fig_17}. 
 On the same picture we also find the 
 point $(\hax)_1$, which marks the region of total stability. 
 This means that for $\xi < \hax < (\hax)_1$ particles are stable for 
 any initial phase $\Omega$. 
 For $(\hax)_1 < \hax < (\hax^*)_+$ only particles which have $\ya$ 
 in the proper range (namely $\Omega$ in the proper interval, 
 see Fig.~\ref{fig1:fig_21b}) are stable. 
 In the laboratory frame, this means 
 that only particles with $\varphi_x+2\varphi_y$ 
 in the same range can be stable. 
 The point $(\hax^*)_+$ becomes the more ``external'' point of stability. 
 We note that for $\xi=1$ the point $1/\sqrt{\xi}$ becomes 1, and the 
 pink dashed area goes to zero. 

 %%%%%%%%%%%%%%%%%%%%%%%%%%%%%%%%%%%%%%%%%%%%%%%%%%%%%%%%%%%%%%%%%%%%%%%5
 \subsubsection{Region $-1/4 < \xi \le 0$}
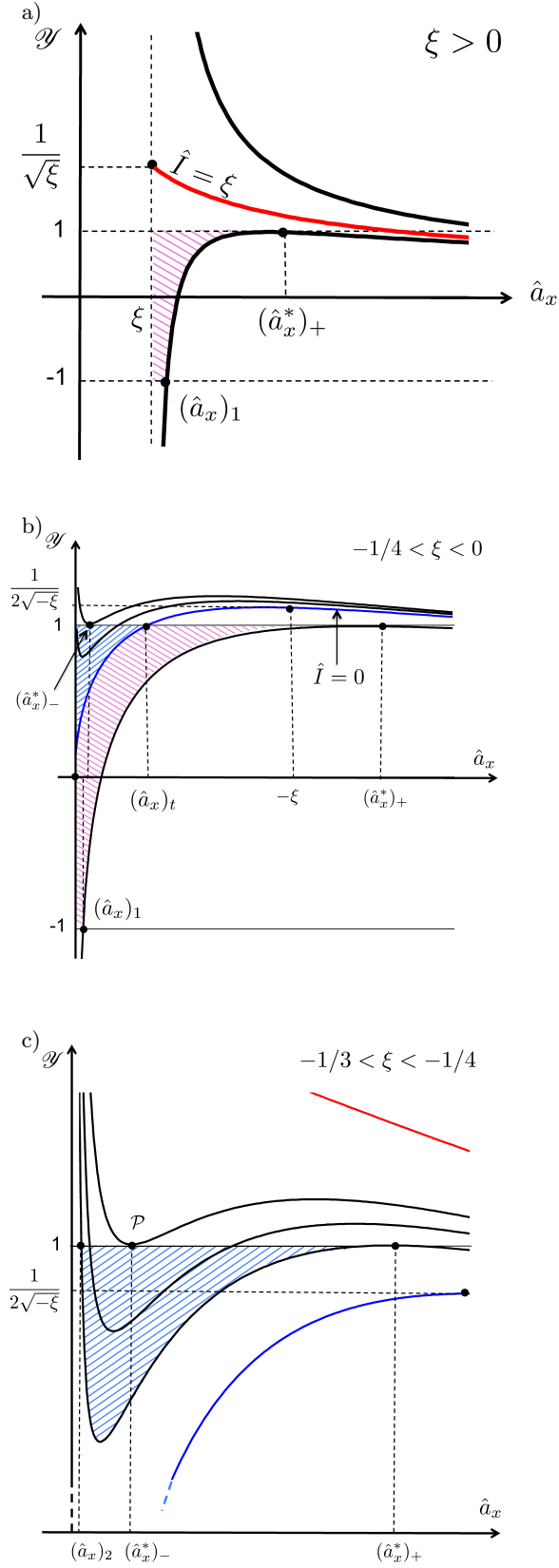
\begin{figure}[H]
 \begin{center}
 \unitlength 0.9mm
 \begin{picture}(80,240)
 \put( 0,160)  {\epsfig{file=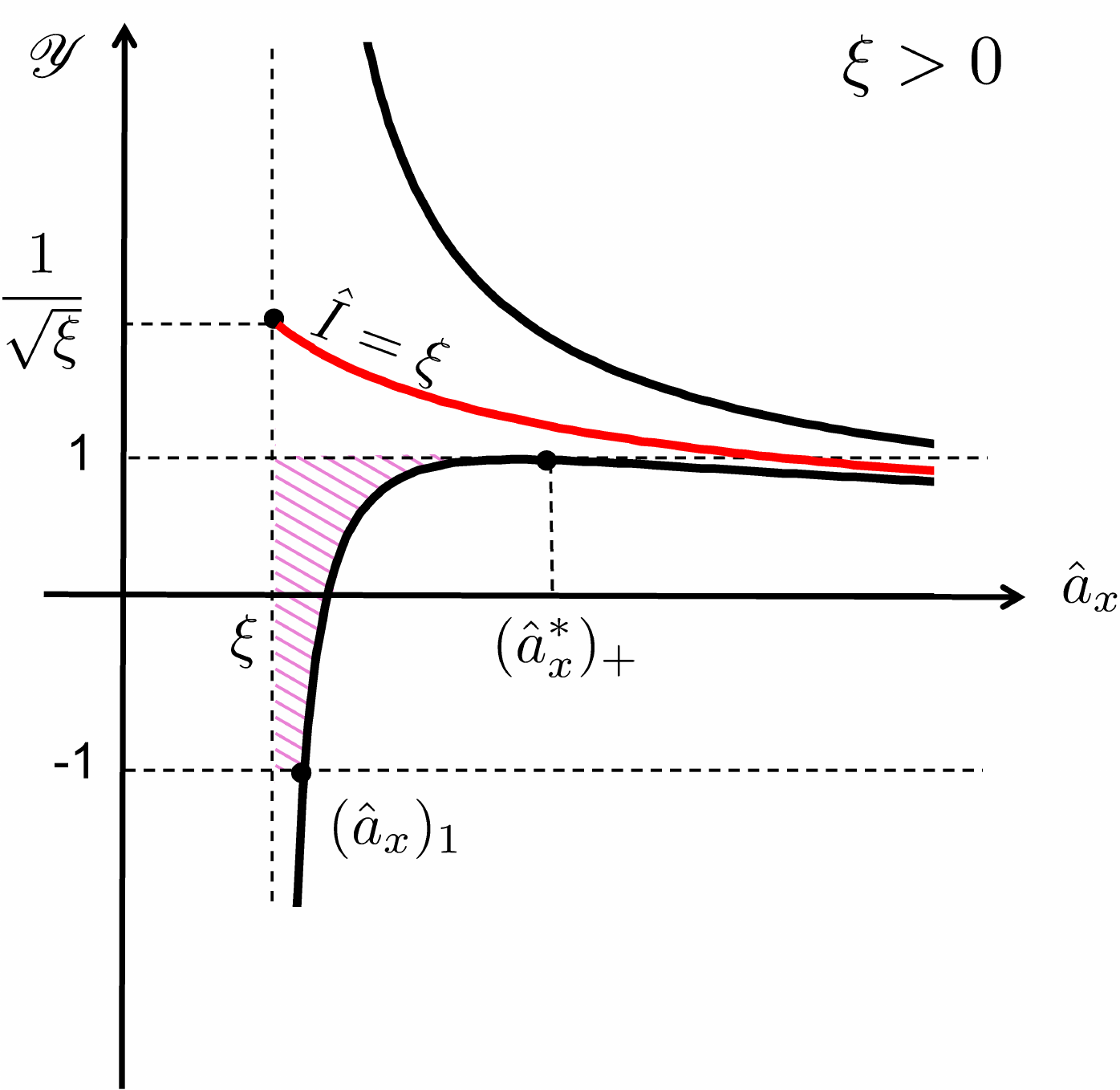,width=75mm}}
 \put( -5,70)  {\epsfig{file=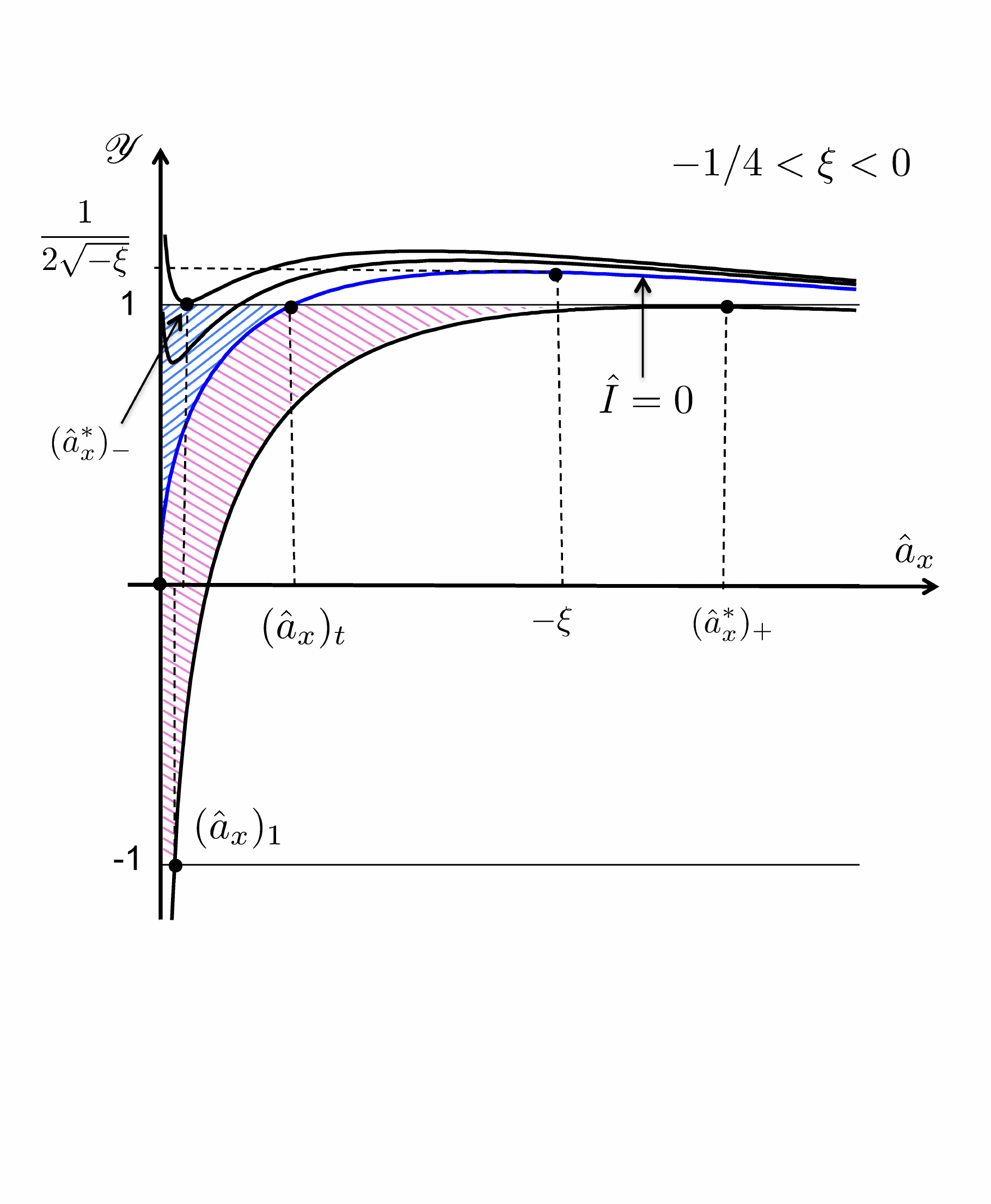,width=75mm}}
 \put( -2, 0)  {\epsfig{file=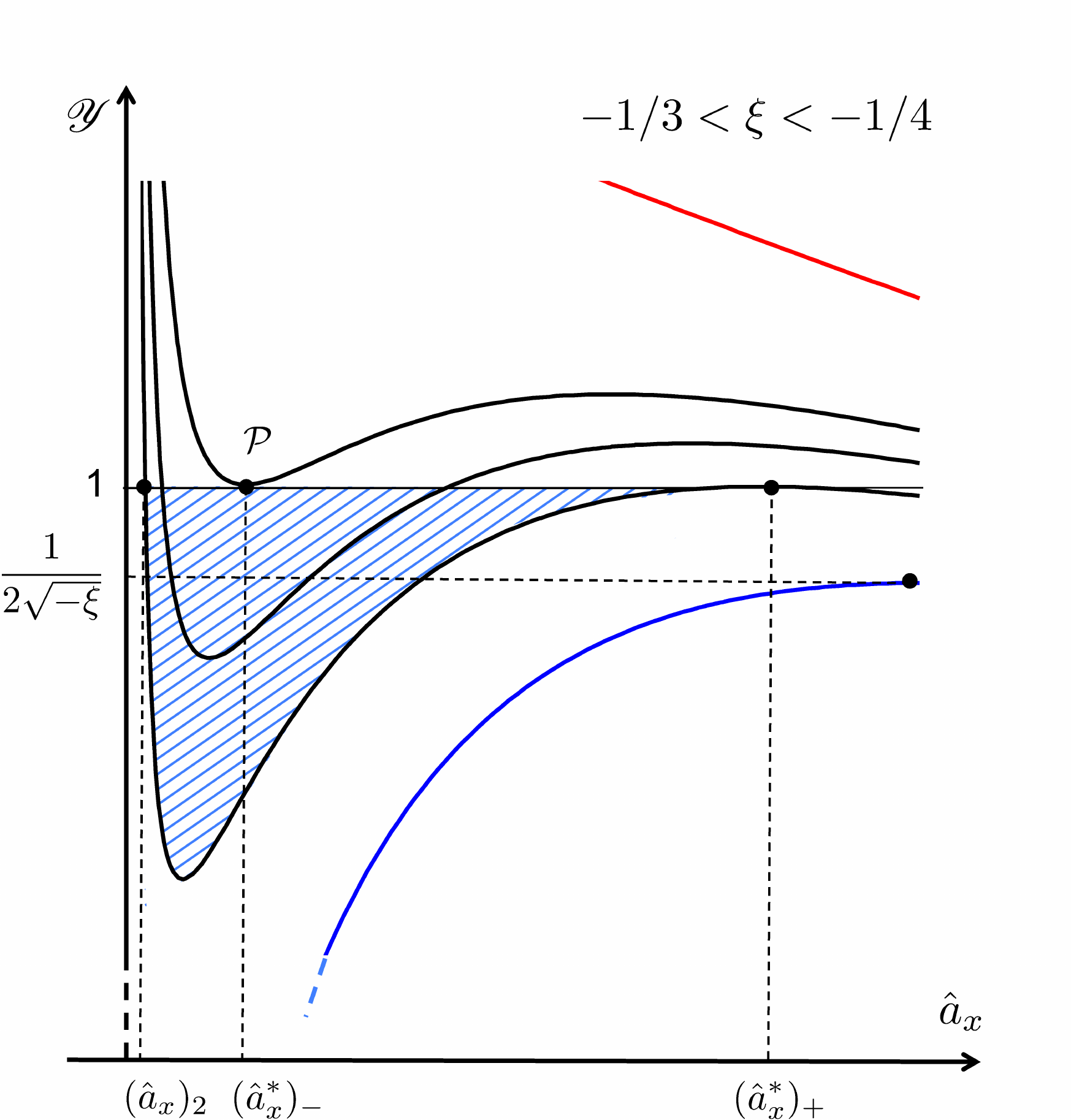,width=75mm}}
 \put( 0,240) {a)}
 \put( 0,160) {b)}
 \put( 0,80)  {c)}
 \end{picture}
 \caption{
 Impact of the level lines on the stability of motion.
 The three pictures refer to the three relevant regimes. 
 Note that in a) the curves stop at $\hax=\xi$ because 
 of the condition of Eq.~\ref{eq:86}. 
 }
 \label{fig1:fig_17}
 \end{center}
 \end{figure}

 The properties of the level lines for $\xi<0$ are discussed and summarized 
 in Fig.~\ref{fig1:fig_27} of the Appendix B. 
 A version of the same picture with the relevant quantities is 
 shown in part b) of Fig.~\ref{fig1:fig_17}. 
 In this region it holds $1/(2\sqrt{-\xi}) > 1$, 
 that means that the blue curve shown in part b) of the Fig~\ref{fig1:fig_17} 
 will always cross the line $\ya=1$ twice. 
 The level lines on the left of (or above of) the blue curve exhibit a minimum 
 and a maximum, while the level lines on the right of (or below of) the 
 blue curve have only one maximum. 
 We find that there are two curves tangent to $\ya=1$: 
 one above the blue curve tangent in $(\hax^*)_-$, and one below the 
 blue curve tangent in $(\hax^*)_+$. 
 In the dashed blue area we find the level lines with a minimum and a maximum: 
 we call for convenience these lines of the ``second order''. 
 The minimum of these lines has coordinate $\hax$ 
 always smaller than $(\hax)_t$, 
 and the dynamics of particles in this region is to be bounded by the
 same curve $\ya=1$. 
 This means that for particles in the dashed blue region 
 their trajectory is always bounded in $0<\ya<1$. 
 The red dashed area characterizes instead the initial conditions where
 the level lines have one maximum, we call for convenience these lines 
 of the ``first order''. 
 These first order lines always cross both lines $\ya=1$, and $\ya=-1$, 
 hence this region is stable and as discussed in the previous section 
 the level line tangent to the line $\ya=1$ identifies a fix-line. 
 Therefore $(\hax^*)_+$ is on the fix-line and characterizes the outer 
 point of stability.

 The point $(\hax)_1$ marks the region of total stability: 
 if $0 < \hax < (\hax)_1$ particles are stable for any initial phase 
 $\Omega$. 
% If $\xi\rightarrow 0^-$, the blue curve of Fig.~\ref{fig1:fig_27} 
% will overlap with the red curve approaching a pattern typical 
% of the region $0 < \xi \le 1$. 

 %
 %
 %
 %
 %
 %

 %%%%%%%%%%%%%%%%%%%%%%%%%%%%%%%%%%%%%%%%%%%%%%%%%%%%%%%%%%%%%%%%%%%%%%%5
 \subsubsection{$-1/3 < \xi \le -1/4$}
 \label{sect:pseudo} 
 In this regime, referring to Fig.~\ref{fig1:fig_27}, we find that the
 maximum of the blue line is $1/(2\sqrt{-\xi}) < 1$, and the
 inflection point of the red line is $1/(\sqrt{-3\xi}) > 1$.  As a
 consequence all level lines below the blue curve describe unstable
 dynamics.  Among the level lines in the region between the red and
 blue line, there are two level lines tangent to the line $\ya=1$.
 The situation is illustrated in part c) of Fig.~\ref{fig1:fig_17}.
 One level line is tangent to $\ya=1$ in $(\hax^*)_+$, this curve also
 intercepts the line $\ya=1$ in $(\hax)_2$. The second level line is
 tangent to $\ya=1$ in $(\hax^*)_-$.  Only the blue dashed area is
 stable as any initial condition in the system of coordinates
 $(\hax,\ya)$ will move on a level line that crosses twice the line
 $\ya=1$.  Interestingly, the point ${\cal P} = ((\hax^*)_-,1)$ is a
 point that does not move because it also identifies a fix-line (the
 level line is tangent to $\ya=1$ in $(\hax^*)_-$), but at the same
 time the level line tangent to ${\cal P}$ has $\ya > 1$.  Therefore
 any point in the neighborhood of ${\cal P}$ with $-1 < \ya < 1$ will
 always remain close to it, and ${\cal P} $ is an equivalent of a fix
 point in this system of coordinates.  We call this a ``stationary''
 point. \label{stationary-point}
  
 Lastly, we also find the point $(\hax)_2$, which is the extreme point 
 of stability, for $\hax < (\hax)_2$ there is no stability.

 %%%%%%%%%%%%%%%%%%%%%%%%%%%%%%%%%%%%%%%%%%%%%%%%%%%%%%%%%%%%%%%%%%%%%%%5
 \subsubsection{$\xi \le -1/3$}
 In this region the condition $\xi \le -1/3$ implies 
 $1/\sqrt{-3\xi} \le 1$.
 This means that the inflection point of the red curve
 in Fig.~\ref{fig1:fig_27} is below or equal to line $\ya=1$.
 Therefore if $\xi \le -1/3$ no point in $\hax>0$, $-1<\ya<1$ 
 can be stable.

 %%%%%%%%%%%%%%%%%%%%%%%%%%%%%%%%%%%%%%%%%%%%%%%%%%%%%%%%%%%%%%%%
 \subsection{Simulation examples}
 In order to confirm the theory as developed in this paper 
 we directly integrated the
 equation of motion of the coordinates $(\hax,\ya)$ and verified that the
 stability of particle is determined as previously discussed. 
 The vicinity of the resonance is explored here on parameters that 
 will be used later in multi-particle simulations. 
 We use the following parameters $\Lambda = -7.44 \times 10^{-3}$ m$^{-1}$,
 and $\Delta_r  = 4.64 \times 10^{-3}$, from which we find the 
 scaling factor of 
 Eq.~\ref{eq:81} equal to $\Delta_r^2/(16\Lambda^2)=0.024$ m$^{2}$, and 
 the $\mu=-1$. 

\begin{figure}[h]
 \begin{center}
 \unitlength 0.9mm
 \begin{picture}(80,100)
 \put(-10, 0) {\epsfig{file=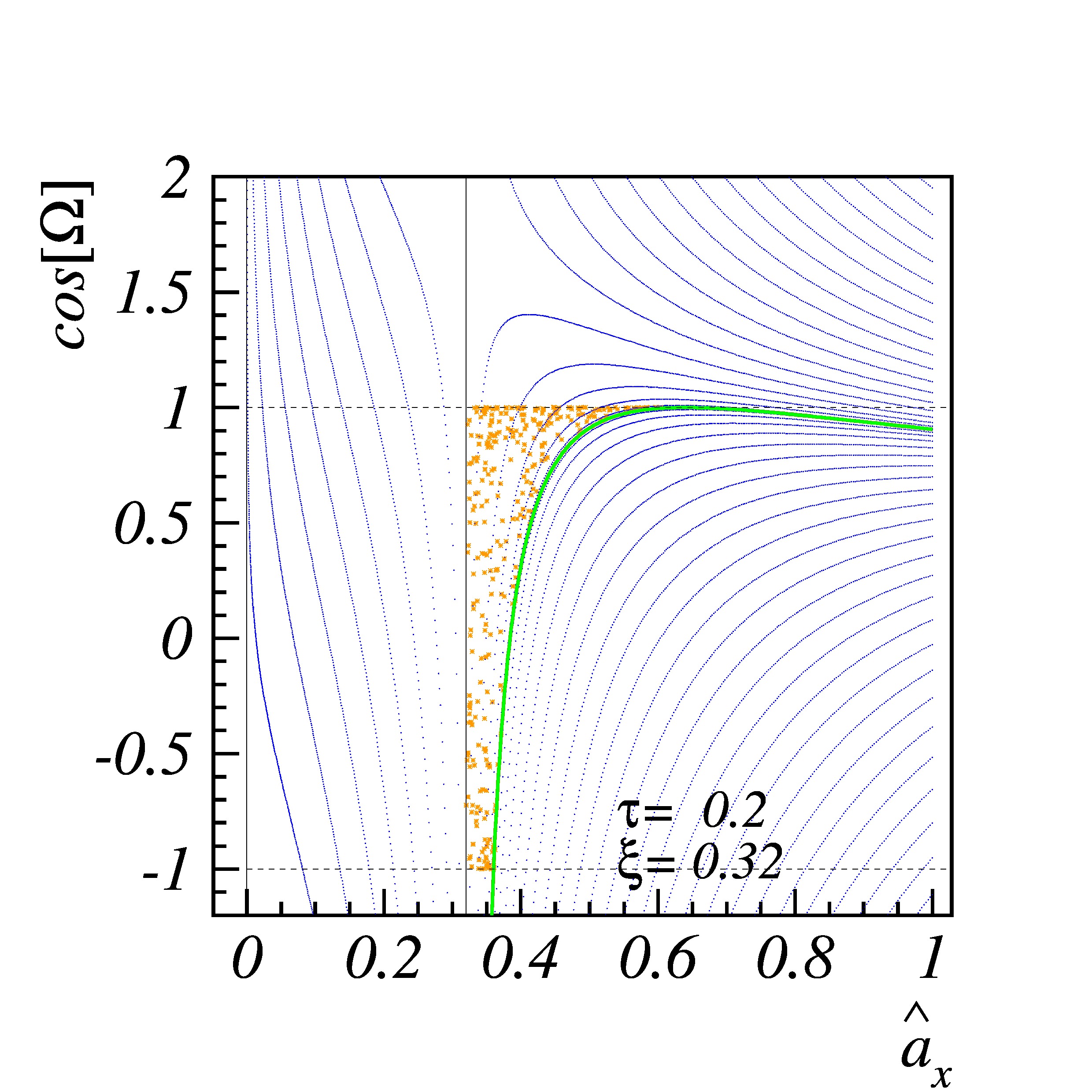,width=100mm}}
 \end{picture}
 \caption{
 Level lines of the invariant. The green curve 
 touches both the lines $\ya=\pm 1$ and bound the stability region 
 marked by the orange dots, which are the set of stable
 initial condition obtained from simulations. 
 }
 \label{fig1:fig_18}
 \end{center}
 \end{figure}

 In Fig.~\ref{fig1:fig_18} we show an example of level lines
 for $\xi = 0.32$, where each curve corresponds to a different $\hat I$. 
 The green curve is the level curve of the fix-line defined by 
 $\tau = 0.2$, the corresponding invariant in the laboratory frame is 
 $C = 0.0155$ m$^2$. 
 The set of orange dots in the picture shows the set of initial
 conditions that are stable.
 It is evident that the green curve bounds the set of stable initial 
 conditions as discussed in the previous section. 

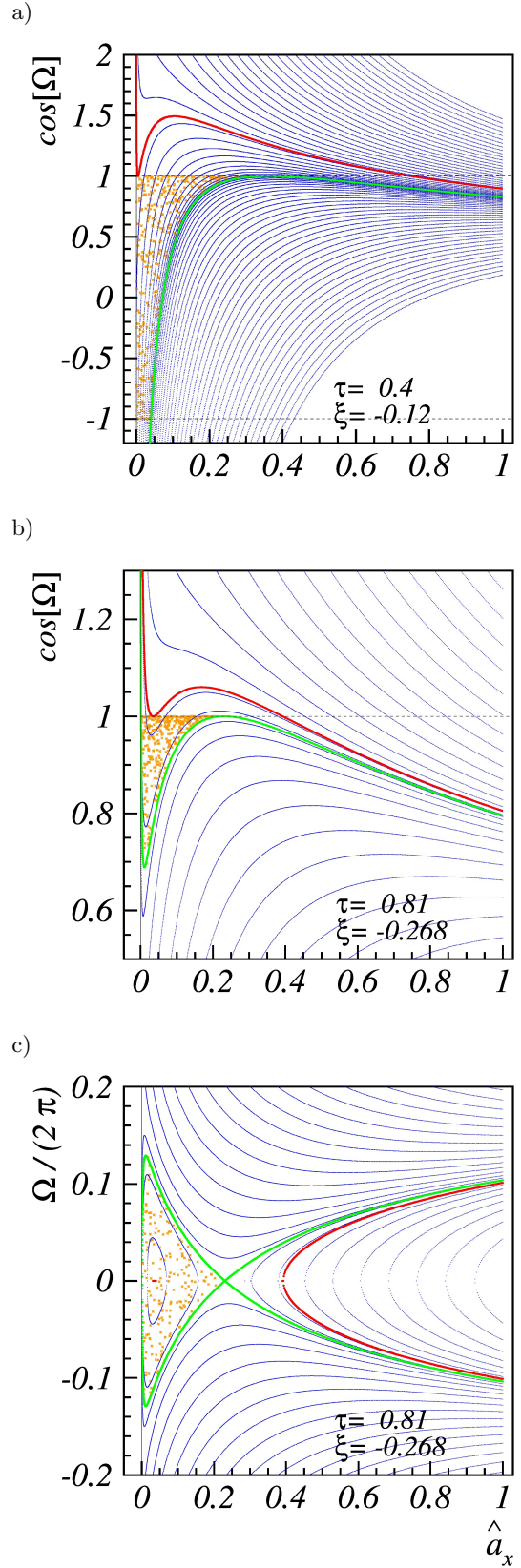
\begin{figure}[H]
 \begin{center}
 \unitlength 0.9mm
 \begin{picture}(80,235)
 \put( 0,160) {\epsfig{file=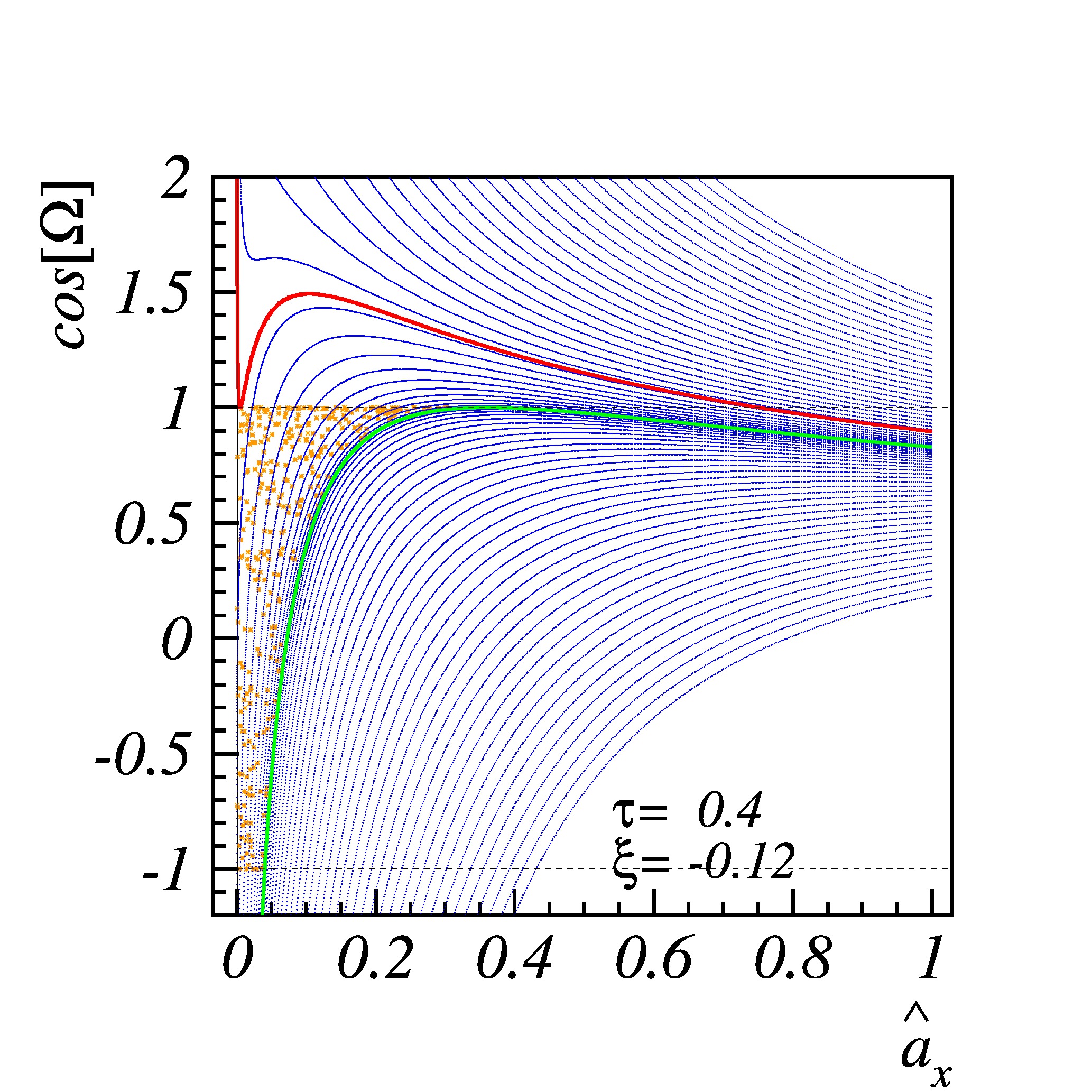,width=80mm}}
 \put( 0,80)  {\epsfig{file=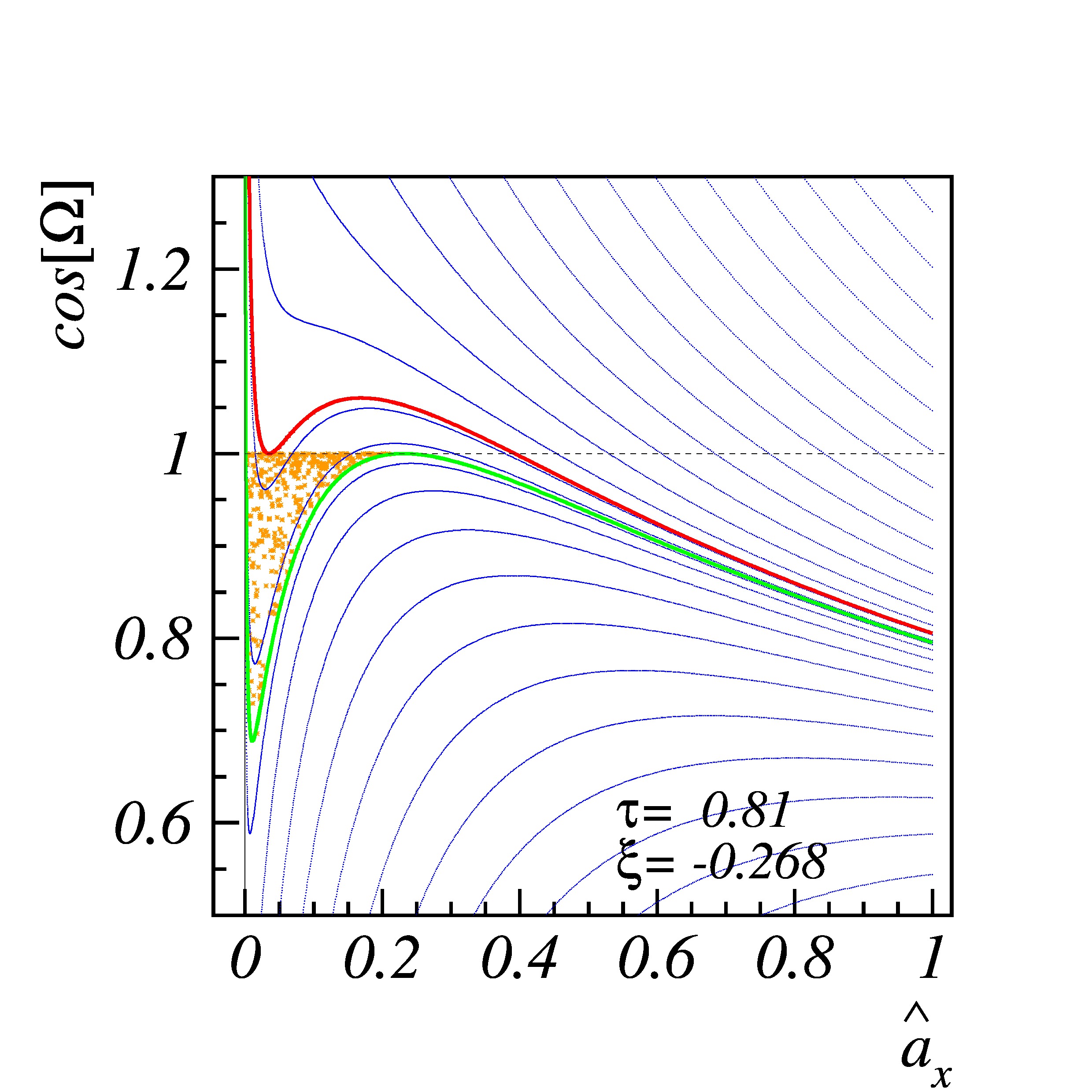,width=80mm}}
 \put( 0, 0)  {\epsfig{file=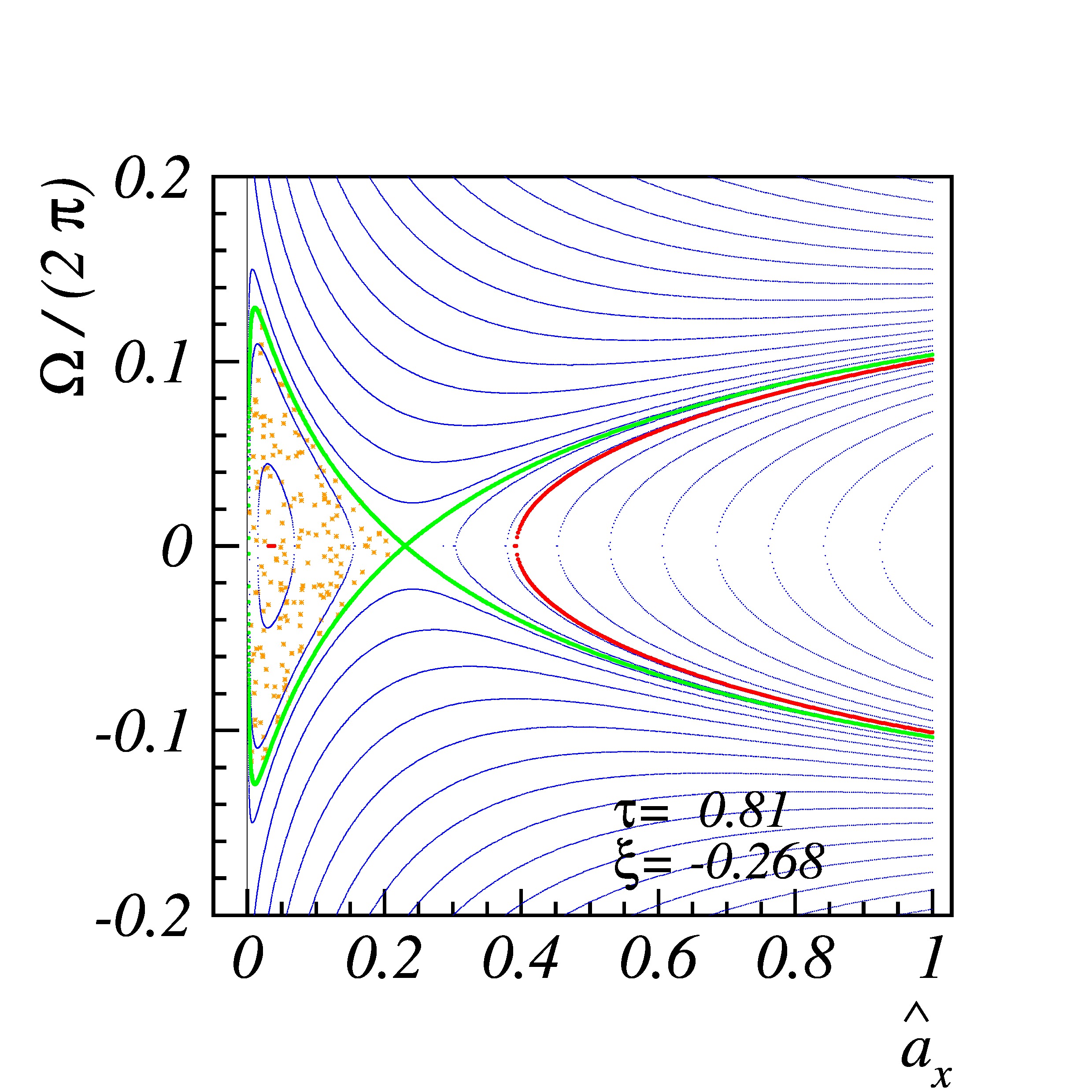,width=80mm}}
 \put( 0,240) {a)}
 \put( 0,160) {b)}
 \put( 0,80)  {c)}
 \end{picture}
 \caption{
 a) $\tau=0.4, \xi=-0.12$. Level curves of the invariant. 
 The two curves correspondent to the fix-lines are in red and green. 
 Note that the level line of the unstable fix-line (green) is bounding the
 stable particles;
 b) $\tau=0.81, \xi=-0.268$. Level curves correspondent to the fix-lines, 
    they are curves of the second order (see text). 
 The orange dots are in the region of stable trajectories;
 c) the same situation as in part b) but now in the $\hax,\Omega$ coordinates.
 }
 \label{fig1:fig_19}
 \end{center}
 \end{figure}

 Part a) of Fig.~\ref{fig1:fig_19} shows simulations 
 in the regime $-1/4<\xi<0$.
 The parameters of the plot are
 $\tau=0.4, \xi=-0.12$.
 Among the level lines, two are of relevance 
 and colored with green and red.
 The green curve is a level curve of the first order. 
 This curve touches $\cos[\Omega]=1$ in one point (fix-line), and the
 set of orange dots on the left of the green curve in $-1\le\ya\le1$
 are all the stable initial conditions. In the picture there is
 also a red line, which is of the second order, and this curve also 
 touches $\cos[\Omega]=1$ in one point.

  In part b) of Fig.~\ref{fig1:fig_19} 
 the parameters of the level line associated to the fix-line are
 $\tau=0.81,\xi=-0.268$, therefore the level lines have pattern 
 typical of the region $-1/3 < \xi < -1/4$. 
 The green curve is a level line of the second order and it bounds the 
 stable initial conditions (orange dots) with the line $\cos[\Omega]=1$. 
 For sake of clarity we show in the part c) of Fig.~\ref{fig1:fig_19} 
 the same scenario in the $\hat a_x,\Omega$ plane. 
 The role of the fix-line is very evident.
 It also becomes clear that  there is a ``stationary'' fix point 
 where the red level line touches the line
 $\Omega=0$, and a single red dot is in the center of an ``island''.

 %
 %
 %

 %%%%%%%%%%%%%%%%%%%%%%%%%%%%%%%%%%%%%%%%%%%%%%%%%%%%%%%%%%%%%%%%%%%%%%%
 \subsection{Summary}
 Because of the resonance the motion of a particle is also affected by
 the time dependence of the variables $\hax,\hay,\tpx,\tpy$.  At any
 time the variables $\hax,\hay$ are forced to lie on the straight line
 $2\hax - \hay = 2\xi$, hence the motion of a particle is bounded, if
 $\hax$ remains bounded.  The stability properties are retrieved as
 follows: given the initial condition $\hax,\hay,\tpx,\tpy$, we find
 $\xi$ and $\Omega$, from which $\ya$ is computed.  The pair
 $(\hax,\ya)$ with $\xi$ allows to identify to which region the motion
 belongs to and correspondingly to which type of stability (discussed
 in Sec.~\ref{SOM} Figs.~\ref{fig1:fig_17}).

 In order to classify the stability properties we will make use of two
 definitions of stability based on the discussion made in
 Sec.~\ref{SOM}: {\it We call {\bf stability of type 1} the stability
   of a particle, when its motion is associated to a level line of the
   first order.  Similarly we define a {\bf stability of type 2} when
   the stability is associated to a level line of the second order.}
 
\begin{figure}[h]
 \begin{center}
 \unitlength 0.7mm
 \begin{picture}(80,95)
 \put(-20,  0) {\epsfig{file=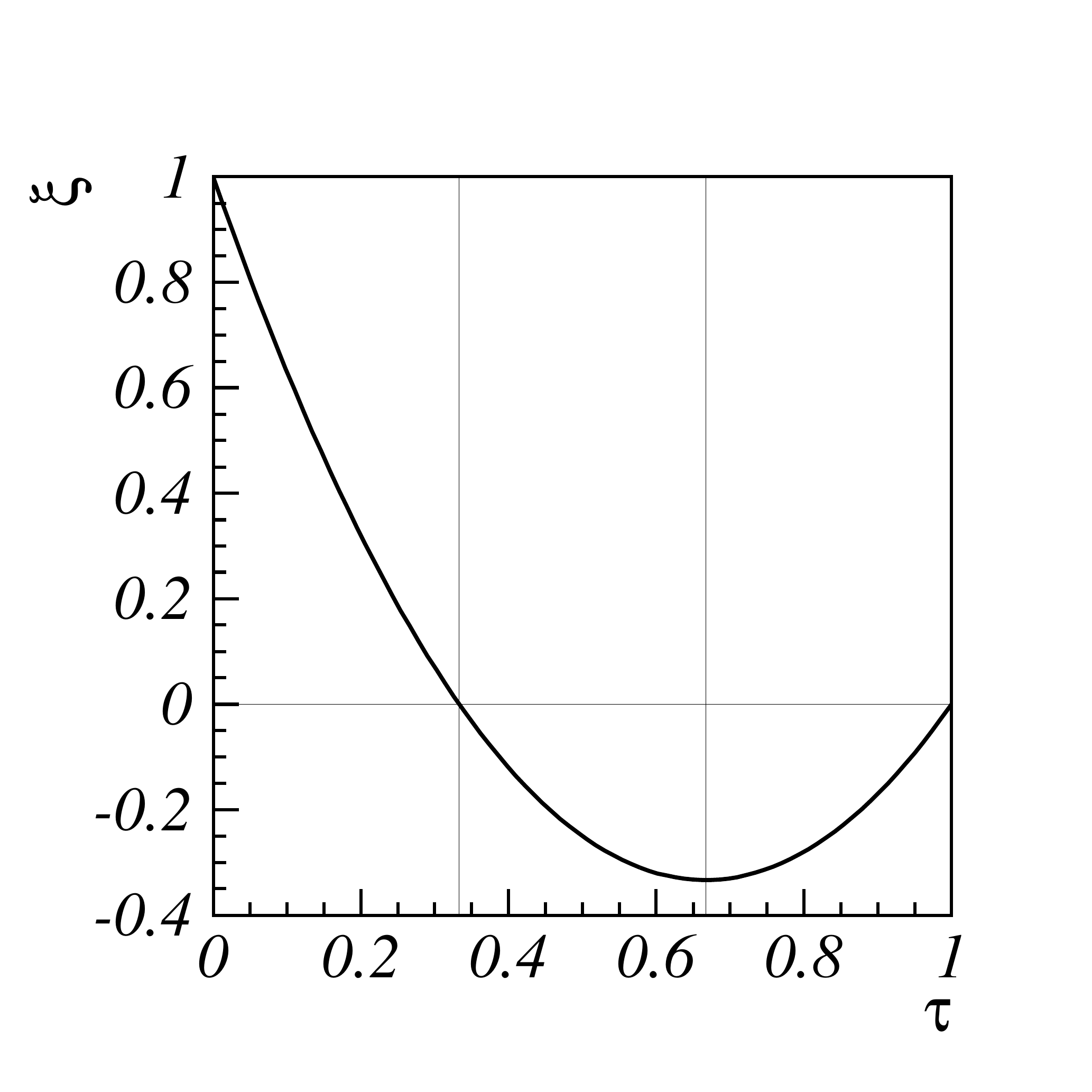,width=90mm}}
 \end{picture}
 \caption{
 $\xi$ as function of $\tau$. 
 The curve shows a symmetry at $\tau=2/3$, because for
 $\tau > 1/3$ the line $2\tax-\tay=C$ crosses the
 fix-line parabolic-like curve twice. 
 }
 \label{fig1:fig_21}
 \end{center}
 \end{figure}

\begin{figure}[H]
 \begin{center}
 \unitlength 0.85mm
 \begin{picture}(80,218)
 \put( -14, -0)  {\epsfig{file=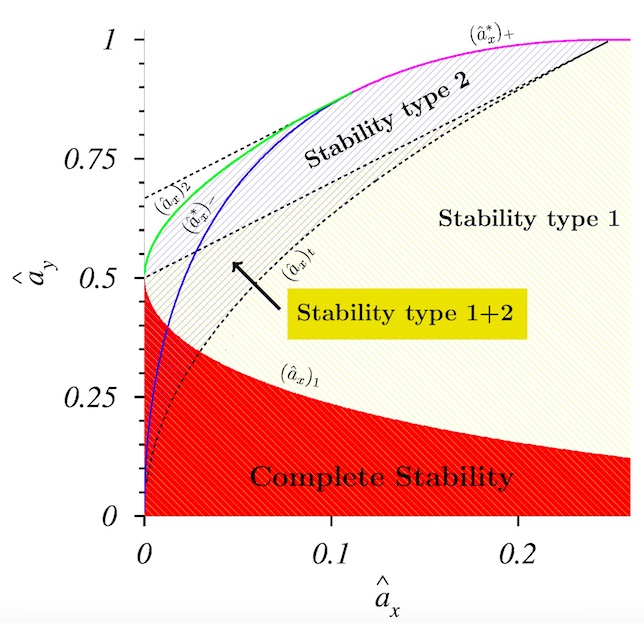,width=87mm}}
 \put( -15, 110)  {\epsfig{file=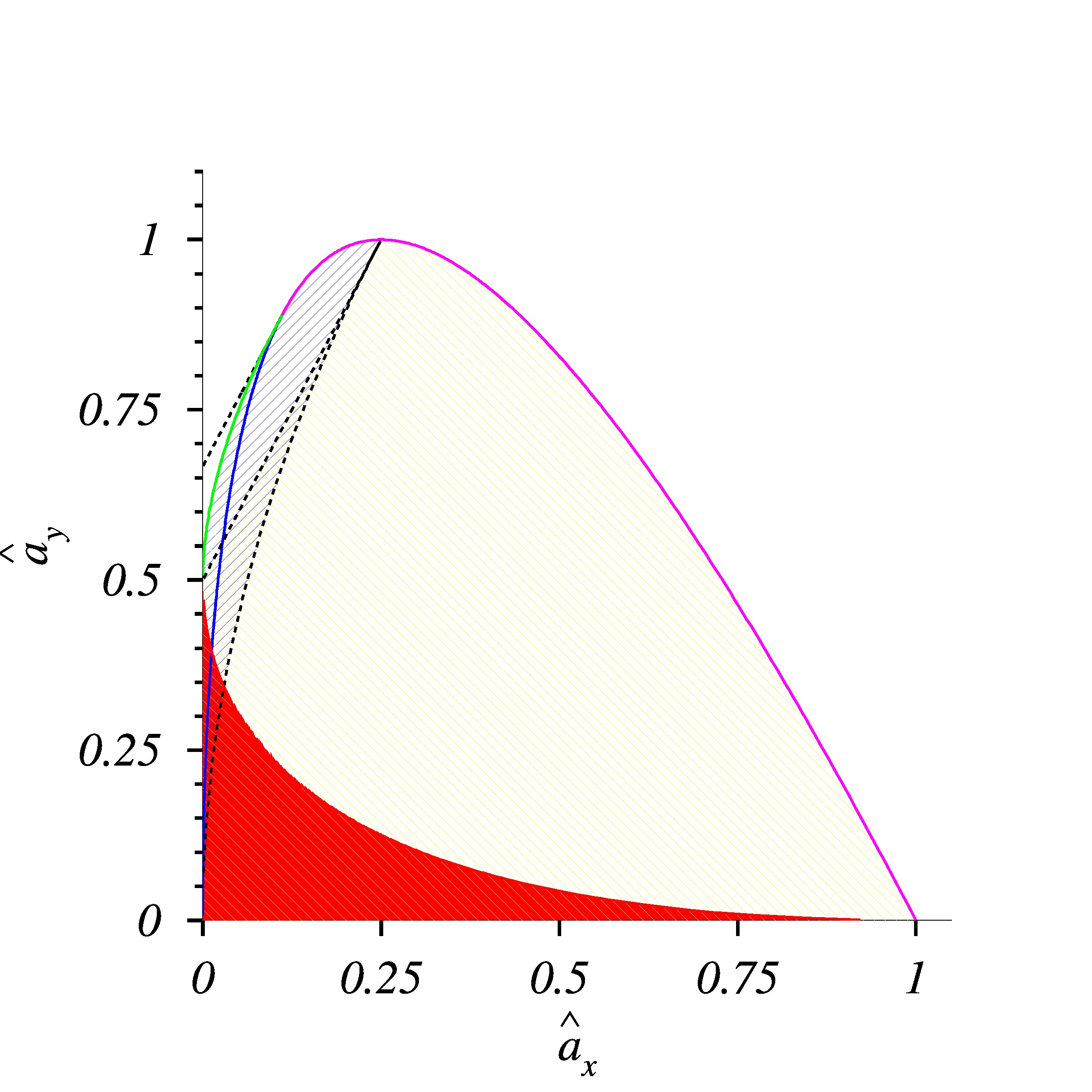,width=105mm}}
 \put(-10,210) {a)}
 \put(-10,100) {b)}
 \end{picture}
 \caption{
 Summary of all the parameters that characterize the stability domain. 
 On the bottom is a zoom for the sake of clarity. 
 The border dividing the red dashed area from the yellow dashed area 
 is the collection of the  $(\hax)_1$ defined in Fig.~\ref{fig1:fig_18} part a) and b). 
% This is also 
% the edge of the red region in the pictures of Fig.~\ref{fig1:fig_22}.
 The green line is the collection of the $(\hax)_2$, i.e. 
 is the part of initial condition that are stable beyond the fix-line.
 The blue line is the collection of the stable fix-line $(\hax^*)_-$, 
 the pink line is the collection of the unstable fix-lines $(\hax^*)_+$. 
 The points $(\hax)_t$ are defined in part b) of Fig.~\ref{fig1:fig_17}. 
 }
 \label{fig1:fig_20}
 \end{center}
 \end{figure}

 The findings on the stability in the proximity of the third order
 coupled resonance $Q_x + 2Q_y = N$ are summarized in
 Fig.~\ref{fig1:fig_20}.  In these pictures the fix-lines define the
 border of the domain of stability (pink line unstable fix-lines, and
 blue line stable fix-lines).  As each fix-line is characterized by $0
 \le \tau \le 1$, $\tau$ parameterizes the border of stability in
 $\hax,\hay$ coordinates as well.  The relation of $\tau$ with $\xi$
 is shown in Fig.~\ref{fig1:fig_21}.

 The stability properties are summarized in Fig.~\ref{fig1:fig_20} 
 in terms of  curves collecting the points 
 $(\hax)_2,(\hax^*)_-,(\hax^*)_+,(\hax)_1,(\hax)_t$ 
 defined in Fig.~\ref{fig1:fig_17}. 
 The bottom picture of Fig.~\ref{fig1:fig_20} provides a zoom with the 
 summary of all the symbols. 
 We can therefore make the following general classification of the 
 regions of stability: 

 \begin{description} 
%%%%%%%%%%%%%%%%%%%%%%%%%%%%

 \item[Complete stability]
 The red wide area in Fig.~\ref{fig1:fig_20} is the region of 
 complete stability, i.e. particles with
 $\hax\hay$ inside this area are stable for any
 $\tpx,\tpy$. The upper edge of the red area is the set of the points 
 $(\hax)_1$ identified in part a),b) of Fig.~\ref{fig1:fig_17};

%%%%%%%%%%%%%%%%%%%%%%%%%%%%%%%%
 \item[Partial stability type 1]
 The yellow dashed area is a region of partial stability that extend in the
 region $-1/4 \le \xi  \le 1$. 
 The allowed angles $\Omega$ are given by 
 \begin{equation}
   |\Omega - \pi(1+\mu)/2 + 2\pi N'| \le \Delta\Omega 
   \label{eq:95_0}
 \end{equation}
 with $N'$ an integer. 
 The angle $\Delta\Omega$ is represented in Fig.~\ref{fig1:fig_21b} for the 
 case of $\xi > 0$, and is obtained directly from Eq.~\ref{eq:87} as 
 \begin{equation}
   \Delta\Omega(\hax,\hay) =
   \textrm{arcos}
   \left[
   \frac{1}{\mu}
         \frac{\hat I_{fl} - \hax}{\sqrt{\hax}(\hax-\xi)}
   \right], 
 \label{eq:95}
 \end{equation}
 with $\hat I_{fl}$ the invariant correspondent to the fix-line defined 
 by $\xi$ (or $\tau$). We here consider $0 \le \Delta\Omega \le \pi$. 

 \begin{figure}[H]
 \begin{center}
 \unitlength 0.7mm
 \begin{picture}(80,95)
 \put(-10,  0) {\epsfig{file=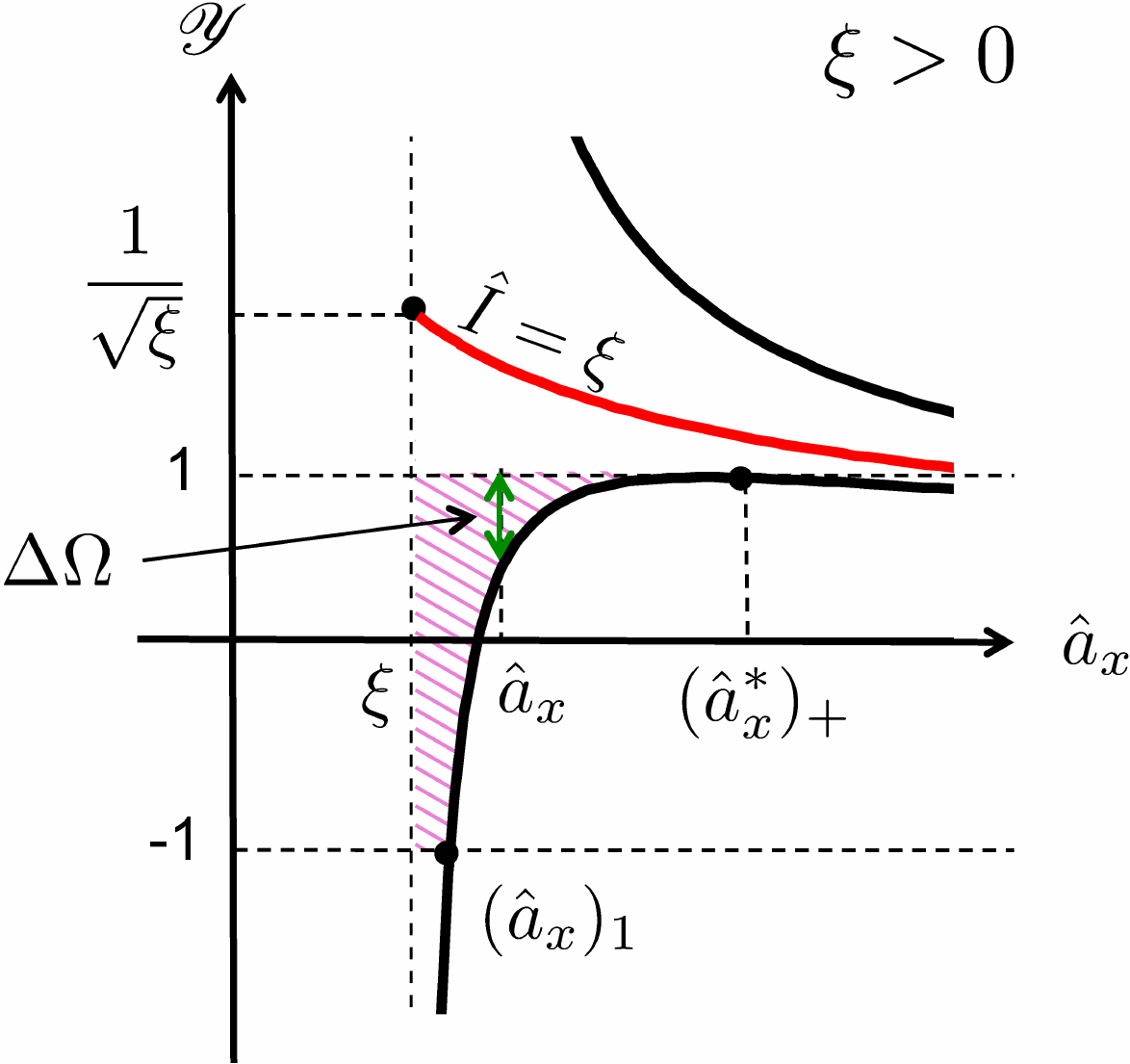,width=74mm}}
 \end{picture}
 \caption{
 Definition of $\Delta\Omega$ for the case $\xi>0$. 
 A similar definition applies for other value of $\xi$ 
 seen in the different parts of  Fig.~\ref{fig1:fig_17}.
 }
 \label{fig1:fig_21b}
 \end{center}
 \end{figure}
 %
 %
 %

%%%%%%%%%%%%%%%%%%%%%%%%%%%%%%%%%%%%%%
 \item[Partial stability type 2]
 This region is found for $-1/3 < \xi < -1/4$, and
 is shown by the blue dashed area in
 Fig.~\ref{fig1:fig_20}, the stability is type 2, and $\Omega$ is bounded
 to oscillate around $\Omega=0$ or $\Omega=2\pi$.
 The green line is the collection of the points $(\hax)_2$ identified in
 part c) of Fig.~\ref{fig1:fig_17} and is the edge of the stability domain.
 This curve is not a collection of fix-lines, which indeed correspond to 
 the blue curve (the stationary points). 
 The stability angles $\Delta\Omega$ are given by Eq.~\ref{eq:95}.

 \end{description}

 \begin{description}

%%%%%%%%%%%%%%%%%%%%%%%%%%%%%%%%%%%%%% 
 \item[Partial stability type 1+2]
 In Fig.~\ref{fig1:fig_20} 
 this region is the overlapping of the region of partial stability 
 type 2 (blue dashed area) with the region of partial stability
 of type 1 (yellow dashed area). 
 This area is characterized by a stability shown in part b) of 
 Fig.~\ref{fig1:fig_17} for $(\hax)_1 < \hax < (\hax)_t$.
 In Fig.~\ref{fig1:fig_20} the bend curve $(\hax)_t$ is the collection 
 of the points $(\hax)_t$ defined in part b) of Fig.~\ref{fig1:fig_17}. 
 In this region of Fig.~\ref{fig1:fig_20}, 
 according to the initial condition a particle may have
 an evolution of $\Omega$ that is bounded to $0$ or $2\pi$  (stability
 type 2, blue dashed area of part b) of Fig.~\ref{fig1:fig_17}), or
 an unbounded evolution of $\Omega$ that spans periodically
 all $[0,2\pi]$ values 
 (stability type 1, red dashed area of part b) of Fig.~\ref{fig1:fig_17}).
 Also in this case the angles that are stable are given by Eq.~\ref{eq:95}. 
 
 \end{description}

\begin{figure}[H]
 \begin{center}
 \unitlength 0.79mm
 \begin{picture}(80,243)
 \put(-5,0){
 \put( 0,160) {\epsfig{file=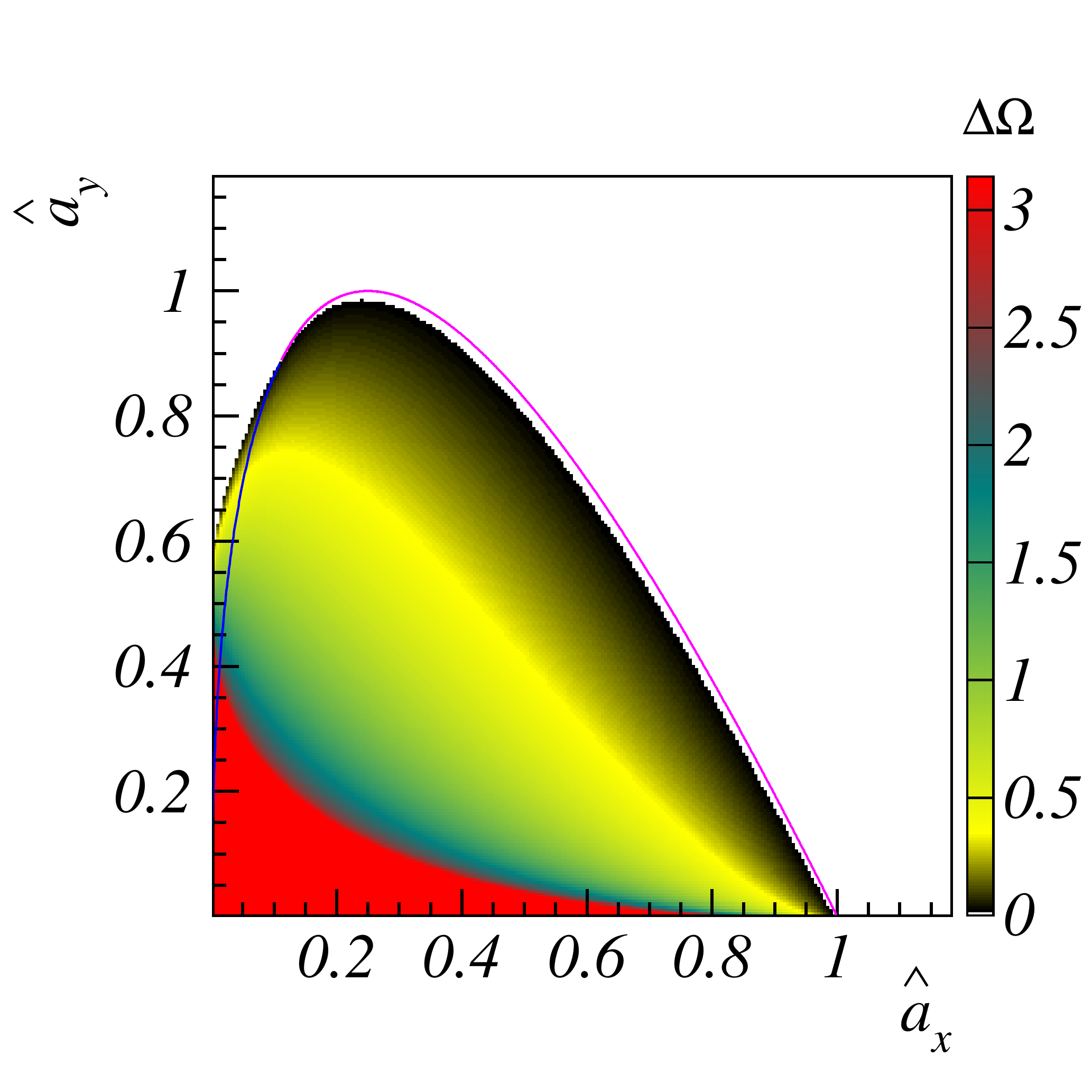,width=68mm}}
 \put( 0, 80) {\epsfig{file=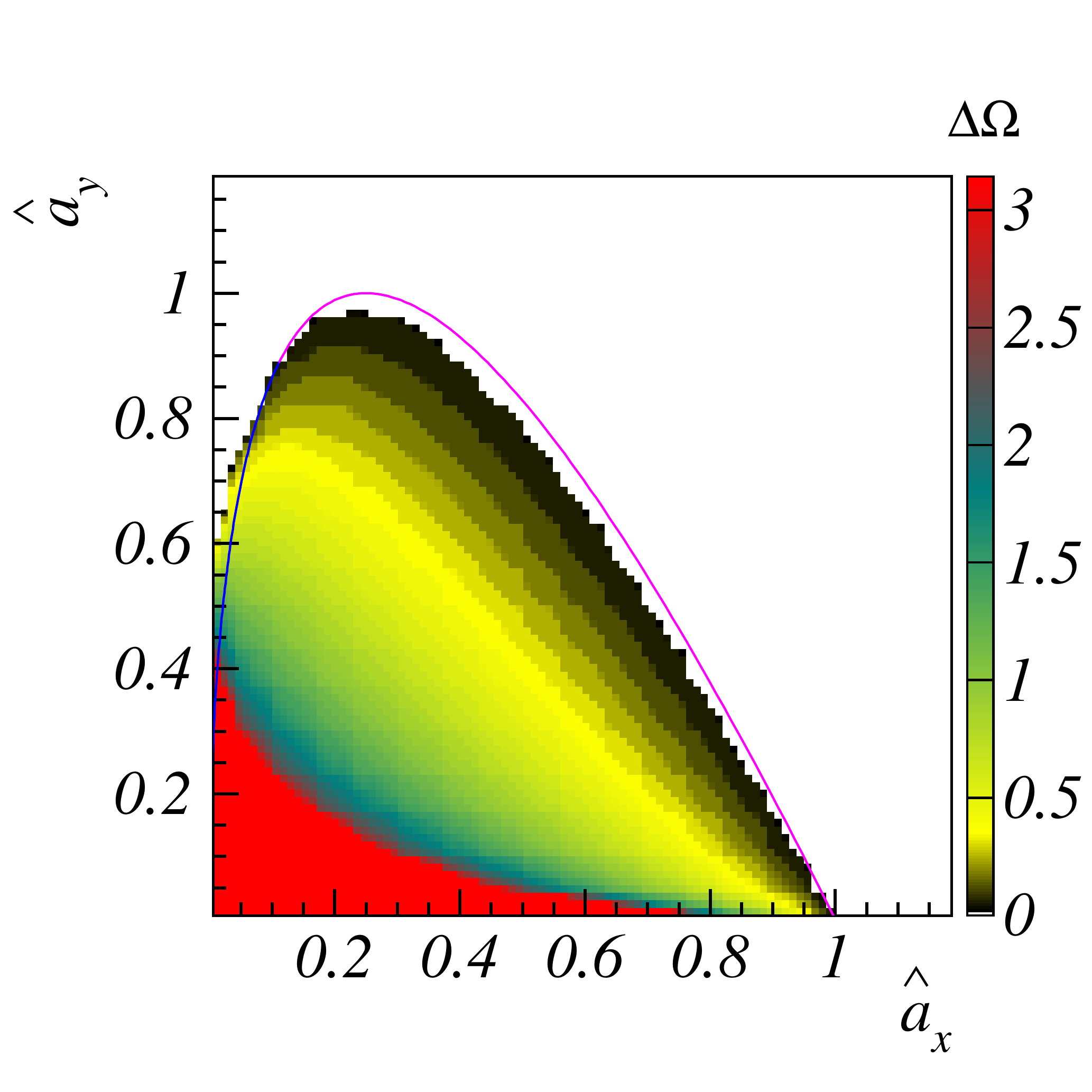,width=68mm}}
 \put( 0,  0) {\epsfig{file=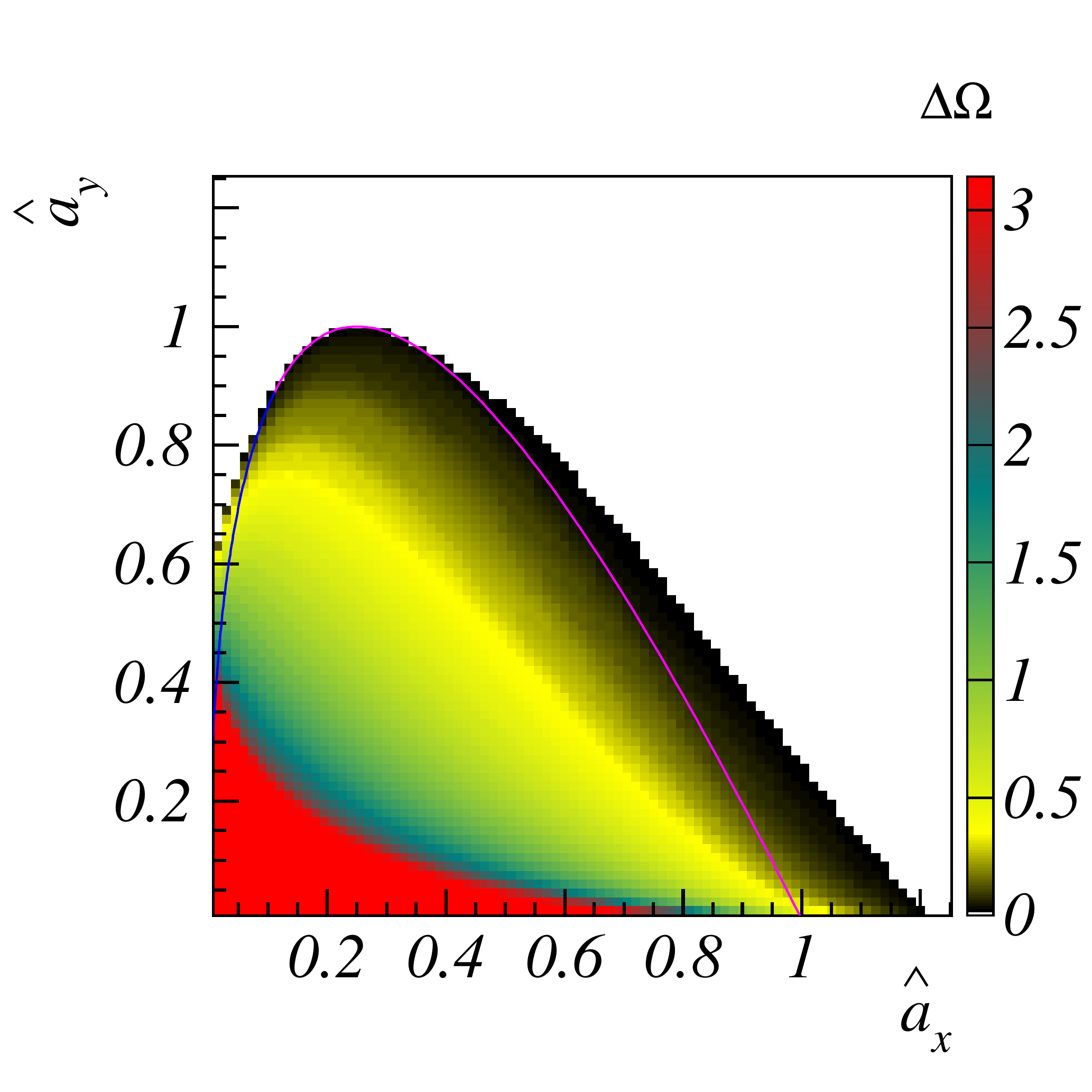,width=68mm}}
 \put( 0,240) {a)}
 \put( 0,160) {b)}
 \put( 0, 80) {c)}
           }
 \end{picture}
 \caption{
 Complete stability domain for a constant focusing circular accelerator. 
 Part a) the stability domain as obtained by the
 analytic theory;
 Part b) the same result obtained by a particle tracking, with 
 a computer code, of 100x100x51x51 initial conditions, 
 and particles are tracked for 1000 turns.
 Part c) shows the stability domain obtained for one
 sextupolar kick (to be explained in the following chapter). 
 The result shows an enlarging of the stability
 domain closes to the $\tax$ axis.
 }
 \label{fig1:fig_22}
 \end{center}
 \end{figure}
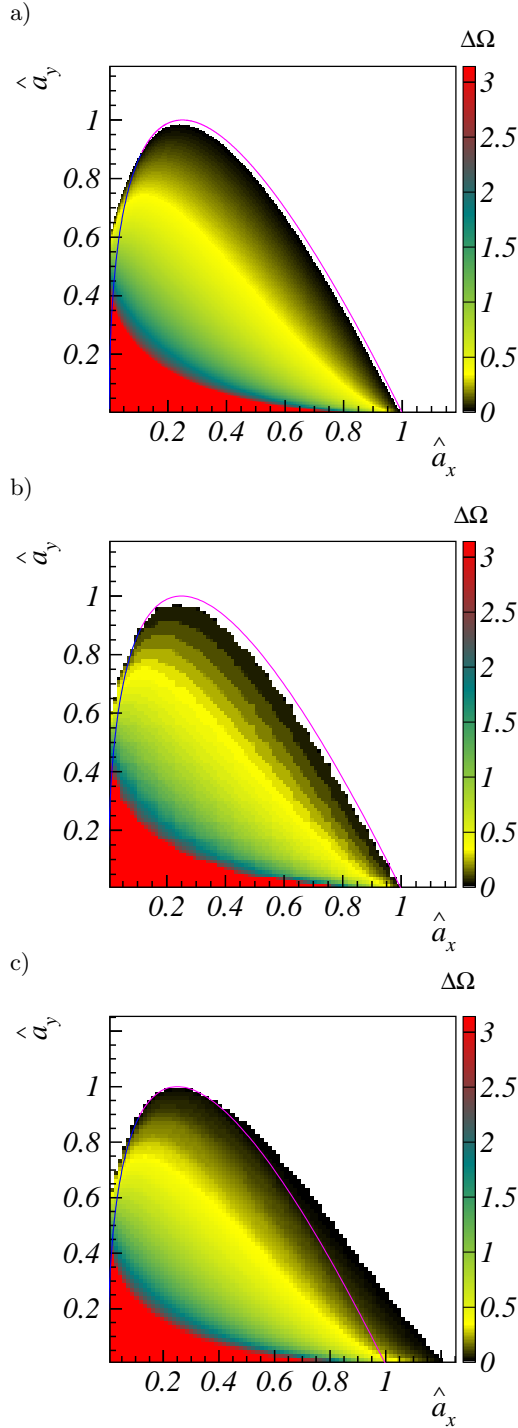

 With this theoretical considerations in mind we compare the stability 
 domain as predicted by our theory with multi-particle simulations.
 In the part a) of the Fig.~\ref{fig1:fig_22} we show the stability 
 domain as obtained from theory, in an $\hax,\hay$ chart. 
 The color code provides the information on the allowed angle $\Delta\Omega$.
 The picture retrieves the discussed properties and shows that the allowed
 range of angles becomes smaller when approaching the unstable fix-lines. 
 The picture also shows the partial stable region beyond the stable 
 fix-lines (blue line from theory). 
 In the red area the motion is stable (complete stability). 
 
 Part b) of the Fig.~\ref{fig1:fig_22} 
 shows the same result obtained by tracking.
 In order to test the theory we considered
 for this simulation a constant focusing lattice where a single harmonics
 was created. The strength of the driving term is
 $\Lambda = -7.44 \times 10^{-3}$ m$^{-1}$, 
 and $\Delta_r  = 4.64 \times 10^{-3}$.
 The stability domain was sampled with 100x100x51x51 initial conditions, and
 the stability was assessed over 1000 turns.
 The pink and blue curves in the picture 
 are the collection of fix-lines as from the theory. 
 These results are in good agreement with the theoretical predictions.

 Part c) of the Fig.~\ref{fig1:fig_22} 
 shows the stability domain obtained when a single sextupolar kick is used
 on a constant focusing lattice. 
 In this case we are not in the ideal condition of the theory, 
 stated in Sect. IV, and some deviation is to be expected, 
 resulting in an enlarging of the stability
 domain closes to the $a_x$ axis.
 For this simulation the parameters are
 $\Delta_r = 4.64\times 10^{-3}$, and 
 $K_2 = 2.52 \times 10^{-5}$ m$^{-2}$ 
 that makes $\Lambda=-2.36\times 10^{-5}$ m$^{-1}$.  

 Note that in spite of the fact that 
 the strength of the sextupole is very small, 
 part c) of Fig.~\ref{fig1:fig_22} respects the scaling expected 
 from the theory (apart from the deviation along $\hax$).

 For more realistic cases in which a lattice with 
 alternating magnet is used, or an arbitrary distribution 
 of sextupolar errors is employed, this scaling will 
 be similar. 
 This subject will be discussed in the following chapters.

 %
 %
 %

 %%%%%%%%%%%%%%%%%%%%%%%%%%%%%%%%%%%%%%%%%%%%%%%%%%%%%%%%%%%%%%%%%%%%%%%%%
 \section{Analysis in an AG circular accelerator}
 Previously we have made the analysis for the constant focusing case
 for a resonance excited by a single sextupole term, we will repeat here
 the analysis in the case of an alternating gradient (AG) structure. 
 We discuss the effect of
 a distribution of errors and use the strategy previously discussed to
 formulate a theory of particle motion and fix-lines stability.  We
 study the onset of stable motion in the proximity of a third order
 coupled resonance.

 %%%%%%%%%%%%%%%%%%%%%%%%%%%%%%%%%%%%%%%%%%%%%%%%%%%%%%%%%%%%%%%%%%%%%%%%%%%%%%
 \subsection{AG structure set-up}
 We consider the equations of motion in an AG structure.
 The equations of motion take the form
 \begin{equation}
 \begin{aligned}
   \frac{d^2 x}{ds^2} + k_x(s) x = f_x(s,x,y), \\
   \frac{d^2 y}{ds^2} + k_y(s) y = f_y(s,x,y). \\
 \end{aligned}
 \label{eq:96}
 \end{equation}
 Here the $k_x(s),k_y(s)$ are the ``time dependent'' focusing/defocusing 
 quadrupole strength.  
 The solutions of these equations for $f_x=f_y=0$ are
 \begin{equation}
 \begin{aligned}
   x(s) = \sqrt{\beta_xa_x}\cos(\phi_x(s)+\varphi_x), \\
   y(s) = \sqrt{\beta_ya_y}\cos(\phi_y(s)+\varphi_y), \\
 \end{aligned}
 \label{eq:97}
 \end{equation}
 where $\beta_x,\beta_y$ are the well known beta function and
 $
   \phi_x(s) = \int_0^s\beta_x(s)^{-1}ds, \quad
   \phi_y(s) = \int_0^s\beta_y(s)^{-1}ds.
 $
 The Hamiltonian of the system has the form
 \begin{equation}
   H_0 = \frac{(x')^2}{2} + k_x(s) \frac{x^2}{2} +
         \frac{(y')^2}{2} + k_y(s) \frac{y^2}{2}, 
 \label{eq:98}
 \end{equation}
 where $x'=dx/ds, y'=dy/ds$, 
 and the canonical equations are
 \begin{equation}
 \begin{aligned}
 \frac{dx}{ds}  =& \frac{\partial H_0}{\partial x'}, & \quad \frac{dx'}{ds} =& -\frac{\partial H_0}{\partial x}, \\
 \frac{dy}{ds}  =& \frac{\partial H_0}{\partial y'}, & \quad \frac{dy'}{ds} =& -\frac{\partial H_0}{\partial y}. \\
 \end{aligned}
 \label{eq:99}
 \end{equation}

 %%%%%%%%%%%%%%%%%%%%%%%%%%%%%%%%%%%%%%%%%%%%%%%%%%%%%%%%%%%%%%%%%%%%%%%%%%%%%%
 \subsection{Equations of the constants}
 Despite the fact that the unperturbed motion is time dependent,
 the equations of the otherwise constants
 $a_x,\varphi_x,a_y,\varphi_y$ for the perturbed system with Hamiltonian
 $H=H_0+H_1$ are as before 
 \begin{equation}
 \begin{aligned}
 \frac{da_x}{ds}       &= -&2 \frac{\partial H_1}{\partial \varphi_x}, & \quad \frac{d\varphi_x}{ds} &=  &2 \frac{\partial H_1}{\partial a_x}, \\
 \frac{da_y}{ds}       &= -&2 \frac{\partial H_1}{\partial \varphi_y}, & \quad \frac{d\varphi_y}{ds} &=  &2 \frac{\partial H_1}{\partial a_y}, \\
 \end{aligned}
 \label{eq:100}
 \end{equation}
 where we set the perturbing Hamiltonian $H_1$ due to distributed 
 sextupoles as 
 \begin{equation}
   H_1 = k_2(s) \left(\frac{x^3}{6} - \frac{xy^2}{2}\right).
 \label{eq:101}
 \end{equation}
 Let's consider a distribution of sextupolar errors localized at
 specific positions $s_j$.
 The function $k_2(s)$ in this case reads
 \begin{equation}
   k_2(s) = \sum_{j,m} K_{2j} \delta(s - s_j + m L), 
 \label{eq:102}
 \end{equation}
 where $L$ is the periodicity of the system, hence the length of the
 accelerator, and $K_{2j}$ is the integrated strength of the sextupolar error 
 located at the position $s_j$. 
 The unperturbed solution reads
 \begin{equation}
 \begin{aligned}
     x = \sqrt{\beta_x a_x}\frac{e^{i A_x} +e^{-i A_x}}{2} \\
     y = \sqrt{\beta_y a_y}\frac{e^{i A_y} +e^{-i A_y}}{2} \\
 \end{aligned}
 \label{eq:103}
 \end{equation}
 with $A_x = \phi_x(s) + \varphi_x, A_y = \phi_y(s) + \varphi_y$.
 We substitute $(x,y)$ into the Hamiltonian $H_1$ and obtain 
 {\small
 \begin{equation}
 \begin{split}
   H_1 =&
    \sum_{j,m} K_{2j} \delta(s - s_j + m L)
   \frac{1}{48}(\beta_x a_x)^{3/2}(e^{i A_x} +e^{-i A_x})^3 - \\
   -&
   \sum_{j,m} K_{2j} \delta(s - s_j + m L)
   \frac{1}{16}
     \sqrt{\beta_x a_x}\beta_y a_y \times \\
   &  (e^{i A_x} +e^{-i A_x})
     (e^{i A_y} +e^{-i A_y})^2. \\
 \end{split}
 \label{eq:104}
 \end{equation}
 }
 Note that the function $\phi_x(s)$ has the property
 \begin{equation}
   \phi_x(mL) = 2\pi Q_x m, 
 \label{eq:105}
 \end{equation}
 with $m$ an integer ($\phi_x$ is defined for the linear case). 
 Therefore we can define the function
 \begin{equation}
   \Dx(s) = \phi_x(s) - 2\pi Q_x \frac{s}{L},
 \label{eq:106}
 \end{equation}
 which is a periodic function with $\Dx(mL) = 0$.
 In a similar way we define $\Dy(s)$.
 Therefore
 \begin{equation}
 \begin{aligned}
   A_x = \Dx(s) + 2\pi Q_x \frac{s}{L} + \varphi_x, \\
   A_y = \Dy(s) + 2\pi Q_y \frac{s}{L} + \varphi_y. \\
 \end{aligned}
 \label{eq:107}
 \end{equation}

 We therefore find that the Hamiltonian Eq.~\ref{eq:104} is composed 
 of the periodic functions $\Dx(s),\Dy(s), \beta_x(s),\beta_y(s)$, which 
 have the periodicity of the accelerator $L$, and also 
 by non-periodic functions like the term $2\pi Q_x s/L$ (etc.). 
 On the other hand the delta function in Eq.~\ref{eq:104} 
 is different from zero only at the 
 longitudinal positions $s_j-mL$, and at these positions 
 the periodic functions $\Dx(s),\Dy(s),\beta_x(s),\beta_y(s)$ 
 assume the value $\Dx(s_j),\Dy(s_j),\beta_x(s_j),\beta_y(s_j)$ 
 independently on $m$. We can therefore replace in all the periodic 
 functions of  Eq.~\ref{eq:104} the coordinate $s$ by $s_j$. 

 To this purpose we call
 \begin{equation}
 \begin{aligned}
   A_{xj} = \Dx(s_j) + 2\pi Q_x \frac{s}{L} + \varphi_x, \\
   A_{yj} = \Dy(s_j) + 2\pi Q_y \frac{s}{L} + \varphi_y, \\
 \end{aligned}
 \label{eq:108}
 \end{equation}
 and $\beta_{xj}=\beta_x(s_j),\beta_{yj}=\beta_y(s_j)$. 
 Hence
 {\small
 \begin{equation}
 \begin{split}
   H_1 =&
    \sum_{j,m} K_{2j} \delta(s - s_j + m L)
   \frac{1}{48}(\beta_{xj} a_x)^{3/2} (e^{i A_{xj}} +e^{-i A_{xj}})^3 - \\
   -&
   \sum_{j,m} K_{2j} \delta(s - s_j + m L)
   \frac{1}{16}
     \sqrt{\beta_{xj} a_x}\beta_{yj} \times \\
     & a_y(e^{i A_{xj}} +e^{-i A_{xj}})(e^{i A_{yj}} +e^{-i A_{yj}})^2. \\
 \end{split}
 \label{eq:109}
 \end{equation}
 }
 We substitute the delta expansion and $A_{xj}, A_{yj}$
 obtaining
 {\small
 \begin{equation}
 \begin{split}
   H_1 =&
    \frac{1}{48L} a_x^{3/2}\sum_{j} \sum_{t} \sum_{q=0}^3  K_{2j}
 \beta_{xj}^{3/2}
 \binom{3}{q} \times \\
 & \exp\{i2\pi\frac{s - s_j}{L} t + i (\Dx(s_j) + 2\pi Q_x \frac{s}{L} + \varphi_x)
 [2q - 3]\} - \\
   -&
 \frac{1}{16L}
   \sqrt{a_x}a_y
   \sum_{j} \sum_{t} \sum_{q=0}^1 \sum_{p=0}^2
 K_{2j}
     \sqrt{\beta_{xj}}\beta_{yj}
 \binom{1}{q}
  \binom{2}{p} \times \\
 &  \times
    \exp{\{i[2\pi\frac{- s_j}{L} t + \Dx(s_j) (2q - 1)+ \Dy(s_j)(2p - 2)]\}}
    \times \\
 &  \times \exp{\{i2\pi [Q_x  (2q - 1)+ Q_y (2p - 2) + t]\frac{s}{L}\}}
    \times \\
 &  \times \exp{\{i[ \varphi_x  (2q - 1)+ \varphi_y(2p - 2) ]\}}. \\
 \end{split}
 \label{eq:110}
 \end{equation}
 }
 As the tunes $Q_x,Q_y$ are close the resonance,
 meaning $\Delta_r = Q_x + 2 Q_y - N$ is small, it follows that 
 between all oscillating terms in Eq.~\ref{eq:110}, 
 only two have a very slow frequency of oscillation, while all the 
 other can be considered as fast oscillating. 
 We neglect the fast oscillating terms and keep just the slowly 
 oscillating ones, 
 that is we select $q,p,t$ in $Q_x  (2q - 1)+ Q_y (2p - 2) + t$ 
 such as to obtain slow frequencies. 
 This happens for $q=1,p=2,t=-N$, and $q=0,p=0,t=N$. 
 Consequently the slowly varying Hamiltonian reads
 {\small
 \begin{equation}
 \begin{split}
   H_{s1} =&
 -\frac{1}{8L} \sqrt{a_x}a_y
   \sum_{j} K_{2j} \sqrt{\beta_{xj}}\beta_{yj} \\
 &
   \cos[2\pi\frac{s_j}{L} N + \Dx(s_j) + 2 \Dy(s_j)
         + 2\pi \Delta_r \frac{s}{L} + \varphi_x  + 2 \varphi_y ], \\
 \end{split}
 \label{eq:111}
 \end{equation}
 }
 we re-write $H_{s1}$ in the following form 
 \begin{equation}
 \begin{split}
   H_{s1} =\sqrt{a_x}a_y \left\{\right.
   &
     \Lambda_c
     \cos\left[\varphi_x+2\varphi_y +2\pi\Delta_r\frac{s}{L}\right] - \\
  -&
     \Lambda_s
     \sin\left[\varphi_x+2\varphi_y +2\pi\Delta_r\frac{s}{L}\right]
     \left. \right\} \\
 \end{split}
 \label{eq:112}
 \end{equation}
 with
 \begin{equation}
 \begin{split}
   \Lambda_c = & -\sum_{j} \frac{1}{8L}K_{2j}  \sqrt{\beta_{xj}}\beta_{yj} 
                 \times \\ 
               & \times \cos\left[2\pi\frac{s_j}{L} N +  \Dx(s_j) + 2 \Dy(s_j) 
                    \right], \\
   \Lambda_s = & -\sum_{j} \frac{1}{8L}K_{2j}  \sqrt{\beta_{xj}}\beta_{yj}
                 \times \\
               & \times \sin\left[2\pi\frac{s_j}{L}N  +  \Dx(s_j) + 2 \Dy(s_j)
                    \right]. \\
 \end{split}
 \label{eq:113}
 \end{equation}
 The coefficients $\Lambda_c,\Lambda_s$ depend only on the distribution
 of the sextupolar errors, on the optics, and on the selected resonance 
 (in this case $Q_x+2Q_y=N$). 
 Next we remove the time dependence in the argument of the trigonometric
 function in Eqs.~\ref{eq:112} with the transformation
 \begin{equation}
 \begin{split}
    a_x       &= \tax \\
    a_y       &= \tay \\
    \varphi_x &=  \tpx - t_x  2\pi \Delta_r \frac{s}{L} \\
    \varphi_y &=  \tpy - t_y  2\pi \Delta_r \frac{s}{L} \\
 \end{split}
 \label{eq:114}
 \end{equation}
 where we require that
 \begin{equation}
    t_x + 2t_y = 1.
 \label{eq:115}
 \end{equation}
 The new Hamiltonian reads 
 \begin{equation}
 \begin{aligned}
   \tilde{H}_{s1} =& \sqrt{\tax}\tay
   \left\{
   \Lambda_c
     \cos\left[\tpx+2\tpy\right]
     -
     \Lambda_s
     \sin\left[\tpx+2\tpy\right]
   \right\} + \\ 
                  +&(\tax t_x + \tay t_y) 2\pi\frac{\Delta_r}{2L}.
 \end{aligned}
 \label{eq:116}
 \end{equation}
 The last term is equivalent to the similar term in the Hamiltonian 
 for the constant focusing Eq.~\ref{eq:31}. It arises from the requirement 
 of making the Hamiltonian time independent. 
 At this point it is convenient to define the angle $\alpha$ as
 \begin{equation}
   \cos\alpha = \frac{\Lambda_c}{\Lambda}, \quad
   \sin\alpha = \frac{\Lambda_s}{\Lambda},
 \label{eq:117}
 \end{equation}
 with $\Lambda=\sqrt{\Lambda_s^2 + \Lambda_c^2}>0$.
 Therefore the final form of the slowly varying Hamiltonian in an 
 AG lattice becomes 
 \begin{equation}
   \tilde{H}_{s1} =\Lambda\sqrt{\tax}\tay\cos(\tpx+2\tpy+\alpha)
 +(\tax t_x + \tay t_y) \frac{2\pi \Delta_r}{2L}.
 \label{eq:118}
 \end{equation}
 The canonical equations are 
 \begin{equation}
 \begin{aligned}
 -\tax' &= 2\frac{\partial \tilde H_{s1}}{\partial \tpx} =
 2\sqrt{\tax}\tay\Lambda\sin(\tpx+2\tpy+\alpha) \\
 \tpx' &=2\frac{\partial \tilde H_{s1}}{\partial \tax} =
 2\frac{1}{2\sqrt{\tax}}\tay\Lambda\cos(\tpx+2\tpy+\alpha)
 +t_x\frac{2\pi \Delta_r}{L}\\
 -\tay' &=2\frac{\partial \tilde H_{s1}}{\partial \tpy} =
 4\sqrt{\tax}\tay\Lambda\sin(\tpx+2\tpy+\alpha) \\
 \tpy' &=2\frac{\partial \tilde H_{s1}}{\partial \tay} =
 2\sqrt{\tax}\Lambda\cos(\tpx+2\tpy+\alpha)
 +t_y \frac{2\pi \Delta_r}{L} \\
 \end{aligned}
 \label{eq:119}
 \end{equation}
 multiplying the first equation by 2 and subtracting the third one 
 we find that 
 \begin{equation}
   2\tax - \tay = C
 \label{eq:120}
 \end{equation}
 where $C$ is a constant. The previous expression is then an invariant of 
 motion.

 %%%%%%%%%%%%%%%%%%%%%%%%%%%%%%%%%%%%%%%%%%%%%%%%%%%%%%%%%%%%%%%%%%%%%%%%%%%%%%
 \subsection{Special solutions}
 Via Eq.~\ref{eq:119} the equations of the fix-line are 
 \begin{equation}
 \begin{aligned}
 -\tax' &= 0 &=\quad&
 \Lambda2\sqrt{\tax}\tay\sin(\tpx+2\tpy+\alpha) \\
 \tpx'  &= 0 &=\quad&
 \Lambda2\frac{1}{2\sqrt{\tax}}\tay\cos(\tpx+2\tpy+\alpha)
 +t_x \frac{2\pi \Delta_r}{L} \\
 - \tay' &= 0 &=\quad&
 \Lambda4\sqrt{\tax}\tay\sin(\tpx+2\tpy+\alpha) \\
 \tpy' &= 0 &=\quad&
 \Lambda2\sqrt{\tax}\cos(\tpx+2\tpy+\alpha)
 +t_y  \frac{2\pi \Delta_r}{L} \\
 \end{aligned}
 \label{eq:121}
 \end{equation}
 The condition $\tax'=\tay'=0$ is obtained from the condition
 \begin{equation}
   \sin(\tpx+2\tpy+\alpha) = 0
 \label{eq:122}
 \end{equation}
 which yields
 \begin{equation}
   \tpx+2\tpy+\alpha = \pi M
 \label{eq:123}
 \end{equation}
 with $M$ an integer.
 That means that the other two equations become 
 \begin{equation}
 \left\{
 \begin{aligned}
 0 &=
 \Lambda2\frac{1}{2\sqrt{\tax}}\tay (-1)^M
 +t_x\frac{2\pi \Delta_r}{L}\\
 0 &=
 \Lambda2\sqrt{\tax} (-1)^M
 +t_y \frac{2\pi \Delta_r}{L} \\
 \end{aligned}
 \right.
 \label{eq:124}
 \end{equation}
 Multiplying the second by 2 and summing them up and  
 recalling Eq.~\ref{eq:115} 
 we find that the equation for the fix-line reads
 \begin{equation}
 0 = \Lambda
 \left(
   \frac{1}{\sqrt{\tax}}\tay + 4 \sqrt{\tax}
 \right) (-1)^M
 +\frac{2\pi \Delta_r}{L}
 \label{eq:125}
 \end{equation}
 and the condition for its existence is that
 \begin{equation}
   \Lambda \Delta_r (-1)^M  < 0.
 \label{eq:126}
 \end{equation}
 Therefore from the equations for the fix-line we write
 \begin{equation}
 \begin{aligned}
 0 &=
 \Lambda \frac{1}{\sqrt{\tax}}\tay(-1)^M\Delta_r
 +\frac{2\pi \Delta_r^2}{L}t_x\\
 0 &=\Lambda 2 \sqrt{\tax} (-1)^M\Delta_r
 +\frac{2\pi \Delta_r^2}{L}t_y \\
 \end{aligned}
 \label{eq:127}
 \end{equation}
 from which we conclude that $t_x \ge 0$ and $t_y \ge 0$.
 That means that $t_x \ge 0$ and  $1-t_x \ge 0$,
 so that we get 
 $
   1 \ge t_x \ge 0.
 $
 We then parameterize $t_x,t_y$ as follows
 \begin{equation}
 \begin{aligned}
 t_x &=  \tau,\\
 t_y &=  \frac{1}{2}(1-\tau),\\
 \end{aligned}
 \label{eq:128}
 \end{equation}
 with $0 \le \tau \le 1$.
 We write the solution for the fix-line as a function of $\tau$, and
 get 
 \begin{equation}
 \begin{aligned}
 \tay &=
 \frac{(2\pi\Delta_r)^2}{4 \Lambda^2L^2}
 \tau(1-\tau),\\
 \tax &=
 \frac{(2\pi\Delta_r)^2}{16 \Lambda^2L^2}(1-\tau)^2. \\
 \end{aligned}
 \label{eq:129}
 \end{equation}
 We find the same solution as for the case of the constant focusing 
 except 
 that the scaling depends on $\Lambda/L$, with $\Lambda$ being 
 the driving term of the resonance as already computed by
 Hagedorn/Schoch~\cite{Hagedorn-I,Hagedorn-II,Schoch}, 
 Guignard~\cite{guign1,guign2}.

 %%%%%%%%%%%%%%%%%%%%%%%%%%%%%%%%%%%%%%%%%%%%%%%%%%%%%%%%%%%%%%%%%%%%%%%%%%%%%%
 \subsection{Fix-line} 
 The analytic form of the fix-line can now be found. 
 From $\tilde\varphi_x'=\tilde\varphi_y'=0$ we get 
 \begin{equation}
   \tilde\varphi_x = \tilde\varphi_{x,0}, \qquad
   \tilde\varphi_y = \tilde\varphi_{y,0},
 \label{eq:130}
 \end{equation}
 and returning back to the coordinates of the Hamiltonian $H_{s1}$ we find
 \begin{equation}
 \begin{aligned}
   \varphi_x = \tilde{\varphi}_{x,0} - t_x 2\pi\Delta_r\frac{s}{L}, \\
   \varphi_y = \tilde{\varphi}_{y,0} - t_y 2\pi\Delta_r\frac{s}{L}, \\
 \end{aligned}
 \label{eq:131}
 \end{equation}
 with 
 $\sin(\tilde{\varphi}_{x,0}+2\tilde{\varphi}_{y,0} + \alpha)=0$.
 The solution has therefore the form
 {\small
 \begin{equation}
 \begin{aligned}
   x  &= \sqrt{\beta_xa_x}
         \cos\left[\phi_x(s)- t_x 2\pi\Delta_r\frac{s}{L} + \tilde{\varphi}_{x,0} \right] \\
   x' &= -\frac{\alpha_x}{\sqrt{\beta_x}}\sqrt{a_x}
         \cos\left[\phi_x(s)- t_x 2\pi\Delta_r\frac{s}{L} + \tilde{\varphi}_{x,0} \right] - \\
      & - \sqrt{\beta_xa_x}
         \sin\left[\phi_x(s)- t_x 2\pi\Delta_r\frac{s}{L} + \tilde{\varphi}_{x,0} \right]
          \left(\frac{1}{\beta_x}- t_x 2\pi\Delta_r\frac{1}{L}\right) \\
   y  &= \sqrt{\beta_ya_y}
         \cos\left[\phi_y(s)- t_y 2\pi\Delta_r\frac{s}{L} + \tilde{\varphi}_{y,0} \right] \\
   y' &= -\frac{\alpha_y}{\sqrt{\beta_y}}\sqrt{a_y}
         \cos\left[\phi_y(s)- t_y 2\pi\Delta_r\frac{s}{L} + \tilde{\varphi}_{y,0} \right]- \\
      & - \sqrt{\beta_ya_y}
         \sin\left[\phi_y(s)- t_y 2\pi\Delta_r\frac{s}{L} + \tilde{\varphi}_{y,0} \right]
          \left(\frac{1}{\beta_y}- t_y 2\pi\Delta_r\frac{1}{L}\right). \\
 \end{aligned}
 \label{eq:132}
 \end{equation}
 }
 This is the equation of the fix-line, which is a closed line in the
 4D phase space.
 
 The Poincar\'e surface of section is identified by the condition 
 $s = L {\cal N}$
 with ${\cal N}$ an integer correspondent to the ${\cal N}$-th turn.
 From the solution of the fix-line equation we have
 $
  \sin(\tilde{\varphi}_{x,0}+2\tilde{\varphi}_{y,0} + \alpha)=0
 $
 and we find
 \begin{equation}
   \tilde{\varphi}_{x,0}+2\tilde{\varphi}_{y,0}+\alpha=\pi M,
 \label{eq:133}
 \end{equation}
 with $M$ an integer properly taken to guarantee the existence of the solution.
 It is straightforward to show that
 \begin{equation}
 \begin{split}
  (Q_x  -  t_x\Delta_r)2\pi{\cal N} +  \tilde{\varphi}_{x,0} + \\
 2(Q_y  -  t_y \Delta_r)2\pi{\cal N} + 2\tilde{\varphi}_{y,0} = \\
 2\pi N {\cal N} +  \pi M - \alpha.   \\
 \end{split}
 \label{eq:134}
 \end{equation}
 
 Therefore, by using Eq.~\ref{eq:134}, we find that on the 
 Poincar\'e surface of section we have 
 \begin{equation}
 \begin{aligned}
   x  &= \sqrt{\beta_xa_x}
         \cos\left[
         - 2(Q_y  -  t_y \Delta_r)2\pi{\cal N} - 2\tilde{\varphi}_{y,0}   
         - \alpha +   \pi M
             \right], \\
   y  &= \sqrt{\beta_ya_y}
         \cos\left[
 (Q_y  -  t_y \Delta_r)2\pi{\cal N} + \tilde{\varphi}_{y,0}
             \right], \\
 \end{aligned}
 \label{eq:135}
 \end{equation}
 with $\tilde\varphi_{y,0}$ an initial phase that sets the starting point.
 Lastly, we find the result that the fix-line can be parameterized as
 \begin{equation}
 \begin{aligned}
   x_t &= \sqrt{\beta_xa_x}\cos[-2(Q_y - t_y\Delta_r)t -\alpha +\pi M ], \\
   y_t &= \sqrt{\beta_ya_y}\cos[(Q_y- t_y\Delta_r)t ]. \\
 \end{aligned}
 \label{eq:136}
 \end{equation}
 The presence of the angle $\alpha$ is the only difference with respect to the
 case of the constant focusing lattice. 
 However, by changing the 
 longitudinal position of the Poincar\'e surface of section, $\alpha$ 
 can be made equal to zero.

 %%%%%%%%%%%%%%%%%%%%%%%%%%%%%%%%%%%%%%%%%%%%%%%%%%%%%%%%%%%%%%%%%%%%%%%%%%%%%%
 \subsection{Stability}
 The discussion of the stability of the fix-line is the same as for 
 the case of the constant focusing system. The conclusion holds as before:
 for $\tau > 2/3$ the fix-line is stable, and for $2/3 < \tau < 1$ the
 fix-line is unstable.

 %%%%%%%%%%%%%%%%%%%%%%%%%%%%%%%%%%%%%%%%%%%%%%%%%%%%%%%%%%%%%%%%%%%%%%%%%%%%%%
 \subsection{Stability of Motion}
 Next we study the onset of unstable motion. 
 From the canonical equation Eqs.~\ref{eq:119},
 we keep the first equation; due to $2\tax - \tay = C$ 
 the third equation is redundant.
 The second and fourth equations can be summed up to yield
 \begin{equation}
 \left\{
 \begin{aligned}
 -\tax' &= 2\sqrt{\tax}\tay\Lambda\sin(\Omega+\alpha) \\
 \Omega' &=
 \left[2\frac{1}{2\sqrt{\tax}}\tay+4\sqrt{\tax}\right]
 \Lambda\cos(\Omega+\alpha)
 +\frac{2\pi \Delta_r}{L}\\
 \end{aligned}
 \right.
 \label{eq:137}
 \end{equation}
 with $\Omega = \tpx+2\tpy$.
 Therefore we re-write the set of equations as
 \begin{equation}
 \left\{
 \begin{aligned}
 -\tax' &= 2\sqrt{\tax}(2\tax-C)\Lambda\sin(\Omega+\alpha) \\
 \Omega' &=
 \left[\frac{1}{\sqrt{\tax}}(2\tax-C)+4\sqrt{\tax}\right]
 \Lambda\cos(\Omega+\alpha)
 +\frac{2\pi \Delta_r}{L}\\
 \end{aligned}
 \right.
 \label{eq:138}
 \end{equation}
 The function
 \begin{equation}
   I(\tax,\Omega) = 2\Lambda\sqrt{\tax}(2\tax-C)\cos(\Omega+\alpha)
   + \tax\frac{2\pi \Delta_r}{L}
 \label{eq:139}
 \end{equation} 
 is the invariant of the system as obtained for the constant
 focusing discussion, except for the presence of the angle $\alpha$ in the
 cosine function, but this is just a shift of the stability diagram. 
 Hence, the conclusions obtained for the case of the constant focusing
 structure hold in the AG structure as well.

 %%%%%%%%%%%%%%%%%%%%%%%%%%%%%%%%%%%%%%%%%%%%%%%%%%%%%%%%%%%%%%%%%%%%%%%%
 \subsection{Comparison with simulations}
 In Fig.~\ref{fig1:fig_23} we compare the reconstruction of the
 stability domain with part a) of Fig.~\ref{fig1:fig_23} that is
 derived from the theory and numerical simulations.  As a practical
 example part b) of Fig.~\ref{fig1:fig_23} shows instead the same
 result for the SIS18 reference lattice equipped with a single
 sextupolar kick.  As for the case of the constant focusing we find an
 enlarging of the stability near the $a_x$ axis.  Part c) of
 Fig.~\ref{fig1:fig_23} shows the stability domain for the SIS18 where
 12 sextupolar errors are randomly excited.  The sextupoles are all
 placed at the beginning of each super-period.  Interestingly the
 results are still close to those predicted by the theory.  A small
 shrinking of the stability is found anywhere close to the axis $a_x$.
 The discrepancy can be attributed to the effect of high order
 harmonics not included in the theory.

 \begin{figure}[H]
 \begin{center}
 \unitlength 0.7mm
 \begin{picture}(80,73)
 \put(-5, 0) {\epsfig{file=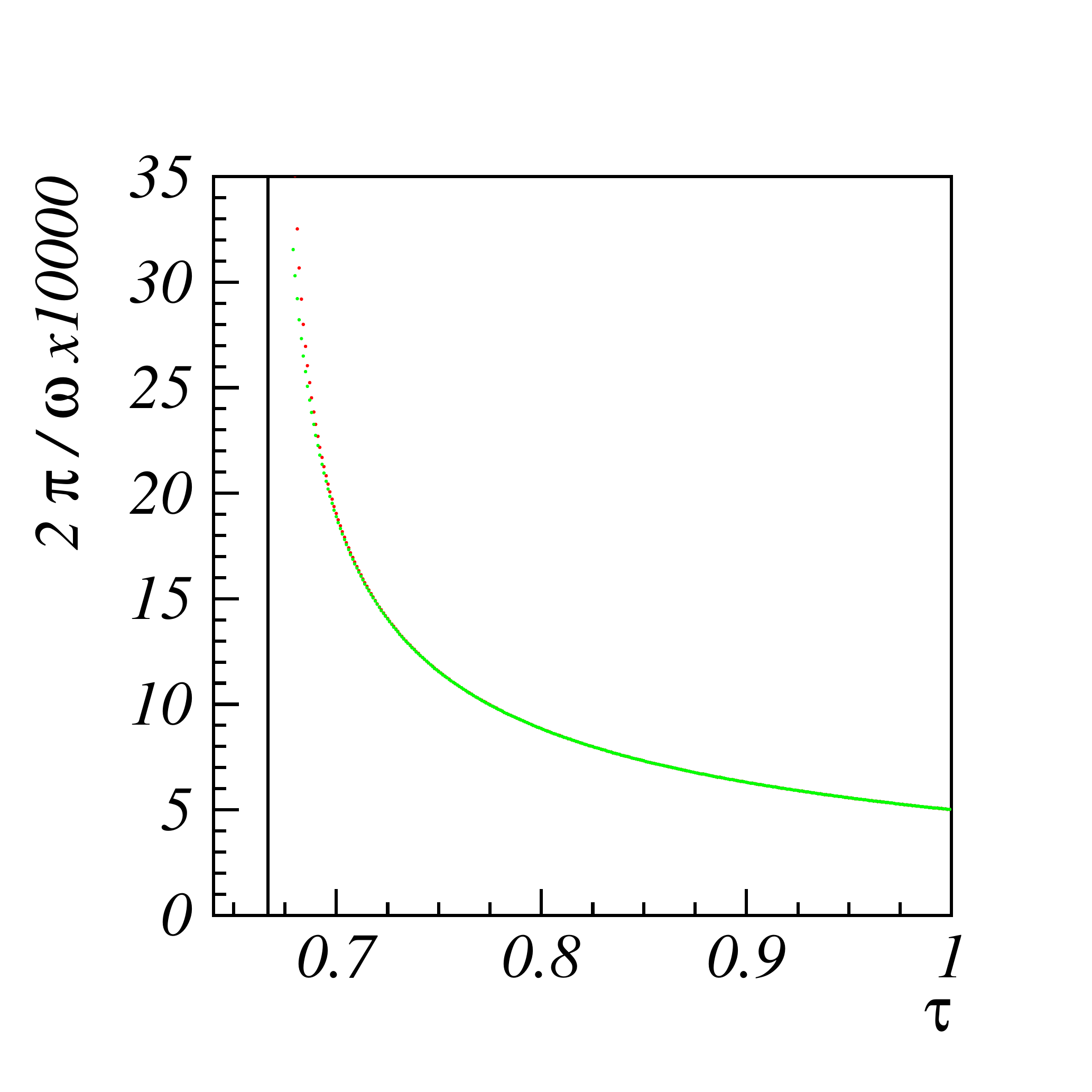,width=60mm}}
 \end{picture}
 \caption{
 Comparison of the wavelength from analytic theory (blue dots) and
 retrieved from simulations (horizontal color black and vertical color red).
 This result is obtained for a FODO cell with one sextupolar
 kick. The agreement is excellent with a discrepancy that
 increases in the proximity of $\tau=2/3$.
 }
 \label{fig1:fig_24}
 \end{center}
 \end{figure}

 We have verified this ansatz by simulations in a FODO structure 
 where we artificially made a single harmonics stronger by placing 
 a sequence of sextupoles with alternating sign of strength.  
 The results are shown in part d) of Fig.~\ref{fig1:fig_23}: 
 the stability domain very well retrieves the theoretical one 
 of part a) of Fig.~\ref{fig1:fig_23}. 

 Lastly, we tested the prediction of the secondary tunes for a
 particle slightly off a fix-line.  These results are shown in
 Fig.~\ref{fig1:fig_24}.  The simulation is made in a FODO cell
 equipped with a single sextupole.  The secondary frequency is
 measured by comparing the side bands to the nominal tunes.  The
 theoretical prediction is obtained from Eq.~\ref{eq:71} where
 $\Delta_r$ is replaced by $2\pi\Delta_r/L$ (green dots).  From the
 simulations we retrieved the wavelength in $x,y$ (red dots).  The
 picture shows an excellent agreement except for slight deviations of
 the side bands where the curves diverges and high order terms become
 relevant (near $\tau\simeq 2/3$).

 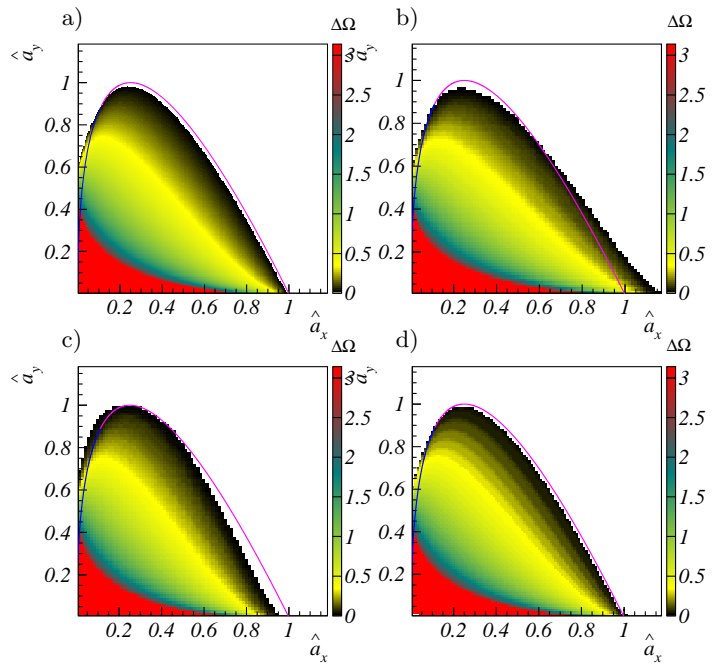
\begin{figure}[H]
 \begin{center}
 \unitlength 0.74mm
 \begin{picture}(80,115)
 \put(3,0){
 \put(-25, 55)   {\epsfig{file=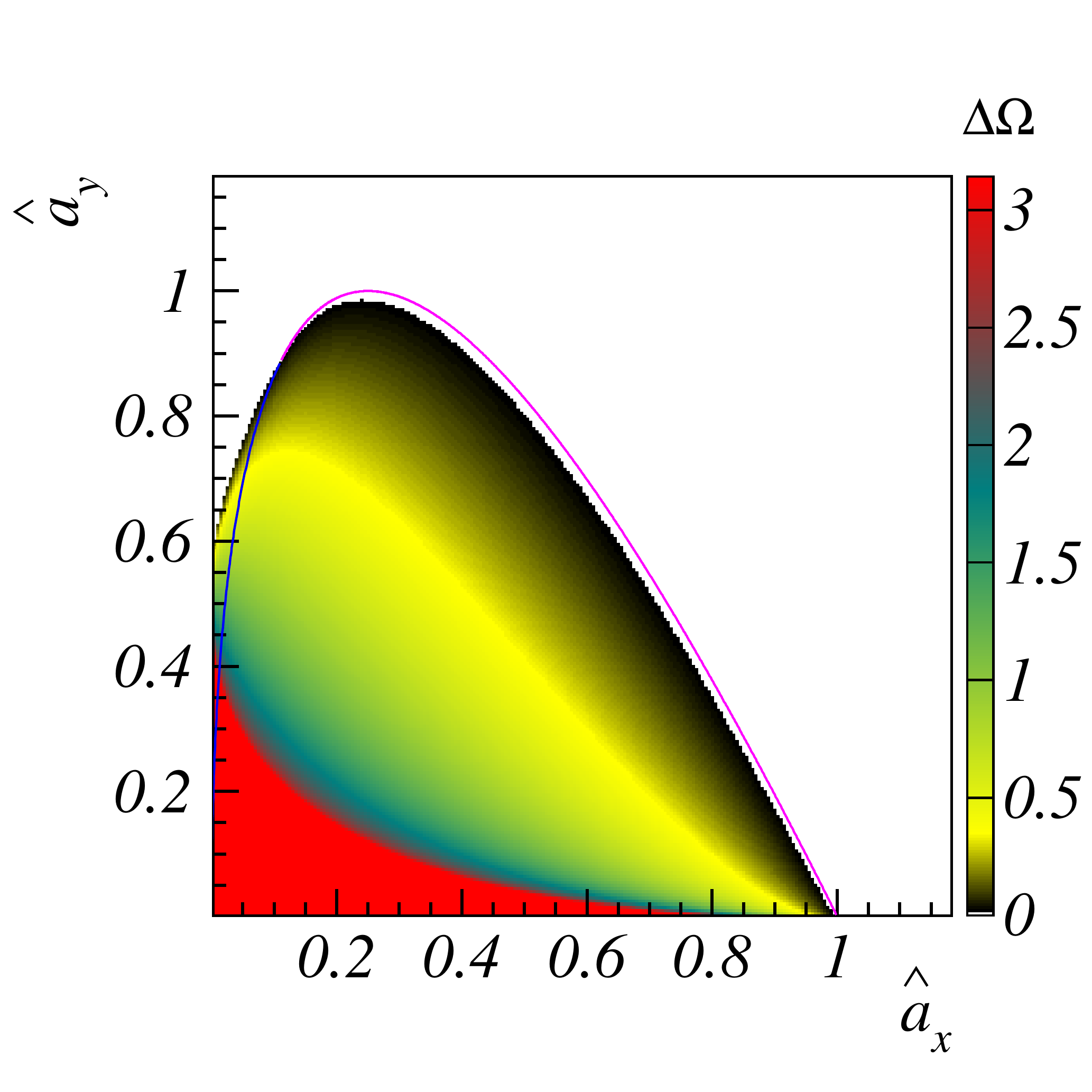,width=49mm}}
 \put( 35, 55)   {\epsfig{file=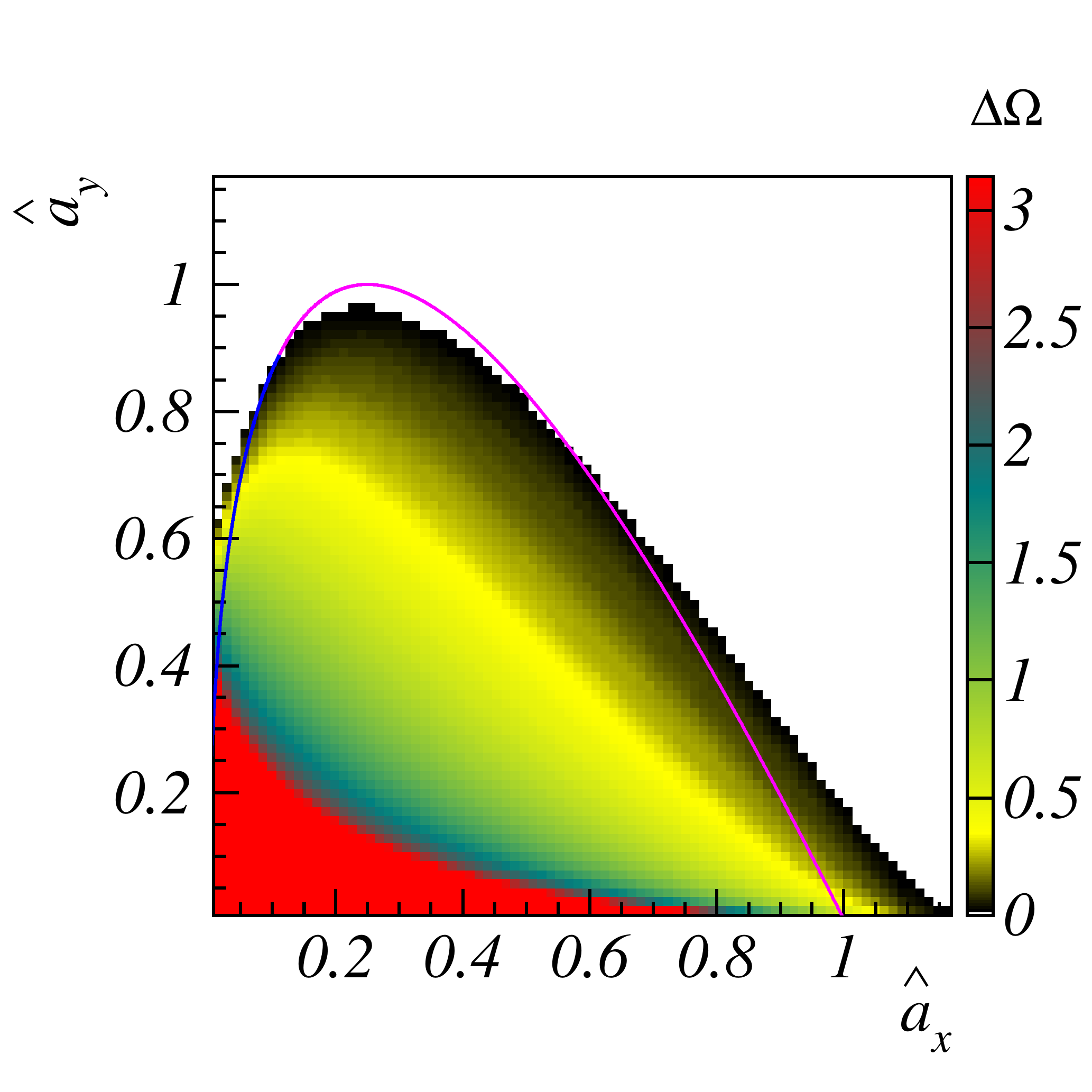,width=49mm}}
 \put(-25,   -3) {\epsfig{file=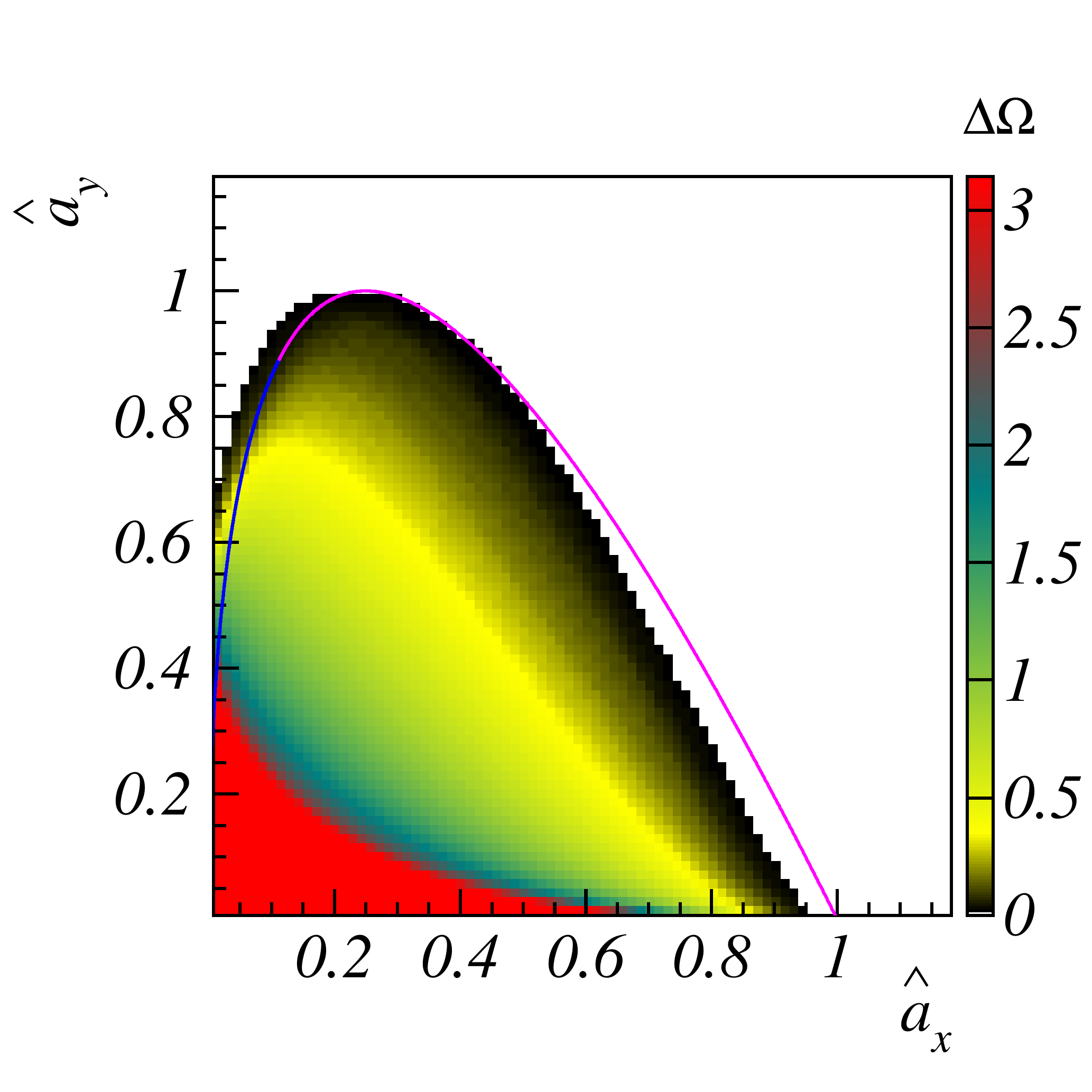,width=49mm}}
 \put( 35,   -3) {\epsfig{file=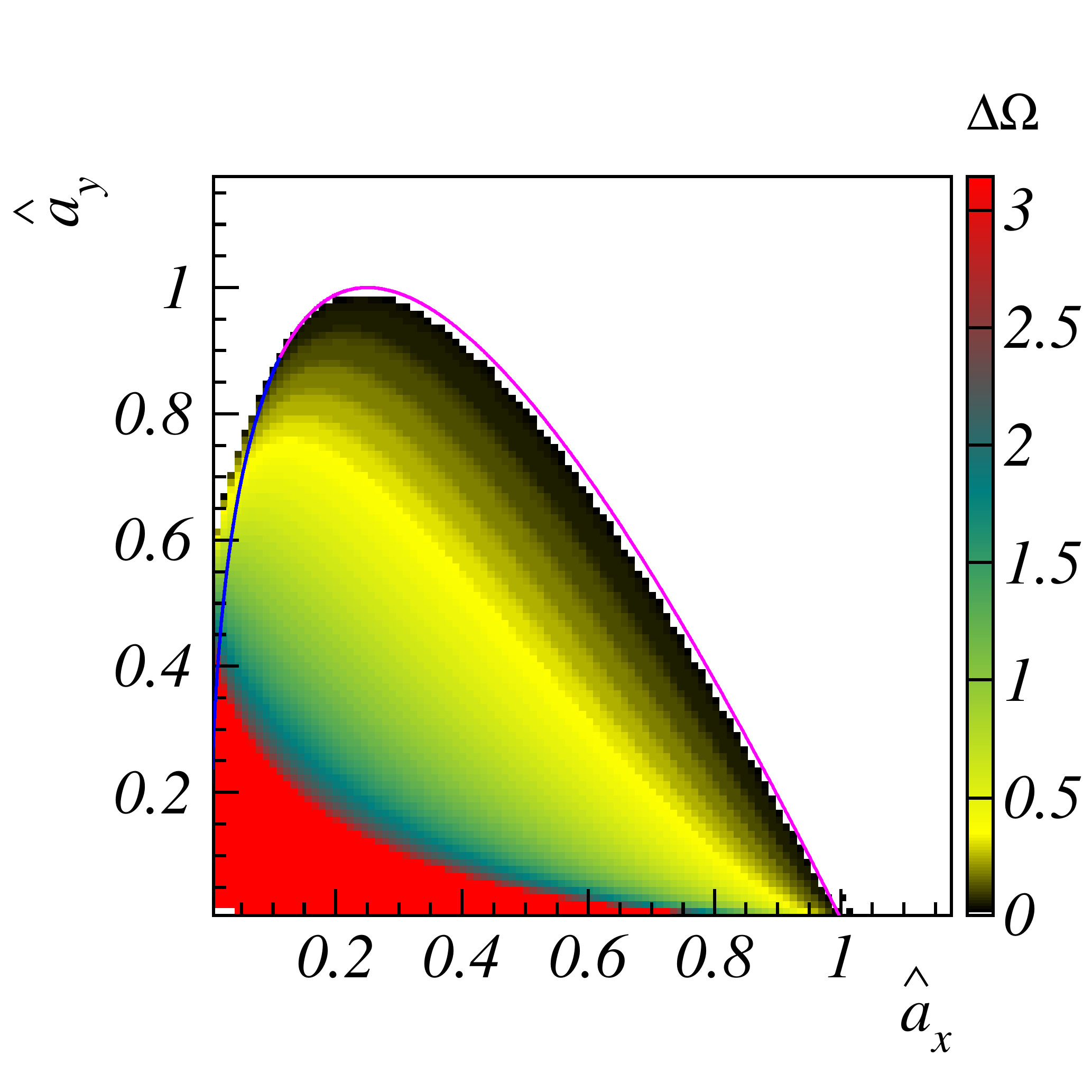,width=49mm}}
 \put(-15,114)   {a)}
 \put( 45,114)   {b)}
 \put(-15,56)    {c)}
 \put( 45,56)    {d)}
          }
 \end{picture}
 \caption{
 Complete stability domain for an AG circular accelerator. 
 Part a) the stability domain as obtained by the
 analytic theory for the SIS18 lattice with a distribution of sextupolar errors.
 Part b) shows the stability domain of SIS18 obtained
 for one sextupolar kick. The result shows a slight enlarging of
 the stability domain close to the $a_x$ axis.
 Part c) the same result obtained by tracking
 of 100x100x51x51 initial conditions, and tracked for 1000 turns. 
 Part d) is an example of a FODO cell with a distribution
 of sextupoles such as to excite only one harmonic. 
 We find a good confirmation of the theory at the edge of the stability
 domain.
 }
 \label{fig1:fig_23}
 \end{center}
 \end{figure}
 %
 %
 %

% %
% %
% %
% \begin{figure}[H]
% \begin{center}
% \unitlength 0.7mm
% \begin{picture}(80,320)
% \put( 0, 240) {\epsfig{file=fl_fig_25a.pdf,width=65mm}}
% \put( 0, 160) {\epsfig{file=fl_fig_25b.pdf,width=65mm}}
% \put( 0,  80) {\epsfig{file=fl_fig_25c.pdf,width=65mm}}
% \put( 0,   0) {\epsfig{file=fl_fig_25d.pdf,width=65mm}}
% \put(-5,320) {a)}
% \put(-5,240) {b)}
% \put(-5,160) {c)}
% \put(-5, 80) {d)}
% %
% \end{picture}
% \caption{
% Complete stability domain for an AG circular accelerator. 
% Part a) the stability domain as obtained by the
% analytic theory for the SIS18 lattice with a distribution of sextupolar errors.
% Part b) shows the stability domain of SIS18 obtained
% for one sextupolar kick. The result shows a slight enlarging of
% the stability domain close to the $a_x$ axis.
% Part c) the same result obtained by tracking
% of 100x100x51x51 initial conditions, and tracked for 1000 turns. 
% Part d) is an example of a FODO cell with a distribution
% of sextupoles such as to excite only one harmonic. 
% We find a good confirmation of the theory at the edge of the stability
% domain.
% }
% \label{fig1:fig_23}
% \end{center}
% \end{figure}
% %
% %
% %

 %%%%%%%%%%%%%%%%%%%%%%%%%%%%%%%%%%%%%%%%%%%%%%%%%%%%%%%%%%%%%%%%%%
 \section{Tori cut}
 In this section we discuss the Poincar\'e surface of section of the tori
 surrounding the fix-line.  We have already obtained those orbits from
 the numerical simulations in Figs.~\ref{fig1:fig_12},
 \ref{fig1:fig_13}, \ref{fig1:fig_14}.  We show here that these orbits
 will be retrieved from the invariants.  In those figures the orbits
 in $x-x'$ are obtained by tracking one single particle for many
 turns, and plotting $x,x'$ at each turn when the condition $y'/y <
 \epsilon$ and $y>0$ are satisfied; here $\epsilon$ is a positive
 small number.  The orbits in $y-y'$ are obtained by making a
 selection of the particle when $x'/x < \epsilon$ and $x>0$.
 
 Without losing generality 
 we make the discussion for the case of the constant focusing structure 
 as presented in Sect. IV, V, VI, VII, VIII. 
 Our starting point is to consider a tori, identified by the invariants
 $I=I(\tax,\Omega)$, and $2\tax - \tay = C$, filled with particles. 
 That means that a particle in the tori have $\tax,\tay,\Omega$ 
 satisfying the two invariants, with phases $\tpx,\tpy$ 
 satisfying $\tpx + 2 \tpy = \Omega$ (the $\Omega$ variable is defined in 
 Eq.~\ref{eq:73}). 
 The situation 
 is equivalent to what is done in the simulations when tracking one
 particle repeatedly for many turns, in the condition that the particle
 explores all the tori.
 
 The spatial coordinates of the particles populating the tori are given by
 \begin{equation}
   \frac{x}{\sqrt{\beta_x}} = \sqrt{a_x}\cos(\psi_x), \quad
   \frac{y}{\sqrt{\beta_y}} = \sqrt{a_y}\cos(\psi_y)
 \label{eq:140}
 \end{equation}
 where we define for convenience
 \begin{equation}
   \psi_x= Q_x\theta+\varphi_x, \quad
   \psi_y= Q_y\theta+\varphi_y,
 \label{eq:141}
 \end{equation}
 where the quantities $a_x,a_y,\varphi_x,\varphi_y$ vary in time.
 For this case (constant focusing) the beta functions are
 $\beta_x = 1/Q_x, \beta_y = 1/Q_y$.
 As discussed before, a specific set of initial 
 particle coordinates is identified by 
 the initial condition $a_{x0,}a_{y0},\varphi_{x0,}\varphi_{y0}$,
 and this condition is used for integrating
 the equation of motions of $a_x',a_y',\varphi_x',\varphi_y'$
 (these are the Eqs.~\ref{eq:34} with ${\cal F}_x={\cal F}_y=0$).
 The momentum coordinates are obtained by differentiation 
 \begin{equation}
 \begin{aligned}
   \left(\frac{x}{\sqrt{\beta_x}}\right)' &=
   \frac{a_x'}{2\sqrt{a_x}}\cos(\psi_x)
   -\sqrt{a_x}\sin(\psi_x) (Q_x + \varphi_x'), \\
   \left(\frac{y}{\sqrt{\beta_y}}\right)' &=
   \frac{a_y'}{2\sqrt{a_y}}\cos(\psi_y)
   -\sqrt{a_y}\sin(\psi_y) (Q_y + \varphi_y'). 
 \end{aligned}
 \label{eq:142}
 \end{equation}
 The variables $a_x',a_y',\varphi_x',\varphi_y'$ can be 
 expressed in terms of 
 $\tax',\tay',\tpx',\tpy'$, by using the transformations
 \begin{equation}
 \begin{aligned}
   a_x&=\tax,  &a_y&=\tay, \\
   \varphi_x&=\tpx-t_x\theta, &\varphi_y&=\tpy-t_y\theta. \\
 \end{aligned}
 \label{eq:143}
 \end{equation}
 and in these variables we know what is $\tax',\tay',\tpx',\tpy'$ 
 from the equation~\ref{eq:34} (with ${\cal F}_x={\cal F}_y=0$).
 By substituting the equations of motion Eqs.~\ref{eq:34}
 in Eqs.~\ref{eq:142},
 and after some algebraic calculations we obtain
 \begin{equation}
 \begin{aligned}
   \left(\frac{x}{\sqrt{\beta_x}}\right)' &=
   \Lambda \tay\sin(-\psi_x+\Omega) -\sqrt{\tax}\sin(\psi_x) Q_x. \\
   \left(\frac{y}{\sqrt{\beta_y}}\right)' &=
   2 \Lambda\sqrt{\tax\tay}\sin(-\psi_y+\Omega) -\sqrt{\tay}\sin(\psi_y) Q_y. \\
 \end{aligned}
 \label{eq:144}
 \end{equation}
 
 At this point we define the ``normalized'' coordinates (suggested by 
 Eq.~\ref{eq:81}) 
 \begin{equation}
 \begin{aligned}
   \ex  =& \frac{x}{\sqrt{\beta_x}} \left|\frac{4\Lambda}{\Delta_r}\right| \\
   \epx =& \left(
                \frac{d}{d\theta}
                \frac{x}{\sqrt{\beta_x}}\right)\beta_x
                \left|\frac{4\Lambda}{\Delta_r}\right|
        =& 
                x' \sqrt{\beta_x} 
                \left|\frac{4\Lambda}{\Delta_r}\right|
 \end{aligned}
 \label{eq:145}
 \end{equation}
 \begin{equation}
 \begin{aligned}
   \ey  =& \frac{y}{\sqrt{\beta_y}} \left|\frac{4\Lambda}{\Delta_r}\right| \\
   \epy =& \left(
                \frac{d}{d\theta}
                \frac{y}{\sqrt{\beta_y}}\right)\beta_y
                \left|\frac{4\Lambda}{\Delta_r}\right|
        =& 
                y' \sqrt{\beta_y} 
                \left|\frac{4\Lambda}{\Delta_r}\right|
 \end{aligned}
 \label{eq:146}
 \end{equation}
 Therefore the particle in the ``normailzed'' coordinates reads 
 \begin{equation}
 \begin{aligned}
   \ex  &= \sqrt{\hax}\cos(\psi_x) \\
   \epx &=
   \frac{\beta_x}{4}\mu\Delta_r\hay\sin(-\psi_x+\Omega)
   -\sqrt{\hax}\sin(\psi_x) \\
   \ey  &= \sqrt{\hay}\cos(\psi_y) \\
   \epy &=
   \frac{\beta_y}{2}\mu\Delta_r\sqrt{\hax\hay}\sin(-\psi_y+\Omega) -
   \sqrt{\hay}\sin(\psi_y) \\
 \end{aligned}
 \label{eq:147}
 \end{equation}
 where the coordinates $\hax,\hay$ are those defined in Eq.~\ref{eq:81}.

 Using the definition
 $
   \tpx + 2\tpy = \Omega,
 $
 we find that $\psi_x,\psi_y$ are bounded by the relation
 $
   \psi_x+2\psi_y=N\theta + \Omega.
 $
 As the Tori is being filled, we make sure which particles satisfy some
 special condition at say $\theta=0$ 
 (we do not need to consider many turns, because the tori
 is completely populated with particles, and after each turn the tori
 remain populated in the same way).
 Hence the tori cut at $\theta=0$ yields 
 \begin{equation}
   \psi_x+2\psi_y= \Omega.
 \label{eq:148}
 \end{equation}

 In order to obtain the orbits in $x-x'$, i.e. in  $\ex-\epx$, 
 we require that we select
 only the particles with $\epy=0$.
 For small $\Delta_r$ this happens for all particles in the Tori,
 which have $\psi_y=\pi m$, with $m$ an integer.
 That means that we choose particles with a phase in the 
 $x$ plane as 
 \begin{equation}
   \psi_x = - 2\pi m + \Omega.
 \label{eq:149}
 \end{equation}
 
 By substituting Eq.~\ref{eq:149} into the Eq.~\ref{eq:147}
 (formula for $\epx$) we find that the coordinates of the orbits obtained 
 by cutting the tori imposing $\epy=0$ are given by the map
 \begin{equation}
 \left\{
 \begin{aligned}
   \ex  &= \sqrt{\hax}\cos(\Omega)   \\
   \epx &= - \sqrt{\hax}\sin(\Omega) \\
 \end{aligned}
 \right.
 \label{eq:151}
 \end{equation}
 where $\hax,\Omega$ must satisfy the invariants that defines
 the tori.
 As we started with the filled tori, any pair $\hax,\Omega$ satisfying
 $\hat I=\hat I(\hax,\Omega), 2\hax-\hay=2\xi$
 will identify a particle initially populating the tori, and can be 
 used in  Eq.~\ref{eq:151}. 
 The condition $y>0$, i.e. $\ey >0$, implies $m=2n$ with $n$ an integer. 
 However, this condition does not change Eq.~\ref{eq:151}. 
 
 We proceed in a similar way for the cut in $\epx=0$, we select the
 particles that satisfies $\psi_x=\pi m$, with $m$ an integer, in 
 addition the condition that $x >0$, i.e., $\ex > 0$, 
 implies $m=2n$, with $n$ an integer. 
 As discussed before we consider the tori at $\theta=0$.
 Therefore once we fix $\Omega$, there are the following phases possible
 \begin{equation}
   \psi_y = \frac{\Omega}{2} - \pi n.
 \label{eq:152}
 \end{equation}
 Substituting it in Eq.~\ref{eq:147} (last two equations)
 we find the following map 
 \begin{equation}
 \left\{
 \begin{aligned}
   \ey  &=  (-1)^n\sqrt{\hay}\cos\left(\frac{\Omega}{2}\right) \\
   \epy &= -(-1)^n\sqrt{\hay}\sin\left(\frac{\Omega}{2}\right) \\
 \end{aligned}
 \right.
 \label{eq:153}
 \end{equation}

 \begin{figure}[H]
 \begin{center}
 \unitlength 0.87mm
 \begin{picture}(80,178)
 \put( -10, 95) {\epsfig{file=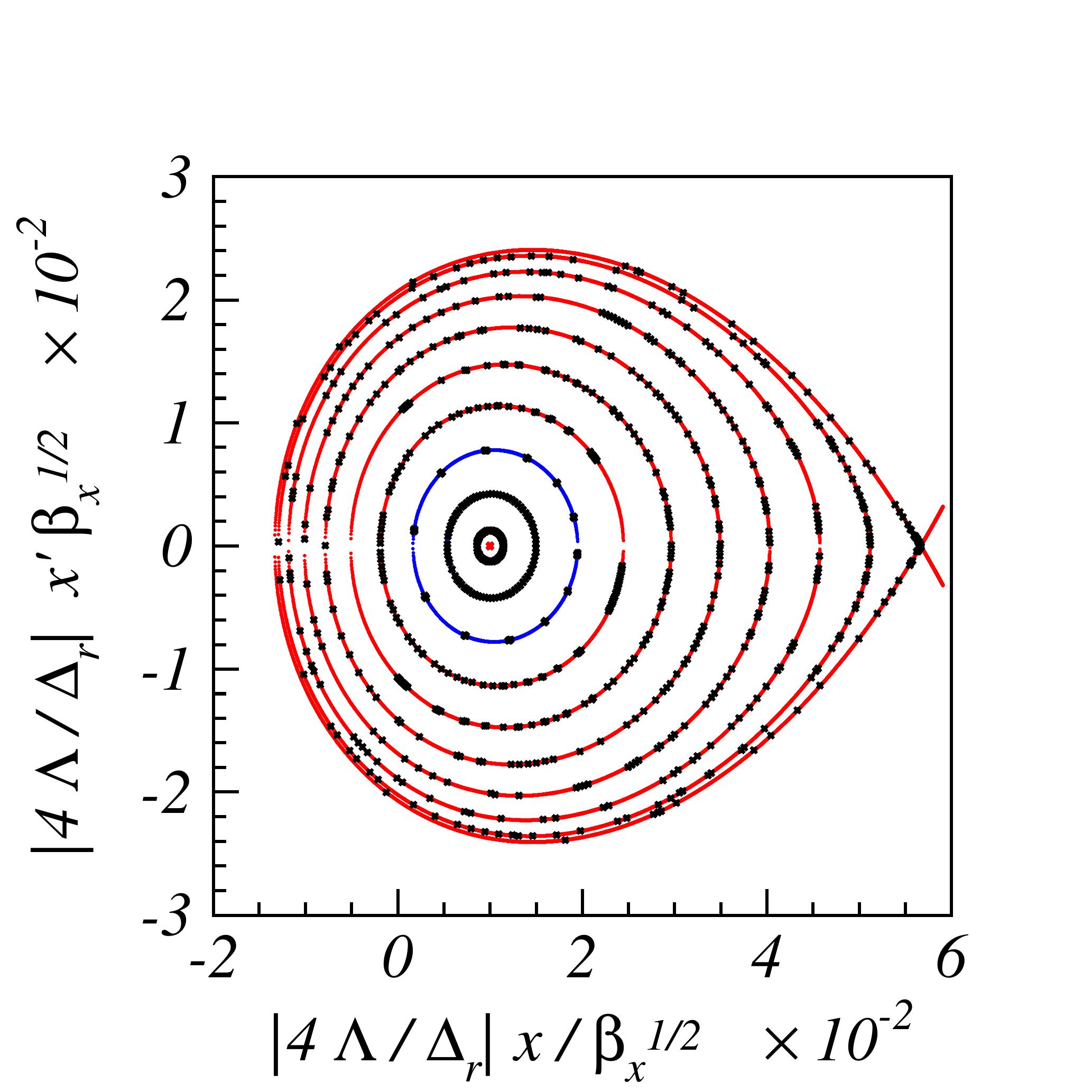,width=90mm}}
 \put( -10,  0) {\epsfig{file=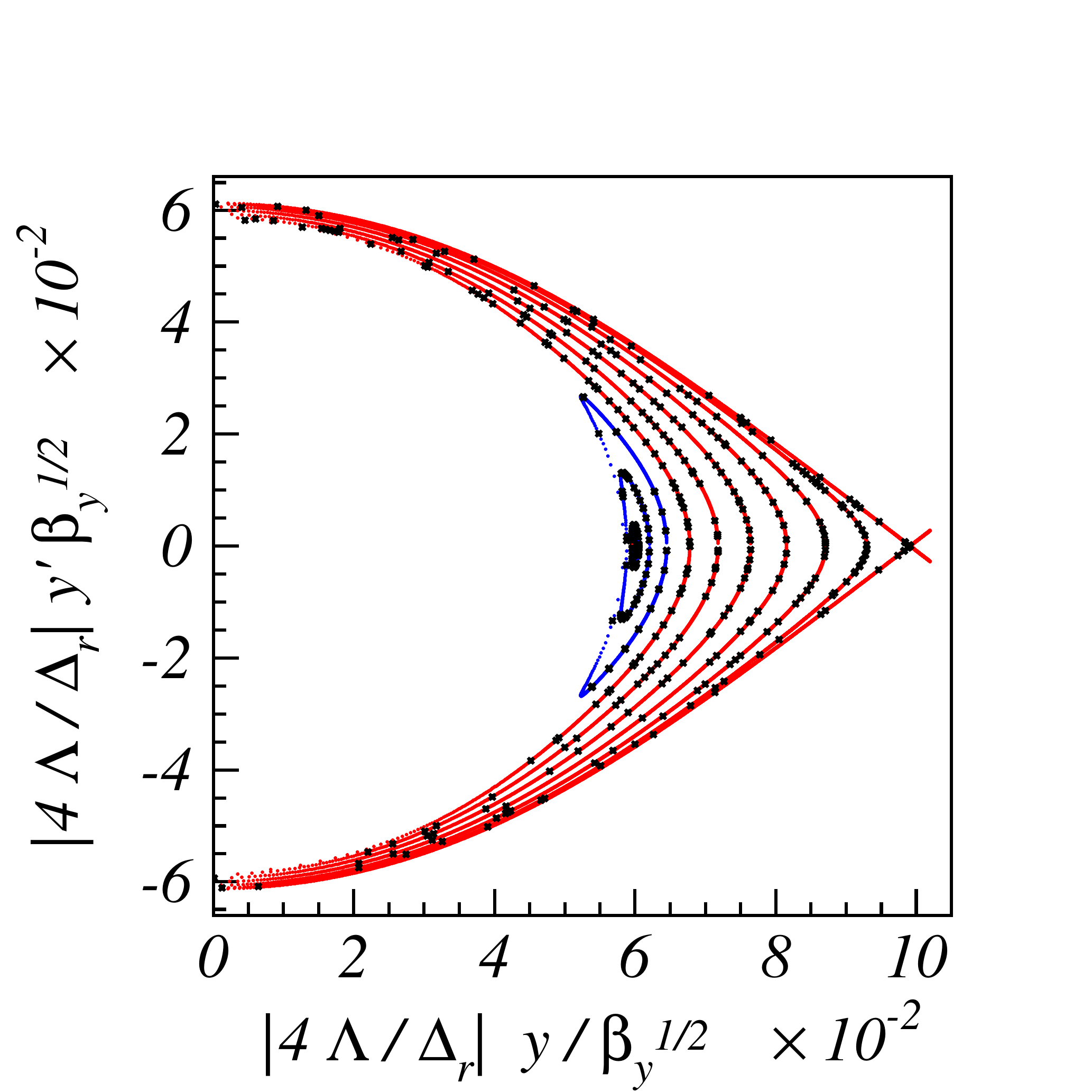,width=90mm}}
 \put(-5,185) {a)}
 \put(-5, 95) {b)}
  
 \end{picture}
 \caption{
 Tori cut for parameters $\Delta_r, \Lambda$ such that $\mu=-1$. 
 The solid curves are the orbits obtained from the analytic theory, 
 while the dots are the coordinates of one particle tracked and plot 
 when it satisfies the tori cut condition. 
 The red curves are the orbits of particles in the red dashed region of part 
 b) of Fig.~\ref{fig1:fig_17} (stability type 1); 
 The blue orbits are for particles with initial condition in the 
 blue dashed region of part b) of Fig.~\ref{fig1:fig_17} 
 (stability type 2); 
 Note in the center of the circular blue orbits one red dot: 
 this is the ``stationary point'' discussed in Sec.~\ref{stationary-point}. 
 }
 \label{fig1:fig_25} 
 \end{center}
 \end{figure}

 As before $\hax,\Omega$ must satisfy the invariants that define the
 tori, i.e. $\hat I=\hat I(\hax,\Omega), 2\hax-\hay=2\xi$.
 We recall that the fix-lines are defined by the level lines of
 $\hat I=\hat I(\hax,\Omega)$ that are tangent to the line $\ya=1$,
 with $\ya=-\mu\cos\Omega$, being $\mu=\pm 1$, the sign depends on the
 combination of the signs of $\Lambda, \Delta_r$.
 That simply means that $\Omega$ of the fix-line is given by $\Omega =
 \Omega_{fl} = \frac{\pi}{2} (1+\mu)$. 
 Substituting this value into Eq.~\ref{eq:153} we find that 
 the fix-line are located at 
 $\ey=0,\epy=-(-1)^n\sqrt{\hay}$ for $\mu=1$, 
 and at $\ey = (-1)^n\sqrt{\hay},\epy=0$ for $\mu=-1$ 
 under the tori cut conditions. 
 Figure~\ref{fig1:fig_25} shows a comparison of the orbits in $x-x'$ and
 $y-x'$ with the particle tracking. For this simulation it holds $\mu
 =-1$. The color of the curves is associated to the stability
 properties of the level lines as discussed in Fig.~\ref{fig1:fig_17}. 
%
% We also observe that the invariants do not change their value
% for the transformation $\hax,\Omega \rightarrow \hax,-\Omega$,
% hence the maps discussed here have to be also applied for $-\Omega$.

 %%%%%%%%%%%%%%%%%%%%%%%%%%%%%%%%%%%%%%%%%%%%%%%%%%%%%%%%%%%%%%%%%%%%%%%%%
 \section{Conclusion}
 We have derived the theory of the fix-lines and the stability domain
 close to a third order coupled resonance.

 We find that the number of fix-lines is infinite, and that they are
 of stable or unstable type.  The set of fix-lines form a continuous
 curve of parabolic shape in the plane of the single particle
 emittances $\tax,\tay$.  The theory suggests a natural
 parametrization of the fix-lines based on the canonical
 transformation that makes the system time independent.
 
 Our theory predicts the behavior of perturbation close to a fix-line
 and we find the property that the perturbation oscillates along a
 line with $2\delta\tax=\delta\tay$.  Our analysis shows that these
 objects define the border of stability when a single harmonics is
 excited.  The structure of the canonical equation of the constants
 suggests a natural symmetry set by the resonance condition.  We find
 that for the variables $\tax,\Omega$ the dynamics imposes an
 invariant, and using this invariant we discuss the complete set of
 topological properties of the level lines.  It turns out that the
 level line associated to the fix-line creates a barrier to particles,
 hence they set the edge of the stability domain.
 
 The extension of the theory to the alternating gradient circular
 structure has been worked out essentially retrieving the driving term
 of Hagedorn/Schoch~\cite{Hagedorn-I,Hagedorn-II,Schoch}
 Guignard~\cite{guign1,guign2}, setting then the relation between the
 fix-lines and the onset of unbounded motion.  The distributions of
 errors creates mainly an angle $\alpha$ that enters into the theory,
 but it does not otherwise change the conclusion found for the
 constant focusing case.
 
 This work will be the foundation for the study of the resonance
 crossing induced by space charge conjoint with a coupled resonance
 line.
 
 %%%%%%%%%%%%%%%%%%%%%%%%%%%%%%%%%%%%%%%%%%%%%%%%%%%%%%%%%%%%%%%%%%%%%%%%%
 \section{Acknowledgment}
 "The research leading to these results has received funding from the
 European Commission under the FP7 Research Infrastructures project EuCARD-2,
 grant agreement no.312453"

 %%%%%%%%%%%%%%%%%%%%%%%%%%%%%%%%%%%%%%%%%%%%%%%%%%%%%%%%%%%%%%%%%%%%%%%%
 \section{Appendix A: The perturbed motion around a fix-line} 
 The starting point are the equations of the perturbed system, 
 Eq.~\ref{eq:53},  that we report here for convenience 
 \begin{equation}
 \begin{aligned}
 \delta\tax' &= - \lambda\delta\tpx -2\lambda \delta\tpy,
 \\
 \delta\tay' &= -2\lambda\delta\tpx -4\lambda\delta\tpy,
 \\
 \delta\tpx' &= A_{xx} \delta\tax + A_{xy} \delta\tay,
 \\
 \delta\tpy' &= A_{yx} \delta\tax + A_{yy} \delta\tay,
 \\
 \end{aligned}
 \label{eq1:53}
 \end{equation}
 after some algebra they become 
 \begin{equation}
 \begin{aligned}
 \delta\tax'' =& -\omega^2\delta\tax + \lambda(A_{xy} + 2 A_{yy}) C_p,  \\
 \delta\tay'' =& -\omega^2\delta\tay - \lambda(A_{xx} + 2 A_{yx}) C_p,  \\
 \delta\tpx'' =& -\lambda(A_{xx} + 2 A_{xy})(\delta\tpx +2\delta\tpy),  \\
 \delta\tpy'' =& -\lambda(A_{yx} + 2 A_{yy})(\delta\tpx +2\delta\tpy),  \\
 \end{aligned}
 \label{eq1:54}
 \end{equation}
 where we defined
 \begin{equation}
   \omega^2=\lambda(A_{xx} + 2 A_{yx} + 2A_{xy} + 4 A_{yy}).
 \label{eq1:55}
 \end{equation}
 The coefficient $C_p$ is a constant obtained from integrating 
 Eqs.~\ref{eq1:53} 
 \begin{equation}
   2\delta\tax = \delta\tay + C_p.
 \label{eq1:56}
 \end{equation}
 From equation Eq.~\ref{eq1:54} we also find
 \begin{equation}
   (\delta\tpx+2\delta\tpy)'' = -\omega^2(\delta\tpx+2\delta\tpy).
 \label{eq1:57}
 \end{equation}
 The general conclusion is that for
 \begin{equation}
   \omega^2 > 0
 \label{eq1:58}
 \end{equation}
 the quantity $\delta\tpx+2\delta\tpy$ is stable and oscillate 
 around zero as well as $\delta\tax$, and $\delta\tay$. 
 We now consider the case $\omega^2>0$. 
 Integrating Eq.~\ref{eq1:57} we find 
 \begin{equation}
  \delta\tpx+2\delta\tpy = A\cos(\omega\theta + \xi) 
 \label{eq1:1}
 \end{equation}
 with $A,\xi$ some constant. 
 Using this result in  Eq.~\ref{eq1:54} we find 
 \begin{equation}
 \begin{aligned}
 \delta\tpx'' =& -\lambda(A_{xx} + 2 A_{xy})A\cos(\omega\theta + \xi),  \\
 \delta\tpy'' =& -\lambda(A_{yx} + 2 A_{yy})A\cos(\omega\theta + \xi).  \\
 \end{aligned}
 \label{eq1:2}
 \end{equation}
 The integration of these equations yields 
 \begin{equation}
 \begin{aligned}
 \delta\tpx =&  \lambda(A_{xx} + 2 A_{xy})
                \frac{A}{\omega^2}\cos(\omega\theta + \xi) 
                + C_x\theta + D_x,  \\
 \delta\tpy =&  \lambda(A_{yx} + 2 A_{yy})
                \frac{A}{\omega^2}\cos(\omega\theta + \xi) 
                + C_y\theta + D_y,  \\
 \end{aligned}
 \label{eq1:3}
 \end{equation}
 with $C_x,C_y,D_x,D_y$ the integration constants to be determined
 later. The first two equations of Eq~\ref{eq1:54} can be integrated
 as well, we find 
 \begin{equation}
 \begin{aligned}
 \delta\tax =& B_x\sin(\omega\theta+\tau_x) 
                 + \frac{\lambda}{\omega^2}(A_{xy} + 2 A_{yy}) C_p,  \\
 \delta\tay =& B_y\sin(\omega\theta+\tau_y)
                 - \frac{\lambda}{\omega^2}(A_{xx} + 2 A_{yx}) C_p,  \\
 \end{aligned}
 \label{eq1:4}
 \end{equation}
with integration constants $B_x,B_y,\tau_x,\tau_y$. 
The equations~\ref{eq1:3}, and~\ref{eq1:4} are obtained from 
Eq.~\ref{eq1:54}, but these solutions must be consistent with 
Eq.~\ref{eq1:53}. We next find the constraint imposed by this condition. 

We use the first two equations of Eq.~\ref{eq1:53}
 \begin{equation}
 \begin{aligned}
 \delta\tax' &= - \lambda\delta\tpx -2\lambda \delta\tpy,
 \\
 \delta\tay' &= -2\lambda\delta\tpx -4\lambda\delta\tpy,
 \\
 \end{aligned}
 \label{eq1:5}
 \end{equation}
and substitute Eq.~\ref{eq1:4} and Eq.~\ref{eq1:1}. 
We find 
 \begin{equation}
 \begin{aligned}
   B_x\omega\cos(\omega\theta+\tau_x) = - \lambda A\cos(\omega\theta + \xi)
 \\
   B_y\omega\cos(\omega\theta+\tau_y) = - 2\lambda A\cos(\omega\theta + \xi)
 \\
 \end{aligned}
 \label{eq1:6}
 \end{equation}
therefore it must be $\tau_x=\tau_y=\xi$, 
$B_x=-\lambda A/\omega$, and $B_y=-2\lambda A/\omega$. 
The solution becomes 
 \begin{equation}
 \begin{aligned}
 \delta\tax =& -\frac{\lambda A}{\omega}\sin(\omega\theta+\xi) 
                 + \frac{\lambda}{\omega^2}(A_{xy} + 2 A_{yy}) C_p,  \\
 \delta\tay =& -\frac{2\lambda A}{\omega}\sin(\omega\theta+\xi)
                 - \frac{\lambda}{\omega^2}(A_{xx} + 2 A_{yx}) C_p,  \\
 \delta\tpx =&  \lambda(A_{xx} + 2 A_{xy})
                \frac{A}{\omega^2}\cos(\omega\theta + \xi) 
                + C_x\theta + D_x,  \\
 \delta\tpy =&  \lambda(A_{yx} + 2 A_{yy})
                \frac{A}{\omega^2}\cos(\omega\theta + \xi) 
                + C_y\theta + D_y,  \\
 \end{aligned}
 \label{eq1:7}
 \end{equation}
 Now we use the last two equations of Eqs.~\ref{eq1:53}
 \begin{equation}
 \begin{aligned}
 \delta\tpx' &= A_{xx} \delta\tax + A_{xy} \delta\tay,
 \\
 \delta\tpy' &= A_{yx} \delta\tax + A_{yy} \delta\tay,
 \\
 \end{aligned}
 \label{eq1:8}
 \end{equation}
and substituting the Eqs.~\ref{eq1:7} we find 
 \begin{equation}
 \begin{aligned}
                C_x  
&= 
2\frac{\lambda}{\omega^2} C_p ( A_{xx}A_{yy} - A_{xy}A_{yx}  ), 
 \\
                C_y 
&= 
-\frac{\lambda}{\omega^2} C_p ( A_{yy}A_{xx} - A_{yx}A_{xy}  ). 
 \\
 \end{aligned}
 \label{eq1:10}
 \end{equation}
Therefore the solutions take the form 
 \begin{equation}
 \begin{aligned}
 \delta\tax =& -\frac{\lambda A}{\omega}\sin(\omega\theta+\xi) 
               + \frac{\lambda}{\omega^2}(A_{xy} + 2 A_{yy}) C_p,  \\
 \delta\tay =& -\frac{2\lambda A}{\omega}\sin(\omega\theta+\xi)
                 - \frac{\lambda}{\omega^2}(A_{xx} + 2 A_{yx}) C_p,  \\
 \delta\tpx =&  \lambda(A_{xx} + 2 A_{xy})
                \frac{A}{\omega^2}\cos(\omega\theta + \xi) \\
             & + 2\frac{\lambda}{\omega^2} C_p ( A_{xx}A_{yy} - A_{xy}A_{yx} )
                  \theta + D_x,  \\
 \delta\tpy =&  \lambda(A_{yx} + 2 A_{yy})
                \frac{A}{\omega^2}\cos(\omega\theta + \xi) \\
             & -\frac{\lambda}{\omega^2} C_p ( A_{xx}A_{yy} - A_{xy}A_{yx} )
                  \theta + D_y.  \\
 \end{aligned}
 \label{eq1:11}
 \end{equation}
As 
 \begin{equation}
  \delta\tpx+2\delta\tpy = A\cos(\omega\theta + \xi) 
 \label{eq1:12}
 \end{equation}
substituting we find 
 \begin{equation}
  D_x + 2D_y = 0
 \label{eq1:13}
 \end{equation}
Therefore the general evolution of a perturbation around a fix-line point 
takes the form 
 \begin{equation}
 \begin{aligned}
 \delta\tax =& -\frac{\lambda A}{\omega}\sin(\omega\theta+\xi) 
                 + \frac{\lambda}{\omega^2}(A_{xy} + 2 A_{yy}) C_p,  \\
 \delta\tay =& -\frac{2\lambda A}{\omega}\sin(\omega\theta+\xi)
                 - \frac{\lambda}{\omega^2}(A_{xx} + 2 A_{yx}) C_p,  \\
 \delta\tpx =&  \lambda(A_{xx} + 2 A_{xy})
                \frac{A}{\omega^2}\cos(\omega\theta + \xi) \\
             & + 2\frac{\lambda}{\omega^2} C_p ( A_{xx}A_{yy} -  A_{xy}A_{yx}  )
                  \theta -2 D_y,  \\
 \delta\tpy =&  \lambda(A_{yx} + 2 A_{yy})
                \frac{A}{\omega^2}\cos(\omega\theta + \xi) \\
             & -\frac{\lambda}{\omega^2} C_p ( A_{xx}A_{yy} - A_{xy}A_{yx} )
                  \theta + D_y.  \\
 \end{aligned}
 \label{eq1:14}
 \end{equation}
The free parameters $A,\xi,C_p,D_y$ are found from Eq.~\ref{eq1:14} 
at $\theta=0$ using the initial conditions 
 \begin{equation}
 \begin{aligned}
 (\delta\tax)_0 =& -\frac{\lambda A}{\omega}\sin(\xi) 
                 + \frac{\lambda}{\omega^2}(A_{xy} + 2 A_{yy}) C_p,  \\
 (\delta\tay)_0 =& -\frac{2\lambda A}{\omega}\sin(\xi)
                 - \frac{\lambda}{\omega^2}(A_{xx} + 2 A_{yx}) C_p,  \\
 (\delta\tpx)_0 =&  \lambda(A_{xx} + 2 A_{xy})
                \frac{A}{\omega^2}\cos(\xi) 
                -2 D_y,  \\
 (\delta\tpy)_0 =&  \lambda(A_{yx} + 2 A_{yy})
                \frac{A}{\omega^2}\cos(\xi) 
                + D_y.  \\
 \end{aligned}
 \label{eq1:15}
 \end{equation}

%%%%%%%%%%%%%%%%%%%%%%%%%%%%%%%%%%%%%%%%%%%%%%%%%%%%%%%%%%%%%%%%%%%%%%%%%%
\subsection{Discussion}

Suppose we consider a fix-line defined by $\tax,\tay$. 
This means that a particle on this fix-line has constant $\tax,\tay$, 
while its phases $\tpx,\tpy$ are also constant, and satisfy the 
condition $\tpx+ 2\tpy = \pi M$. 
Taking any other pair of $\tpx,\tpy$ that satisfy $\tpx+ 2\tpy = \pi M$ 
simply means to take another point on the fix-line, but again 
$\tpx,\tpy$ remain constant. 
Therefore any phases $\tpx,\tpy$ satisfying $\tpx+ 2\tpy = \pi M$ 
are equally good.

By saying that $\tax,\tay$ identify a fix-line we intend 
the identification of the fix-line without any reference to the phases 
of a particle on it. 

Consider now at $\theta=0$ a point of coordinates $\tax,\tay,\tpx,\tpy$ 
lying on the fix-line identified by $\tax,\tay$. 
Consider now a perturbation $\delta\tax,\delta\tay,\delta\tpx,\delta\tpy$ 
from that point, and consider the particle with coordinates 
\begin{equation}
  \tax+\delta\tax,
  \tay+\delta\tay,
  \tpx+\delta\tpx,
  \tpy+\delta\tpy. 
  \label{eq1:21}
\end{equation}
The evolution of the coordinates of the particle is given by 
$
  \tax  = const, 
  \tay  = const,
  \tpx = const, 
  \tpy = const, 
$
and by the evolution of the perturbation 
$\delta\tax,\delta\tay,\delta\tpx,\delta\tpy$,  
which is given by the equations Eq.~\ref{eq1:14}. 
We note that if 
$
   C_p = 2(\delta\tax)_0 - (\delta\tay)_0 \ne 0, 
$
the oscillation of the perturbation $\delta\tax,\delta\tay$ 
is not around $\tax,\tay$, 
while the perturbations $\delta\tpx,\delta\tpy$ are unbounded. 
The perturbative analysis seems to collapse. 
On the other hand the quantity $\delta\tpx+2\delta\tpy$ 
(Eq.~\ref{eq1:12}) remains bounded around zero! 

Lets now consider the point with coordinates 
\begin{equation}
 \begin{aligned}
  \tax^{c} =& \tax + \delta\tax^c = 
             \tax + \frac{\lambda}{\omega^2}(A_{xy} + 2 A_{yy}) C_p, \\
  \tay^{c} =& \tay + \delta\tay^c = 
             \tay - \frac{\lambda}{\omega^2}(A_{xx} + 2 A_{yx}) C_p, \\
 \tpx^c =& \tpx + \delta\tpx^c = \tpx - 2 \alpha,   \\
 \tpy^c =& \tpy + \delta\tpy^c = \tpy +   \alpha,   \\
 \end{aligned}
 \label{eq1:22} 
\end{equation}
where we defined for convenience 
$\delta\tax^c,\delta\tay^c,\delta\tpx^c,\delta\tpy^c$. 
The parameter $\alpha$ is an arbitrary parameter that defines the point.   
We observe that the coordinates $\tax+\delta\tax,\tay+\delta\tay$ 
oscillate around $\tax^c,\tay^c$. 

By construction the coordinates, 
$\tax^c,\tay^c,\tpx^c,\tpy^c$ do not change in time. 
We will prove that they identify a fix-line point of a new fix-line 
identified by $\tax^c,\tay^c$. 

We start by writing the equation of the fix-line in the general form 
\begin{equation}
  \frac{\partial \tilde H_{s1}}{\partial \tax} + 
  2\frac{\partial \tilde H_{s1}}{\partial \tay} = 0, 
  \label{eq1:24} 
\end{equation}
this equation is Eq.~\ref{eq:41} (top) expressed in terms of the slowly 
varying Hamiltonian $\tilde H_{s1}$. 
We now check if $\tax^{c},\tay^{c}$ satisfies the previous equation 
(Eq.~\ref{eq1:24}).  
We find 
\begin{widetext}
\begin{equation}
 \begin{aligned}
  \frac{\partial \tilde H_{s1}}{\partial \tax}(\tax^{c},\tay^{c}) + 
 2\frac{\partial \tilde H_{s1}}{\partial \tay}(\tax^{c},\tay^{c},) =& 
  \frac{\partial \tilde H_{s1}}{\partial \tax}(\tax,\tay) + 
  \frac{\partial^2 \tilde H_{s1}}{\partial \tax^2}(\tax,\tay) \delta\tax^c + 
  \frac{\partial^2 \tilde H_{s1}}{\partial \tax\partial\tay}(\tax,\tay) 
   \delta\tay^c + \\
  +& 
  2\frac{\partial \tilde H_{s1}}{\partial \tay}(\tax,\tay) + 
  2\frac{\partial^2 \tilde H_{s1}}{\partial \tay\partial\tax}(\tax,\tay) 
  \delta\tax^c + 
  2\frac{\partial^2 \tilde H_{s1}}{\partial \tay\partial\tay}(\tax,\tay) 
   \delta\tay^c. \\
 \end{aligned} 
 \label{eq1:25} 
\end{equation}
\end{widetext}
Taking into account that $\tax,\tay$ is a fix-line, then 
\begin{widetext}
\begin{equation}
 \begin{aligned}
  \frac{\partial \tilde H_{s1}}{\partial \tax}(\tax^{c},\tay^{c}) + 
 2\frac{\partial \tilde H_{s1}}{\partial \tay}(\tax^{c},\tay^{c},) =& 
  \frac{\partial^2 \tilde H_{s1}}{\partial \tax^2}(\tax,\tay) \delta\tax^c + 
  \frac{\partial^2 \tilde H_{s1}}{\partial \tax\partial\tay}(\tax,\tay) 
   \delta\tay^c + \\
  +& 
  2\frac{\partial^2 \tilde H_{s1}}{\partial \tay\partial\tax}(\tax,\tay) 
  \delta\tax^c + 
  2\frac{\partial^2 \tilde H_{s1}}{\partial \tay^2}(\tax,\tay) 
   \delta\tay^c. 
 \end{aligned} 
 \label{eq1:26}
\end{equation}
 \end{widetext}
Using the definitions for $A_{xx},A_{xy},A_{yx},A_{yy}$ the previous equation 
becomes 
\begin{equation}
\begin{aligned}
  & \frac{\partial \tilde H_{s1}}{\partial \tax}(\tax^{c},\tay^{c}) + 
 2\frac{\partial \tilde H_{s1}}{\partial \tay}(\tax^{c},\tay^{c},) = \\ 
 &  \frac{1}{2}
  [
  (A_{xx} + 2A_{yx}) \delta\tax^c 
+ (A_{xy} + 2A_{yy}) \delta\tay^c  ]. 
  \label{eq1:27}
\end{aligned}
\end{equation}
Now we substitute $\delta\tax^c,\delta\tay^c$ from Eq.~ \ref{eq1:22}, 
and we find 
\begin{equation}
\begin{aligned}
  & \frac{\partial \tilde H_{s1}}{\partial \tax}(\tax^{c},\tay^{c}) + 
 2\frac{\partial \tilde H_{s1}}{\partial \tay}(\tax^{c},\tay^{c},) = \\
  & \frac{1}{2} \frac{\lambda}{\omega^2} C_p
  [
  (A_{xx} + 2A_{yx}) (A_{xy} + 2 A_{yy})  \\
  & - (A_{xy} + 2A_{yy}) (A_{xx} + 2 A_{yx}) ] 
  = 0. 
  \label{eq1:28}
\end{aligned}
\end{equation}
We conclude that $\tax^c,\tay^c$ 
satisfy the equation for the fix-line, and therefore it identifies a fix-line. 
The phases $\tpx^c,\tpy^c$ also satisfy the condition of a fix-line, 
in fact we find 
\begin{equation}
  \tpx^c+2\tpy^c = 
  \tpx - 2 \alpha + 2 \tpy + 2 \alpha = \tpx + 2 \tpy = 
  \pi M. 
  \label{eq1:31a}
\end{equation}
Therefore $\tax^c,\tay^c,\tpx^c,\tpy^c$ is a point on another fix-line 
identified by $\tax^c,\tay^c$.

The initial perturbation with respect to this new point belonging to 
another fix-line reads 
\begin{equation}
\begin{aligned}
    (\delta\tax)_1 =& \tax + (\delta\tax)_0 - \tax^{c} = 
     (\delta\tax)_0-  \frac{\lambda}{\omega^2}(A_{xy} + 2 A_{yy}) C_p,\\ 
    (\delta\tay)_1 =& \tay + (\delta\tax)_0 - \tay^{c} = 
     (\delta\tax)_0  + \frac{\lambda}{\omega^2}(A_{xx} + 2 A_{yx}) C_p, \\
    (\delta\tpx)_1 =& \tpx + (\delta\tpx)_0 - \tpx^{c} = 
     (\delta\tpx)_0 + 2 \alpha,\\ 
    (\delta\tpy)_1 =& \tpy + (\delta\tpx)_0 - \tpy^{c} = 
     (\delta\tpx)_0 - \alpha.\\
\end{aligned}
\label{eq1:31b} 
\end{equation}
We use the index $1$ to denote the initial condition of the 
original perturbation now re-written with respect to the new fix-line. 
Therefore 
\begin{equation}
 C'_p =   2(\delta\tax)_1 - (\delta\tay)_1 = 0 
  \label{eq1:32}
\end{equation}
This means that the initial perturbation with respect to the point 
belonging to the new fix-line 
has the correspondent $C'_p$ equal to zero, and the evolution of the initial 
perturbation with respect to the new point reads 
\begin{equation}
 \begin{aligned}
 \delta\tax =& -\frac{\lambda' A'}{\omega'}\sin(\omega'\theta+\xi'),  \\
 \delta\tay =& -\frac{2\lambda' A'}{\omega'}\sin(\omega'\theta+\xi'),  \\
 \delta\tpx =&  \lambda'(A'_{xx} + 2 A'_{xy})
                \frac{A'}{\omega^{'2}}\cos(\omega'\theta + \xi') -2 D'_y,  \\
 \delta\tpy =&  \lambda'(A'_{xy} + 2 A'_{yy})
                \frac{A'}{\omega^{'2}}\cos(\omega'\theta + \xi') + D'_y,  \\
 \end{aligned}
 \label{eq1:33} 
\end{equation}
where now $A'_{xx},A'_{xy},A'_{yx},A'_{yy},\lambda',\omega'$ 
are evaluated at $\tax^c,\tay^c$, 
and the constants $A',\xi',D'_y$ are computed from the initial condition 
$(\delta\tax)_1,(\delta\tay)_1$, $(\delta\tpx)_1,(\delta\tpy)_1$. 
We find 
 \begin{equation}
 \begin{aligned}
 (\delta\tax)_1 =& -\frac{\lambda' A'}{\omega'}\sin(\xi'),  \\
 (\delta\tay)_1 =& -\frac{2\lambda' A'}{\omega'}\sin(\xi'),  \\
 (\delta\tpx)_1 =&  \lambda'(A'_{xx} + 2 A'_{xy})
                \frac{A'}{\omega^{'2}}\cos(\xi') -2 D'_y,  \\
 (\delta\tpy)_1 =&  \lambda'(A'_{xy} + 2 A'_{yy})
                \frac{A'}{\omega^{'2}}\cos(\xi') + D'_y.  \\
 \end{aligned}
 \label{eq1:34} 
 \end{equation}
Now we use the definition of 
$(\delta\tax)_1, (\delta\tay)_1$, $(\delta\tpx)_1, (\delta\tpy)_1$ and 
get 
\begin{equation} 
\begin{aligned}
    (\delta\tax)_1 =&  -\frac{\lambda A}{\omega}\sin(\xi), \\
    (\delta\tay)_1 =& -\frac{2\lambda A}{\omega}\sin(\xi), \\
    (\delta\tpx)_1 =&  
   \lambda(A_{xx} + 2 A_{xy}) \frac{A}{\omega^2}\cos(\xi) -2 D_y + 2 \alpha, \\
    (\delta\tpy)_1 =&  
   \lambda(A_{xy} + 2 A_{yy}) \frac{A}{\omega^2}\cos(\xi) + D_y -\alpha.  \\
\end{aligned}
\label{eq1:35}
\end{equation} 
Substituting Eq.~\ref{eq1:35} into Eq.~\ref{eq1:34} we find that 
 \begin{equation}
 \begin{aligned}
    & -\frac{\lambda A}{\omega}\sin(\xi) =  
    -\frac{\lambda' A'}{\omega'}\sin(\xi'),\\
 & \lambda(A_{xx} + 2 A_{xy}) \frac{A}{\omega^2}\cos(\xi) -2 D_y + 2 \alpha
 = \\  
 & =\lambda'(A'_{xx} + 2 A'_{xy})\frac{A'}{\omega^{'2}}\cos(\xi') -2 D'_y,  \\
 &\lambda(A_{xy} + 2 A_{yy}) \frac{A}{\omega^2}\cos(\xi) + D_y -\alpha
 = \\
 & = \lambda'(A'_{xy} + 2 A'_{yy}) \frac{A'}{\omega^{'2}}\cos(\xi') + D'_y. \\
 \end{aligned}
 \label{eq1:37}
 \end{equation}
The parameters on the L.H.S. are determined by the 
perturbation with respect to the original fix-line point. 
The parameters on the R.H.S $A',\xi', D'_y$ are 
the new parameters that identify the 
original perturbation with respect to the new fix-line point. 
Multiplying the last by 2 and summing with the second equation we find 
\begin{equation}
  A\cos(\xi) = A'\cos(\xi'), 
  \label{eq1:40}
\end{equation} 
which combined with the first of Eq.~\ref{eq1:37} yields 
\begin{equation}
\left\{
\begin{aligned}
  A'\cos(\xi')   =& A\cos(\xi) \\
  A'\sin(\xi') =& \frac{\lambda \omega'}{\lambda'\omega}A\sin(\xi). \\
\end{aligned} 
\right.
  \label{eq1:38a}
\end{equation} 
From this system we obtain $A',\xi'$. 

Now we use the free parameter $\alpha$ to require that $D'_y=0$. 
We take the second equation of Eq.~\ref{eq1:37} and we find 
\begin{equation}
 \alpha = 
  \lambda(A_{xy} + 2 A_{yy}) \frac{A}{\omega^2}\cos(\xi) + D_y 
- \lambda'(A'_{xy} + 2 A'_{yy}) \frac{A'}{\omega^{'2}}\cos(\xi')
  \label{eq1:38b} 
\end{equation} 
substituting this value into the first equation of Eq.~ \ref{eq1:37} 
we find 
\begin{equation}
   - 2 D'_y = A \cos(\xi) - A' \cos(\xi') = 0. 
   \label{eq1:39}
\end{equation}
Therefore the parameter $\alpha$ defined in Eq.~\ref{eq1:38b} 
yields $D'_y=0$ in both second and third equations of Eq.~\ref{eq1:37}.

We conclude that the original perturbation around the 
new fix-line point identified by $\alpha$ with the value 
in Eq.~\ref{eq1:38b} evolves according to 
 \begin{equation}
 \begin{aligned}
 \delta\tax =& -\frac{\lambda' A'}{\omega'}\sin(\omega'\theta+\xi'),  \\
 \delta\tay =& -\frac{2\lambda' A'}{\omega'}\sin(\omega'\theta+\xi'),  \\
 \delta\tpx =&  \lambda'(A'_{xx} + 2 A'_{xy})
                \frac{A'}{\omega^{'2}}\cos(\omega'\theta + \xi'),  \\
 \delta\tpy =&  \lambda'(A'_{yx} + 2 A'_{yy})
                \frac{A'}{\omega^{'2}}\cos(\omega'\theta + \xi').  \\
 \end{aligned}
 \label{eq1:41} 
 \end{equation}
That is, the initial perturbation is oscillating around the point 
$\tax^c,\tay^c,\tpx^c,\tpy^c$ of the new fix-line identified by 
$\tax^c,\tay^c$. 
In fact, in Eq.~\ref{eq1:14} we simply replace all constant with 
the primed ones and set $C_p=0$. 
This results let's us proceed with the most simple case, i.e. with 
$C_p=0$.

 %%%%%%%%%%%%%%%%%%%%%%%%%%%%%%%%%%%%%%%%%%%%%%%%%%%%%%%%%%%%%%%%%%%%%%%%
 \section{Appendix B: Properties of the level lines}
 We address here the properties of the level lines as defined by 
 Eq.~\ref{eq:88}. 
 We distinguish the following cases.
 
 %%%%%%%%%%%%%%%%%%%%%%%%%%%%%%%%%%%%%%%%%%%%%%%%%%%%%%%%%%%%%%%%%%%%%%%5
 \subsection{$\xi>0$}
\begin{figure}[h]
 \begin{center}
 \unitlength 0.9mm
 \begin{picture}(80,80)
 \put( 0,0) {\epsfig{file=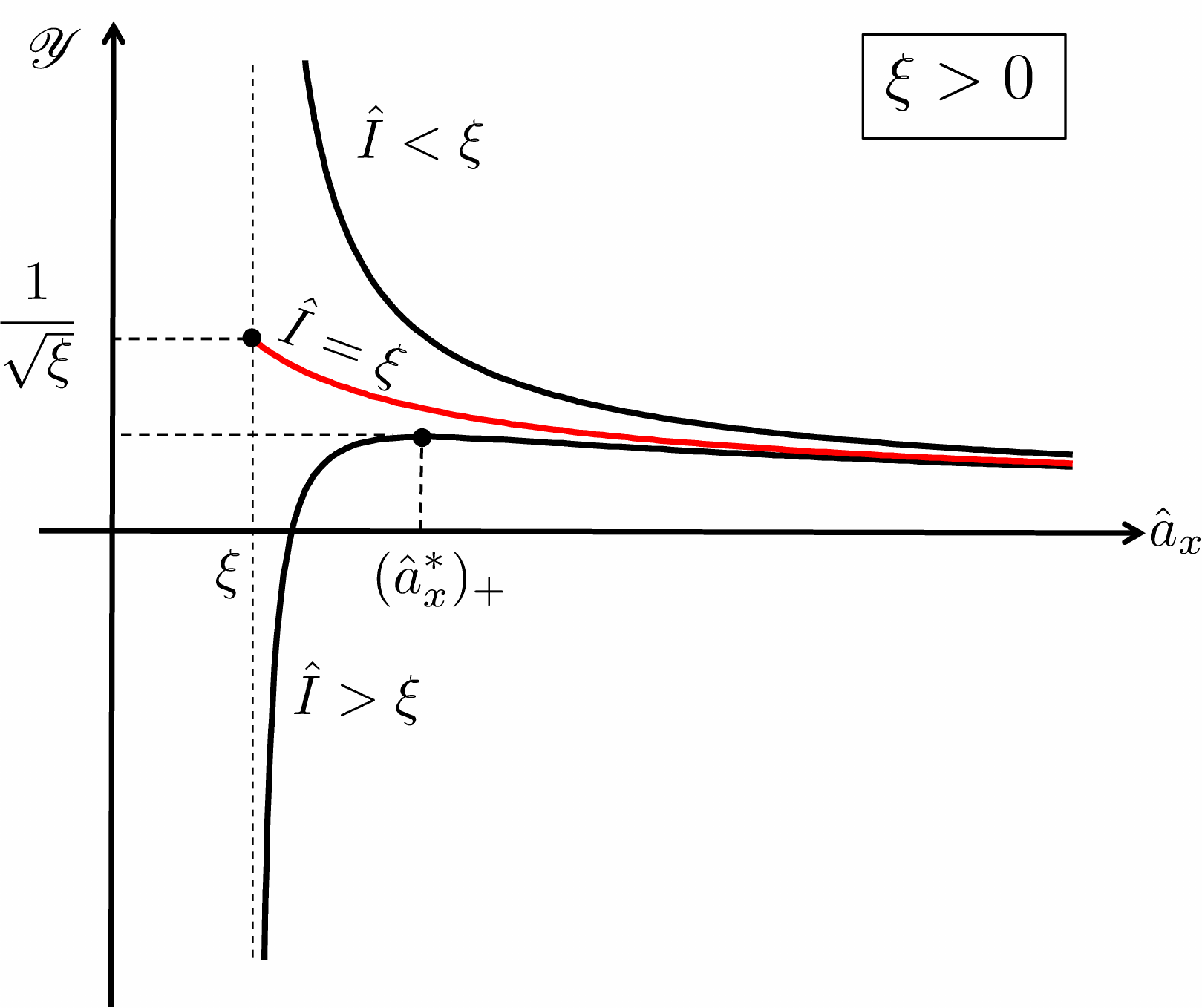,width=80mm}}
 \end{picture}
 \caption{
 Summary of the general behaviour of the level curves in the case
 $\xi>0$.
 }
 \label{fig1:fig_26}
 \end{center}
 \end{figure}

 The level lines have the following properties:
 
 \begin{itemize}
 \item[1)]
 The range of $\hax$ is $\xi < \hax<\infty$, because $\hay \ge 0$, 
 and at $\hax=\xi$ the function $\ya$ is not defined. 
 
 \item[2)]
 For $\hax\rightarrow +\infty$ we always find $\ya(\hax)\rightarrow 0$.
 
 \item[3)]
 Given $\epsilon$ a positive, very small number, we find that 
 for $\hax=\xi+\epsilon$ the following approximation holds 
 \begin{equation}
   \ya(\xi+\epsilon) = -\frac{\hat I - \xi}{\sqrt{\xi}\epsilon}. 
 \label{eq:159}
 \end{equation}
 Therefore the asymptotic behaviour of the curve $\ya$ for 
 $\hax\rightarrow\xi^+$ changes according whether $\hat I \gtreqless \xi$. 
 We find 
 \begin{equation}
 \hax\rightarrow\xi^+ 
 \Longrightarrow \quad 
 \left\{
 \begin{aligned}
  \ya \rightarrow  -\infty & \quad\textrm{for} & \hat I > \xi \\
  \ya \rightarrow  +\infty & \quad\textrm{for} & \hat I < \xi \\
 \end{aligned}
 \right.
 \label{eq:160}
 \end{equation}
 If $\hat I = \xi$ then $\ya$ takes the special form 
 \begin{equation}
   \ya(\hax) =  \frac{1}{\sqrt{\hax}}, 
 \label{eq:161}
 \end{equation}
 and for $\hax\rightarrow \xi^+$ we find 
 $\ya \rightarrow\ya = \frac{1}{\sqrt{\xi}}$. 
 Therefore the level line for $\hat I=\xi$ separates the level lines 
 into two classes of curves: one class where the curves diverge to $+\infty$ 
 when $\hax\rightarrow\xi^+$, while in the other class 
 the curves diverges to $-\infty$ when  $\hax\rightarrow\xi^+$. 
 This level line is shown by the red line in Fig.~\ref{fig1:fig_26}. 
 Other two curves for $\hat I < \xi$ and $\hat I > \xi$ show the 
 divergence property of $\ya$. 
 
 \item[4)]
 If $\hax = \hat I$ and $\hat I \ne \xi$ we always find $\ya=0$.
 Therefore for $\hat I > \xi$ the level curve identified by $\hat I$
 will always cross $\ya=0$.
 
 \item[5)]
 The curves identified by the invariant $\hat I$ can have:
 1) no maximum and no minimum;
 2) one maximum;
 3) both maximum and  minimum.
 
 The solution of $\ya'(\hax^*)=0$ for a level line defined by $\hat I$
 satisfies the relation
 \begin{equation}
   \hat I = \frac{(\hax^*)^2+\hax^*\xi}{3\hax^*-\xi}
 \label{eq:162}
 \end{equation}
 and the value of $\ya(\hax^*)$ is given by
 \begin{equation}
   \ya(\hax^*) = \frac{2\sqrt{\hax^*}}{3\hax^* - \xi}. 
 \label{eq:163}
 \end{equation}

 \item[6)]
 We observe that
 \begin{equation}
 \ya(\xi+\epsilon) \ya(\hax^*) 
 = 
 \left\{
 \begin{aligned}
  < 0 & \quad\textrm{for} & \hat I > \xi \\
  > 0 & \quad\textrm{for} & \hat I < \xi \\
 \end{aligned}
 \right.
 \label{eq:164}
 \end{equation}
 That means that for $\hat I > \xi$ the point $\ya(\hax^*)$ is always
 on the opposite plane of where $\ya$ diverges.  This situation is
 shown in Fig.~\ref{fig1:fig_26}.  The level line below the red curve
 has a maximum in $(\hax^*)_+$, of value $\ya((\hax^*)_+)$.  The point
 $(\hax^*)_+, \ya((\hax^*)_+)$, is located in the half plane $\ya
 >0$. In the other half the curve diverges for $\hax\rightarrow\xi^+$.

 \item[7)]
 The points of maximum or minimum are obtained by the equation 
 $\ya'(\hax) =0$, which reads 
 \begin{equation}
   \hax^2 + (\xi-3\hat I)\hax + \hat I\xi =0, 
 \label{eq:165}
 \end{equation}
 and the values of the invariant for which the level curves 
 can have a  maximum or minimum are
 \begin{equation}
     \xi \ge 9 \hat I  \quad \textrm{or} \quad
    \xi \le \hat I. 
 \label{eq:166}
 \end{equation}
 
 \item[8)]
 If $\hat I$ is in the ranges of  Eqs.~\ref{eq:166} 
 the two solutions of Eq.~\ref{eq:165} are
 \begin{equation}
   (\hax^*)_{\pm} =
   \frac{-\xi+3\hat I \pm \sqrt{(\xi - 3\hat I)^2 - 4\hat I\xi}}{2}. 
 \label{eq:167}
 \end{equation}
 We next have to control which of these solutions is consistent 
 with the point 1). 
 That defines the criteria if the solution is acceptable or not, 
 (i.e. the criteria Eq.~\ref{eq:86}). 
 We accept only the solution satisfying $\xi < \hax < \infty$. 
 According to the values of $\hat I$, and $\xi$ the solution of 
 Eq.~\ref{eq:167} have the following properties
 \begin{equation}
 \hat I \le \frac{\xi}{9} \Longrightarrow
 \left\{
 \begin{aligned}
 (\hax^*)_+ &< \xi & \textrm{ not acceptable} \\
 (\hax^*)_- &<   0 & \textrm{ not acceptable} \\
 \end{aligned}
 \right.
 \label{eq:168}
 \end{equation}
 \begin{equation}
 \hat I \ge \xi \Longrightarrow
 \left\{
 \begin{aligned}
 (\hax^*)_+ & >   \xi & \textrm{ acceptable} \\
 (\hax^*)_+ & =   \xi & \textrm{ not acceptable} \\
 (\hax^*)_- & \le \xi & \textrm{ not acceptable} \\
 \end{aligned}
 \right.
 \label{eq:169}
 \end{equation}

 \end{itemize}
 
 We conclude that the level lines in the region $\hat I > \xi$
 have a maximum at $(\hax^*)_+$ and a minimum at $\hax \rightarrow \infty$.
 For $\hat I \le \xi$ the level curves have no maximum.
 The general properties of the level lines for $\xi > 0$ are summarized
 in the Fig.~\ref{fig1:fig_26}.
 The red curve separates all the level lines into two classes:
 one class of lines without any maximum, and a second class of
 lines with a maximum.
 %
 %
 %

 %%%%%%%%%%%%%%%%%%%%%%%%%%%%%%%%%%%%%%%%%%%%%%%%%%%%%%%%%%%%%%%%%%%%%%%5
 \subsection{$\xi=0$}
\begin{figure}[H]
 \begin{center}
 \unitlength 0.9mm
 \begin{picture}(80,80)
 \put( 0, 0) {\epsfig{file=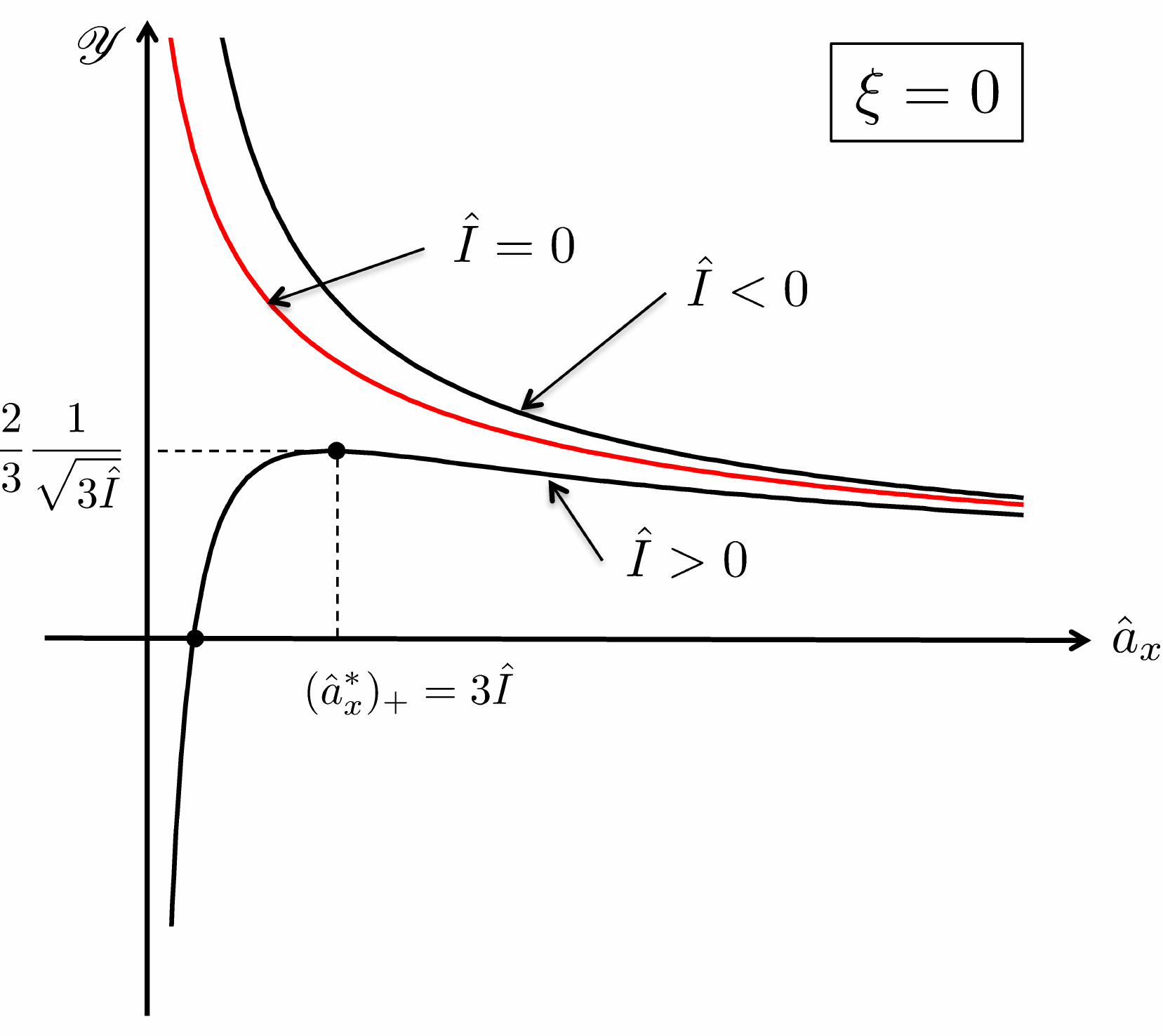,width=80mm}}
 \end{picture}
 \caption{
 Summary of the general behaviour of the level curves in the case
 $\xi=0$.
 }
 \label{fig1:fig_28}
 \end{center}
 \end{figure}

 The level lines have the following properties:
 
 \begin{itemize}
 \item[1)]
 The range of $\hax$ is $0 < \hax$, because on $\hax=0$ the function 
 $\ya$ is not defined. 
 
 \item[2)]
 For $\hax\rightarrow +\infty$ we consistently find
 $\ya(\hax)\rightarrow 0$.
 
 \item[3)]
 Let's study $\ya$ in proximity of $\hax=0$.  We take $\epsilon$ an
 arbitrarily small, positive number.  For $\hax=\epsilon$ we find
 \begin{equation}
   \ya(\epsilon) = -\frac{\hat I}{\epsilon^{3/2}}. 
 \label{eq:177}
 \end{equation}
 with $\hat I \ne 0$. 
 The curves defined by $\hat I$ are distinguished into two 
 classes of behaviour for the limit $\hax\rightarrow 0^+$ according to the 
 value of $\hat I$: 
 \begin{equation}
 \hax\rightarrow 0^+ 
 \Longrightarrow 
 \left\{
 \begin{aligned}
 \ya \rightarrow -\infty & \quad\textrm{for} & \hat I > 0 \\
 \ya \rightarrow +\infty & \quad\textrm{for} & \hat I < 0 \\
 \end{aligned}
 \right..
 \label{eq:178}
 \end{equation}
 If $\hat I = 0$ the behaviour of $\ya$ is different, in fact 
 $\ya$ takes the form 
 \begin{equation}
   \ya(\hax) =  \frac{1}{\sqrt{\hax}}, 
 \label{eq:179}
 \end{equation}
 and for $\hax\rightarrow 0^+$ we find $\ya \rightarrow \infty$. 
 This particular level line is shown in Fig.~\ref{fig1:fig_28} 
 by the red curve. The curves below the red curve diverge to $-\infty$ 
 for $\hax\rightarrow 0^+$, while the curves above the red curve will 
 diverge to $+\infty$ for $\hax\rightarrow 0^+$. 
 
 \item[4)]
 If $\hax = \hat I$ then $\ya=0$.
 Therefore for $\hat I > 0$ the level curve identified by $\hat I$
 will definitely cross $\ya=0$. 
 
 \item[5)]
 The equation of $\hax^*$, Eq.~\ref{eq:165}, becomes now simply 
 $(\hax^*)_\pm=3\hat I (1 \pm |\hat I|/\hat I)/2$, with solutions for
 $-\infty < \hat I < \infty$.

 \item[6)]
 Now we check if the solutions of Eq.~\ref{eq:165}, i.e. of the 
 point 5) satisfy the
 acceptance criteria expressed in Eq.~\ref{eq:86}, which in 
 1) reads $0 < \hax$. 
 According to the value of $\hat I$ we find the following cases 
 \begin{equation}
 \hat I < 0 \Longrightarrow
 \left\{
 \begin{aligned}
 (\hax^*)_+ &  = 0       & \textrm{not acceptable} \\
 (\hax^*)_- &  = 3\hat I & \textrm{not acceptable} \\
 \end{aligned}
 \right.
 \label{eq:180}
 \end{equation}

  \begin{equation}
 \hat I > 0 \Longrightarrow
 \left\{
 \begin{aligned}
 (\hax^*)_+ & = 3\hat I & \textrm{ acceptable} \\
 (\hax^*)_- & = 0       & \textrm{not acceptable} \\
 \end{aligned}
 \right.
 \label{eq:181}
 \end{equation}

  \begin{equation}
 \hat I = 0 \Longrightarrow
 \left\{
 \begin{aligned}
 (\hax^*)_+ & = 0       & \textrm{not acceptable} \\
 (\hax^*)_- & = 0       & \textrm{not acceptable} \\
 \end{aligned}
 \right.
 \label{eq:182}
 \end{equation}
 
 \end{itemize}

 We conclude that in the region $\hat I < 0$ the level curves have no
 maximum; in the interval $\hat I > 0$ the level curves have a maximum
 at $(\hax^*)_+ = 3\hat I$ which value is
 $\frac{2}{3}\frac{1}{\sqrt{3\hat I}}$.  The description of this
 behaviour is shown in Fig.~\ref{fig1:fig_28}.
 
 %%%%%%%%%%%%%%%%%%%%%%%%%%%%%%%%%%%%%%%%%%%%%%%%%%%%%%%%%%%%%%%%%%%%%%%5
 \subsection{$\xi<0$}
\begin{figure}[H]
 \begin{center}
 \unitlength 0.9mm
 \begin{picture}(80,80)
 \put( 0, 0) {\epsfig{file=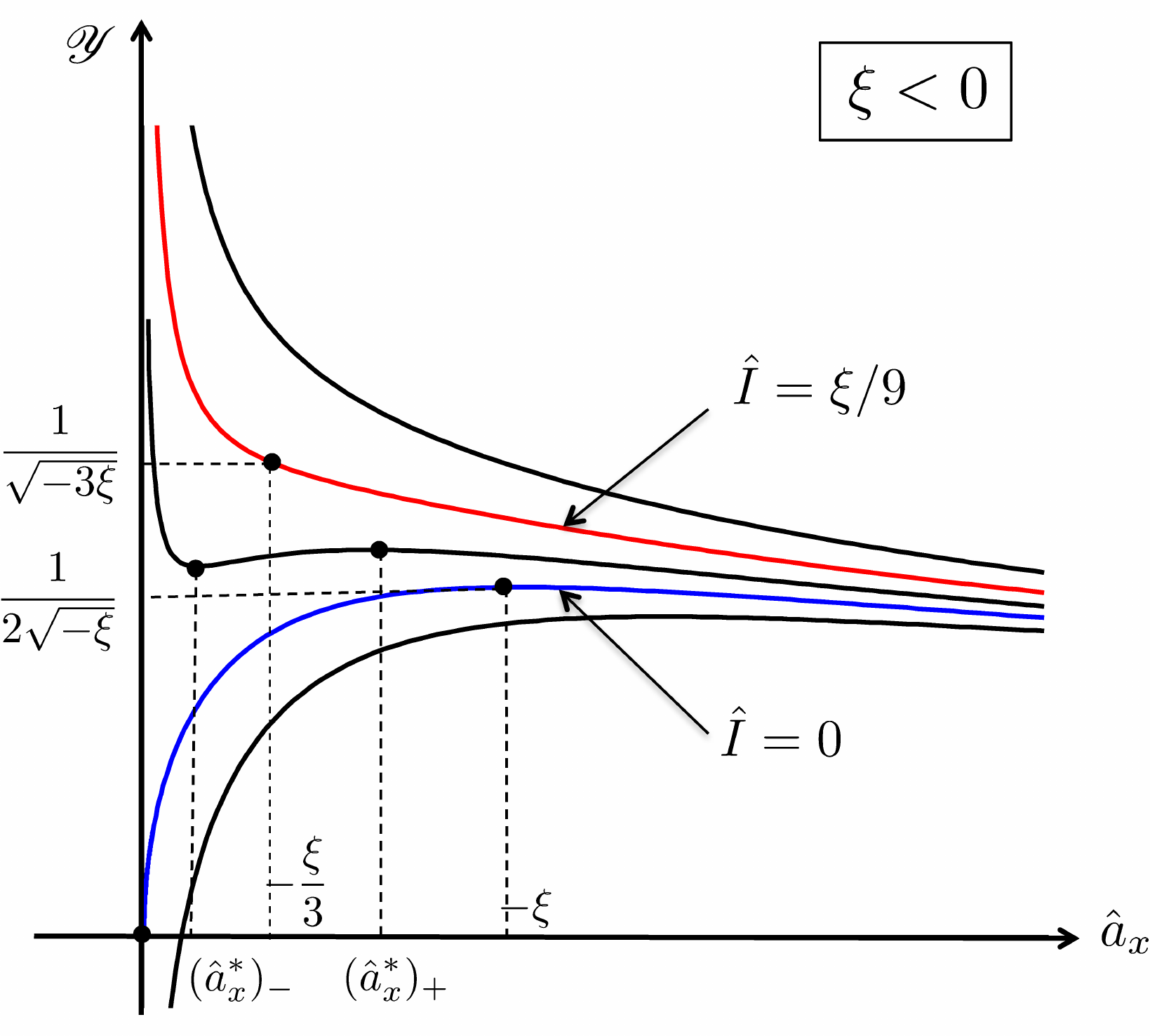,width=80mm}}
 \end{picture}
 \caption{
 Summary of the general behaviour of the level curves in the case
 $\xi<0$.
 }
 \label{fig1:fig_27}
 \end{center}
 \end{figure}

 The level lines have the following properties:
 
 \begin{itemize}
 \item[1)]
 The range of $\hax$ is $0 < \hax$, because at $\hax=0$ the function 
 $\ya$ is not defined. 
 
 \item[2)]
 For $\hax\rightarrow +\infty$ we find $\ya(\hax)\rightarrow 0$.
 
 \item[3)]
 Let's study $\ya$ in proximity of $\hax=0$.  We take $\epsilon$ an
 arbitrarily small, positive number.  For $\hax=\epsilon$ we find
 \begin{equation}
   \ya(\epsilon) = \frac{\hat I}{\sqrt{\epsilon} \xi}. 
 \label{eq:170}
 \end{equation}
 Therefore the curves defined by $\hat I$ are distinguished into two 
 classes of behaviour for the limit $\hax\rightarrow 0^+$ according to the 
 value of $\hat I$: 
 \begin{equation}
 \hax\rightarrow 0^+ 
 \Longrightarrow 
 \left\{
 \begin{aligned}
 \ya \rightarrow -\infty & \quad\textrm{for} & \hat I > 0 \\
 \ya \rightarrow +\infty & \quad\textrm{for} & \hat I < 0 \\
 \end{aligned}
 \right..
 \label{eq:171}
 \end{equation}
 If $\hat I = 0$ the behaviour of $\ya$ is different, in fact 
 $\ya$ takes the form 
 \begin{equation}
   \ya(\hax) =  \frac{\sqrt{\hax}}{\hax-\xi}
 \label{eq:172}
 \end{equation}
 and for $\hax\rightarrow 0^+$ we find
 $\ya \rightarrow 0$. 
 This particular level line is shown in Fig.~\ref{fig1:fig_27} 
 by the blue curve. The curves below the blues curved 
 line diverges to $-\infty$ 
 for $\hax\rightarrow 0^+$, while the curves above the blue curve will 
 diverge to $+\infty$ for $\hax\rightarrow 0^+$. 
 
 \item[4)]
 If $\hax = \hat I$ then $\ya=0$.
 Therefore for $\hat I > 0$ the level curve identified by $\hat I$
 will cross $\ya=0$. 
 
 \item[5)]
 The equation of $\hax^*$, Eq.~\ref{eq:165}
 allows solutions for $\hat I$ in the range
 \begin{equation}
    \hat I \le \xi   \quad \textrm{or} \quad
    \frac{\xi}{9} \le \hat I. 
 \label{eq:173}
 \end{equation}

 \item[6)]
 Now we check if the solutions of Eq.~\ref{eq:165} satisfy the
 acceptance criteria expressed in Eq.~\ref{eq:86}, which 
 in 1) reads $\hax > 0$. 
 According to the value of $\hat I$, and $\xi$ 
 we find the following cases 
 \begin{equation}
 \hat I \le \xi \Longrightarrow
 \left\{
 \begin{aligned}
 (\hax^*)_+ &<  0 & \textrm{ not acceptable} \\
 (\hax^*)_- & \le  \xi & \textrm{ not acceptable} \\
 \end{aligned}
 \right.
 \label{eq:174}
 \end{equation}
 \begin{equation}
 \frac{\xi}{9} \le \hat I < 0 \Longrightarrow
 \left\{
 \begin{aligned}
 (\hax^*)_+ &> 0  & \textrm{ acceptable} \\
 (\hax^*)_+ & = 0 & \textrm{not acceptable} \\
 (\hax^*)_- &> 0  & \textrm{ acceptable} \\
 \end{aligned}
 \right.
 \label{eq:175}
 \end{equation}
 \begin{equation}
 0 \le \hat I  \Longrightarrow
 \left\{
 \begin{aligned}
 (\hax^*)_+ &> 0 & \textrm{ acceptable} \\
 (\hax^*)_- & \le 0 & \textrm{ not acceptable} \\
 \end{aligned}
 \right.
 \label{eq:176}
 \end{equation}
 
 \end{itemize}

 We conclude that in the region $\hat I > 0$ the level curves have
 only a maximum $(\hax^*)_+$; in the interval $\xi/9 < \hat I < 0$ the
 level curves have a minimum $(\hax^*)_-$ and a maximum $(\hax^*)_+$;
 for $\hat I < \xi/9$ the level curves have no maximum, but a minimum
 at $\hax \rightarrow \infty$.  The description of this behaviour is
 shown in Fig.~\ref{fig1:fig_27}, the red curve is the curve for $\hat
 I = \xi/9$.  In this curve the maximum and minimum overlap at
 $\hax=-\xi/3$, which is an inflection point with
 $\ya=1/\sqrt{-3\xi}$.
 
 %%%%%%%%%%%%%%%%%%%%%%%%%%%%%%%%%%%%%%%%%%%%%%%%%%%%%%%%%%%%%%%%%%%%%%%%%

 \end{document}